\documentclass
[superscriptaddress,secnumarabic,amssymb,amsmath,nobibnotes,aps,prd,nofootinbib]{revtex4}%
\usepackage{graphicx}
\usepackage{epsf}
\usepackage{bm}
\usepackage{amsmath}
\usepackage{amsfonts}
\usepackage{amssymb}
\usepackage{epstopdf}
\usepackage[colorlinks=true,             linkcolor=blue,
urlcolor=blue,            citecolor=blue]{hyperref}%
\providecommand{\U}[1]{\protect\rule{.1in}{.1in}}
\newcommand{\be}{\begin{equation}}
\newcommand{\ee}{\end{equation}}

\newcommand{\mincir}{\raise
-3.truept\hbox{\rlap{\hbox{$\sim$}}\raise4.truept\hbox{$<$}\ }}
\newcommand{\magcir}{\raise
-3.truept\hbox{\rlap{\hbox{$\sim$}}\raise4.truept\hbox{$>$}\ }}

\begin{document}

\title{Cosmological Evolution of Two-Scalar fields Cosmology in the Jordan frame}
\author{Alex Giacomini}
\email{alexgiacomini@uach.cl}
\affiliation{Instituto de Ciencias F\'{\i}sicas y Matem\'{a}ticas, Universidad Austral de
Chile, Valdivia, Chile}
\author{Genly Leon}
\email{genly.leon@ucn.cl}
\affiliation{Departamento de Matem\'aticas, Universidad Cat\'olica del Norte, Avda. Angamos
0610, Casilla 1280 Antofagasta, Chile.}
\author{Andronikos Paliathanasis}
\email{anpaliat@phys.uoa.gr}
\affiliation{Institute of Systems Science, Durban University of Technology, PO Box 1334,
Durban 4000, RSA}
\author{Supriya Pan}
\email{supriya.maths@presiuniv.ac.in}
\affiliation{Department of Mathematics, Presidency University, 86/1 College Street, Kolkata
700073, India}

\begin{abstract}
In the present article we study the cosmological evolution of a two-scalar field gravitational theory
defined in the Jordan frame. Specifically, we assume one of the scalar fields to be
minimally coupled to gravity, while the second field which is the Brans-Dicke scalar field is
nonminimally coupled to gravity and also coupled to the other scalar field. In the Einstein frame
this theory reduces to a two-scalar field theory where the two fields can
interact only in the potential term, which means that the quintom theory is recovered. The cosmological evolution is studied by analyzing the equilibrium
points of the field equations in the Jordan frame. We find that the theory can describe the cosmological evolution in large scales, while inflationary  solutions are also provided.

\end{abstract}
\maketitle

\section{Introduction}

\label{sec1}

Observational data since last twenty years \cite{dataacc1,dataacc2,data1,data2,Hinshaw:2012aka,Ade:2015xua,Aghanim:2018eyx} have consistently reported that our universe is currently passing through a phase of accelerating expansion. This has been one of the mysterious discoveries of modern cosmology that thrilled the entire scientific community with a big question mark.  Within the paradigm of standard cosmology, it is impossible to fit such accelerating phase since it demands something more in terms of some exotic component having negative pressure enabling the expansion of the universe in an accelerating manner. Usually two distinct but well known approaches are used to introduce such exotic components. Both the approaches essentially need the introduction of new terms in terms of the new degrees of freedom into the Einstein's field equations. The simplest approach to introduce such new terms  is  the modifications of the matter distribution of the universe within the context of Einstein gravity, and these new exotic components are usually dubbed as dark energy (DE) fluids
\cite{Ratra,Barrow,Linder,Copeland,Overduin,Basil,Kame,nr2,nr3,nr5}. Alternatively, one could introduce such new additional terms through the modifications of the
Einstein-Hilbert action
\cite{cl1,cl2,Brans,Buda,odin1,Ferraro,Ferraro2,pa1,pa2,mo1,mo2,mo3}. The resulting exotic terms coming from the gravity modifications are usually designated as the geometrical dark energy (GDE) fluids.
According to latest observational estimations \cite{Aghanim:2018eyx}, the percentage of such exotic dark energy or geometrical dark energy fluids whatever the universe has  within it,  is  around 68\% of its total energy budget. Hence, the investigations towards this direction is an essential part of modern cosmological research and we still are looking for an actual cosmological theory fitting the observational results.

The theory of scalar fields have played a significant role in the two different approaches
proposed by the cosmologists. When scalar fields are introduced in the
gravitational action integral, they provide new degrees of freedom which
drives the dynamics of the field equations in a way to explain the observable
phenomena and the cosmological history \cite{sot1}. Scalar fields can be
minimally or nonminimally coupled to gravity. When the scalar fields are
introduced in such a way where they are minimally coupled to gravity, then we are in Einstein's
General Relativity, where the contribution of the scalar fields is on the energy momentum tensor. The most common known minimally coupled scalar fields
are, for instance, the quintessence, phantom, quintom and Chiral cosmological models
\cite{Ratra,sigm0,chir3,ndim,anm1,re1903,quin00,deHaro:2016hpl,deHaro:2016cdm,Haro:2019gsv,Haro:2019peq}.

Alternatively, scalar fields can be introduced in the gravitational action such that they nonminimally
couple to gravity. In these models, the contribution of the
scalar field in the field equations includes the terms which correspond to the
energy momentum tensor, and in addition, dynamical terms of geometric origin. In
these models, usually, the gravitational constant $k$, is not a constant
anywhere, rather it becomes variable. A nonminimally coupled scalar field was introduced by Carl H. Brans and Robert H. Dicke to provide a gravitational theory which
satisfies the \textit{Machian Principle~}\cite{Brans}. Brans-Dicke theory is a special case of the so called scalar-tensor theory \cite{fbook}. A main
characteristic of the Brans-Dicke field is the parameter $\omega~$ known as the
Brans-Dicke parameter. Indeed, for small values of $\omega$~the contribution
of the scalar field in the dynamics of the field equations is significant,
while on the other hand, when $\omega$ is large, the main contribution in the dynamics is
followed by the tensor part. Someone would expect that as $\omega~$reaches
infinity, the theory of General Relativity will be recovered, however, that
is not true, which means that General Relativity and Brans-Dicke gravity are
fundamentally different theories \cite{omegaBDGR}. Other well-known nonminimally
coupled models are the Galileon and the Hordenski theories
\cite{gal1,gal2,gal3,hor1,hor3,hor4}.

The nonminimally coupled scalar fields can describe modified theories
of gravity by attributing new degrees of freedom in terms of geometric invariants which are introduced to modify the
Einstein-Hilbert gravitational action. Indeed, the fourth-order $f\left(
R\right)  $-gravity is dynamical and physically equivalent to the Brans-Dicke
theory in the limit where $\omega$ is zero, such that Brans-Dicke theory
reduces to O'Hanlon gravity \cite{Hanlon}. For more details in this direction we refer the reader
to \cite{s01,s02,s03,s04,s05} and the references therein.

As we discussed, Brans-Dicke theory is a different theory compared to Einstein's
General Relativity, however, there is a mathematical connection between these two theories. Indeed, the two theories are conformally equivalent
\cite{fbook,con1,con2}. The latter equivalency is a mathematical
equivalence between the Jordan and the Einstein frame in the solution space of
the field equations while in general the physical properties of the solutions
do not remain invariant under the conformal transformation. For instance, a singular universe in the Einstein frame, under a conformal transformation can
become a non-singular universe in the Jordan frame which is described by a
scalar tensor theory. There is a plethora of studies where the physical
quantities have been studied between the Einstein and the Jordan frame for
various exact solutions \cite{jd1,jd2,jd3,jd5}. In addition, the question of
which is the preferred frame is still unanswered \cite{fr1}.

In this work we considered a two-scalar field cosmological model where one of
the fields is the Brans-Dicke field which is nonminimally coupled to gravity, and the
second scalar field is minimally coupled to gravity. The model of our consideration has the Action Integral of the Brans-Dicke Action where the
Brans-Dicke field is coupled to the Lagrangian of the quintessence scalar
field. Consequently, the two fields interact in the kinetic
and in the potential parts. Such a model was investigated before in \cite{angrg}.
Specifically, in \cite{angrg} the integrability of the cosmological model
was investigated and some exact solutions were determined. While our model is defined in the
Jordan frame, when we perform a conformal transformation in order to pass to the Einstein frame,
we end up with a two-scalar field theory which includes the quintom cosmological model.

Here we study the cosmological evolution of this specific cosmological model in the
Jordan frame by investigating the stability of the equilibrium points. For the background
space we assume this to be described by the spatially flat Friedmann-Lema\^{i}tre-Robertson-Walker (FLRW) universe. Under this minimal assumption we
shall be able to understand the physical evolution of the universe as it is provided by the
field equations and also we will be able to discuss the viability of the cosmological model.
Such an analysis has been applied in various cosmological theories with many
interesting results
\cite{dn1,dn2,dn3,dn4,dn5,Leon:2019iwj,dn6,dn7,dn8,dn9,dn10,dn11,dn12,dyn13,dyn14}. Any
equilibrium point, also known as stationary or critical point, describes a specific phase,
i.e., an exact solution, in the evolution of the generic solution. The
stability of the critical point indicates the stability of the solution, from
where we can draw conclusions for the cosmological evolution.

The plan of the paper is follows. In Section \ref{sec2}, we present the cosmological model of our consideration.
We derive the field equations and we discuss the mathematical relation of the
theory between the Jordan and the Einstein frame, where we show that our model
is equivalent with a two-scalar field theory, such is the quintom cosmology,
in the Einstein frame. In Section \ref{sec3}, we write the field equations in
dimensionless variables by using the $H$-normalization. We end with a
five-dimensional algebraic-differential dynamical system. Section \ref{sec4},
includes the main analysis of our work, where we investigate the critical
points and their stabilities for the model under consideration in the vacuum
case, and in presence of an additional dust fluid source. The scalar field
potentials have been selected in order to reduce the number of equations of
the dimensionless system under study. However, such selection can also be seen
as the limit for other potential forms too. Finally, in Section \ref{sec5}, we
summarize our results with the main findings.

\section{Cosmological model}
\label{sec2}

The gravitation model of our consideration is described by the following
Action Integral%
\begin{equation}
S=\int dx^{4}\sqrt{-g}\left[  \phi\left(  \frac{R}{2}-\frac{1}{2}g^{\mu\nu
}\psi_{;\mu}\psi_{;\nu}+V\left(  \phi,\psi\right)  \right)  -\frac{1}{2}%
\frac{\omega}{\phi}g^{\mu\nu}\phi_{;\mu}\phi_{;\nu}\right], \label{bd2.01}%
\end{equation}
which actually describes a two-scalar field cosmology
defined in the Jordan frame. More specifically, the scalar field $\phi\left(
x^{\mu}\right)  $ is the Brans-Dicke field with $\omega\neq\frac{2}{3},$ which is
nonminimally coupled to gravity and to the second scalar field $\psi\left(
x^{\mu}\right)  $ as well, while the second scalar field $\psi\left(  x^{\mu}\right)
$ is minimally coupled to gravity. The limit $\omega=\frac{2}{3}$
corresponds to a special case where the Brans-Dicke field does not introduce
any real degree of freedom and can be neglected by a scale transformation in  the metric $g_{\mu\nu}$. Therefore, in the following we shall assume that
$\omega\neq\frac{2}{3}$.

An equivalent way to write the Action Integral (\ref{bd2.01}) is by using the
Lagrange function for the quintessence, that means,%
\begin{equation}
S=\int dx^{4}\sqrt{-g}\left[  \phi L_{Q}\left(  R,\psi,\psi_{;\mu}\right)
-\frac{1}{2}\frac{\omega}{\phi}g^{\mu\nu}\phi_{;\mu}\phi_{;\nu}+W\left(
\phi,\psi\right)  \right], \label{bd2.02}%
\end{equation}
where $L_{Q}\left(  R,\psi,\psi_{;\mu}\right)  $ is the Lagrangian function
for the minimally coupled scalar field theory, that is,%
\begin{equation}
L_{Q}=\frac{R}{2}-\frac{1}{2}g^{\mu\nu}\psi_{;\mu}\psi_{;\nu}-\bar{V}\left(
\psi\right), \label{bd2.03}%
\end{equation}
and $W\left(  \phi\left(  x^{\mu}\right)  ,\psi\left(  x^{\mu}\right)
\right)  =V\left(  \phi\left(  x^{\mu}\right)  ,\psi\left(  x^{\mu}\right)
\right)  +\phi\left(  x^{\mu}\right)  \bar{V}\left(  \psi\left(  x^{\mu
}\right)  \right)  \,$. Consequently, when $\phi\left(  x^{\mu}\right)
=\phi_{0}$, the gravitational Action Integral (\ref{bd2.01}) reduces to that of a
minimally coupled scalar field model.

From (\ref{bd2.01}), one can obtain the gravitational field equations by varying this action with respect to
the metric tensor $g_{\mu\nu}$ as follows \cite{angrg}%
\begin{equation}
\phi G_{\mu\nu}=\frac{\omega}{\phi}\left(  \phi_{;\mu}\phi_{;\nu}-\frac{1}%
{2}g_{\mu\nu}g^{\kappa\lambda}\phi_{;\kappa}\phi_{;\lambda}\right)  -\left(
g_{\mu\nu}g^{\kappa\lambda}\phi_{;\kappa\lambda}-\phi_{;\mu}\phi_{;\nu
}\right)  -g_{\mu\nu}V\left(  \phi,\psi\right)  +\phi\left(  \psi_{;\mu}%
\psi_{;\nu}-\frac{1}{2}g_{\mu\nu}g^{\kappa\lambda}\psi_{;\kappa}\psi
_{;\lambda}\right)  , \label{bd2.03b}%
\end{equation}
while the variation of (\ref{bd2.01}) with respect to $\phi\left(  x^{\mu}\right)$, and $\psi\left(
x^{\mu}\right) $ respectively provide with the equations of motion of the two fields as follows, %
\begin{equation}
g^{\kappa\lambda}\phi_{;\kappa\lambda}-\frac{1}{2\phi}g^{\mu\nu}\phi_{;\mu
}\phi_{;\nu}+\frac{ \phi}{2\omega}g^{\mu\nu}\psi_{;\mu}\phi_{;\nu}+\frac{\phi
}{2\omega}\left(  R-2V_{,\phi}\right)  =0, \label{bd2.03c}%
\end{equation}%
\begin{equation}
g^{\kappa\lambda}\psi_{;\kappa\lambda}+\frac{1}{\phi}g^{\mu\nu}\phi_{;\mu}%
\psi_{;\nu}+V_{,\psi}=0. \label{bd2.03d}%
\end{equation}
It is important to mention here that while someone will expect that for  $\phi_{;\kappa}=0$, the limit of General Relativity with a minimally coupled
scalar field will be recovered, but that is not true, because of the constraint
equation (\ref{bd2.03c}) which becomes $R-2V_{,\phi_{0}}=0$.

In order to find an equivalent theory to (\ref{bd2.01}) in the Einstein frame,
we consider the conformal transformation $g_{\mu\nu}=\phi\bar{g}_{\mu\nu}$,
that is, for a four-dimensional manifold, the Ricci scalar $R$ becomes
\cite{hawb}
\begin{equation}
R=\phi^{-2}\bar{R}-6\phi^{-3}\phi_{;\mu\nu}\bar{g}^{\mu\nu}. \label{bd2.04}%
\end{equation}
Replacing (\ref{bd2.04}) and integrating by pats we end up with the gravitational
Action Integral%
\begin{equation}
\bar{S}=\int dx^{4}\sqrt{-g}\left[  \frac{\bar{R}}{2}-\frac{1}{2}\bar{g}%
^{\mu\nu}\psi_{;\mu}\psi_{;\nu}+U\left(  \Phi,\psi\right)  -\frac{1}{2}\bar
{g}_{;\mu}^{\mu\nu}\Phi_{;\mu}\Phi_{;\nu}\right], \label{bd2.05}%
\end{equation}
where the new scalar field, $\Phi\left(  x^{\mu}\right)$, is a functional
form of $\phi\left(  x^{\mu}\right),$ that is, $\Phi=\Phi\left(  \phi\left(
x^{\mu}\right)  \right)$, and specifically, $\Phi\left(  x^{\mu}\right)
\simeq\phi\left(  x^{\mu}\right)$, while the new scalar field
potential~$U\left(  \Phi\left(  \phi\right), \psi\right)  $ is defined as
$U\left(  \phi,\psi\right)  \simeq\phi^{-2}V\left(  \phi,\psi\right)
.~$Consequently, either if we select $V\left(  \phi,\psi\right)  =V\left(
\psi\right)  $ in the Einstein frame, the two scalar fields interact.
Therefore, when $U\left(  \phi,\psi\right)  =V\left(  \psi\right)$, the
Action Integral describes a two-scalar field minimally coupled cosmological
model, and such model has been studied earlier in the literature \cite{lu1}.

Moreover, if in the Jordan frame, we select $V\left(  \phi,\psi\right)
=\phi^{2}F\left(  \psi\right)  +Z\left(  \phi\right)$, then in the Einstein
frame it follows that the two scalar fields are not interacting neither in the
kinetic terms nor in the potential terms. In addition, in the latter scenario, by doing the
change $\psi\left(  x^{\mu}\right)  =i\psi\left(  x^{\mu}\right)  $, the
Action integral (\ref{bd2.05}) takes the form of the quintom theory  \cite{q2,q3}.

For the background geometry of our universe we assume that in the large scales it is well described by the
spatially flat Friedmann--Lema\^{\i}tre--Robertson--Walker (FLRW) spacetime characterized by the following line element%
\begin{equation}
ds^{2}=-N^{2}\left(  t\right)  dt^{2}+a^{2}\left(  t\right)  \left(
dx^{2}+dy^{2}+dz^{2}\right)  , \label{bd2.06}%
\end{equation}
where $N\left(  t\right)  $ is the lapse function and $a\left(  t\right)  $ is
the expansion scale factor of the universe, i.e. the Hubble function of this universe is defined as
$H\left(  t\right)  =\frac{1}{N}\frac{\dot{a}}{a}$. In the following we assume
the co-moving observer $u^{\mu}=\frac{1}{N}\delta_{t}^{\mu},~u^{\mu}u_{\mu}=-1$.

For the line element (\ref{bd2.06}), the gravitational field equations
(\ref{bd2.03b}), (\ref{bd2.03c}) and (\ref{bd2.03d}) can be expressed as
\cite{angrg}%
\begin{align}
&  6\phi H^{2}+6H\dot{\phi}+\frac{\omega}{\phi}\dot{\phi}^{2}+\phi\dot{\psi
}^{2}+2V\left(  \phi,\psi\right)  =0.\label{bd2.07}\\
&  2\phi\dot{H}+3\phi H^{2}+2H\dot{\phi}-\frac{\omega}{2}\frac{\dot{\phi}^{2}%
}{\phi}-\frac{1}{2}\phi\dot{\psi}^{2}+\ddot{\phi}+V\left(  \phi,\psi\right)
=0,\label{bd2.08}\\
&  3\dot{H}+6H^{2}+3\omega H\frac{\dot{\phi}}{\phi}-\frac{1}{2}\dot{\psi}%
^{2}-\frac{\omega}{2}\left(  \frac{\dot{\phi}^{2}}{\phi^{2}}-2\frac{\ddot
{\phi}}{\phi}\right)  +V_{,\phi}\left(  \phi,\psi\right)  =0,\label{bd2.09}\\
&  \phi\ddot{\psi}+\left(  3H\phi+\dot{\phi}\right)  \dot{\psi}+V_{,\psi
}\left(  \phi,\psi\right)  =0. \label{bd2.010}%
\end{align}
where without any loss of generality we selected $N\left(  t\right)  =1$, while
additionally, we assumed that the two scalar fields $\phi\left(  x^{\mu}\right)
,~\psi\left(  x^{\nu}\right)  $ inherit the symmetries of the spacetime, that
means, $\phi\left(  x^{\mu}\right)  =\phi\left(  t\right)$ and $\psi\left( x^{\mu}\right)  =\psi\left(  t\right)$.

Let us note that equation (\ref{bd2.07}) is the constraint equation, while equations
(\ref{bd2.08}),\ (\ref{bd2.09}) and (\ref{bd2.010}) are the second-order
differential equations describing the dynamics of the universe for the two scalar fields
$\phi\left(  t\right)  ,~\psi\left(  t\right)  $ and the scale factor
$a\left(  t\right)$.

{ Now, in presence of an additional matter source, and more specifically, in the presence of  a dust fluid term minimally coupled to the Brans-Dicke field
$\phi\left(  x^{\mu}\right)$ in the gravitational model, the field equations (\ref{bd2.08}),
(\ref{bd2.09}), (\ref{bd2.010}) will remain same, except the constraint equation (\ref{bd2.07}) which becomes}%
\begin{equation}
6\phi H^{2}+6H\dot{\phi}+\frac{\omega}{\phi}\dot{\phi}^{2}+\phi\dot{\psi}%
^{2}+2V\left(  \phi,\psi\right)  =2\rho_{m}, \label{bd2.011}%
\end{equation}
while the continuity equation for the dust fluid is $\dot{\rho}_{m}+3H\rho
_{m}=0$, from where it follows that, $\rho_{m}=\rho_{m0}a^{-3}$, where $\rho_{m0}$
is an integration constant physically which represents the energy density of the dust fluid at current epoch. We investigate first the vacuum case ($\Omega
_{m}=0$), and after that we investigate the matter case.

The gravitational field equations for this particular cosmological model are
described by a point-like Lagrangian, however, such an analysis is not
necessary in the present work. The purpose of this work is to study the dynamical evolution for this cosmological model. In \cite{angrg}, the
specific forms of the scalar field potential were determined such that the
field equations admit linear and quadratic conservation laws in the
momentum and they are integrable. In addition, exact solutions
of special interests, such that the singular solution $a\left(  t\right)
=a_{0}t^{p}$ and the de Sitter solution $a\left(  t\right)  =a_{0}e^{H_{0}t},$ were
determined in \cite{angrg}.

With the analysis of this work, we will be able to understand the
general behavior of the cosmological solution as well as we will be also able  to identify the
attractors of the cosmological model and to study the stability of the particular
solutions. At the same time, we will be able to investigate the similarities and differences of
two-scalar field gravitational theories defined in the Einstein and Jordan frames.

Making the selection of the potential for the two scalar field $V\left(  \phi,\psi\right)  =V\left(  \psi\right)
W\left(  \phi\right)$, we solve for the higher derivatives (in presence of an additional matter source and more specifically in presence of a dust
fluid which is minimally coupled to the Brans-Dicke field) and obtain the field
equations:%
\begin{align}
&  \dot{H}=-\frac{3(\omega-2)H^{2}}{2\omega-3}+\frac{\omega H\dot{\phi}%
}{(2\omega-3)\phi}+\frac{V(\psi)W^{\prime}(\phi)}{2\omega-3}+\frac{\omega
V(\psi)W(\phi)}{(3-2\omega)\phi}+\frac{(\omega-1){\dot{\psi}}^{2}}{4\omega
-6}+\frac{(\omega-1)\omega{\dot{\phi}}^{2}}{(4\omega-6)\phi^{2}},\\
&  \ddot{\phi}=-\frac{6(\omega-1)H\dot{\phi}}{2\omega-3}+\frac{3H^{2}\phi
}{3-2\omega}+\frac{2\phi)V(\psi))W^{\prime}(\phi)}{3-2\omega}+\frac
{3V(\psi)W(\phi)}{2\omega-3}+\frac{\omega{\dot{\phi}}^{2}}{(6-4\omega)\phi
}+\frac{\phi\dot{\psi}^{2}}{6-4\omega},\\
&  \ddot{\psi}=-3H\dot{\psi}-\frac{W(\phi)V^{\prime}(\psi)}{\phi}-\frac
{{\dot{\psi}}\dot{\phi}}{\phi},\\
&  \dot{\rho}_{m}=-3H\rho_{m},
\end{align}
with first integral equation (\ref{bd2.011}).

\section{Dimensionless system}
\label{sec3}

In this section we proceed by writing the field equations with the use of
dimensionless variables. In particular, we make use of the $H-$normalization
\cite{cop1}, which has been applied also in Brans-Dicke theory
\cite{bcop1,bcop2,bcop3,bcop4} and in multi scalar field theories
\cite{mfh01,mfh02,mfh04,mfh06,mfh07,mfh08}. As far as the scalar
field potential is concerned, we make the selection $V\left(  \phi,\psi\right)
=V\left(  \psi\right)  W\left(  \phi\right)  $.

The new dimensionless variables are%
\begin{equation}
x=\frac{\dot{\phi}}{\sqrt{6} H\phi}~,~y=\frac{\dot{\psi}}{\sqrt{6}H}%
~,~z=\frac{V\left(  \psi\right)  W\left(  \phi\right)  }{3\phi~H^{2}},
\label{bd2.012}%
\end{equation}
or equivalently,%
\begin{equation}
\dot{\phi}=\sqrt{6}x\phi H~,~\dot{\psi}=\sqrt{6}yH~,~V\left(  \psi\right)
=3z\frac{\phi}{W\left(  \phi\right)  }H^{2}. \label{bd2.013}%
\end{equation}

In the new variables, the field equations can be expressed as the following
algebraic-differential first-order system%
\begin{align}
\frac{dx}{d\tau}  &  =\frac{-6x^{3}(\omega-1)\omega+\sqrt{6}x^{2}%
(6-7\omega)-6x\left(  (\omega-1)y^{2}+\omega+\mu z-\omega z\right)  -\sqrt
{6}\left(  y^{2}+2\mu z-3z+1\right)  }{4\omega-6},\label{bd2.014}\\
\frac{dy}{d\tau}  &  =\frac{z\left(  \sqrt{6}\lambda(3-2\omega)+6y(\omega
-\mu)\right)  -6y(\omega-1)\left(  x\left(  x\omega+\sqrt{6}\right)
+y^{2}+1\right)  }{4\omega-6},\label{bd2.015}\\
\frac{dz}{d\tau}  &  =\frac{z\left(  -6x^{2}(\omega-1)\omega+\sqrt{6}%
x(\mu(2\omega-3)-4\omega+3)-6y^{2}(\omega-1)+\sqrt{6}\lambda y(2\omega
-3)+6(\omega-\mu z+\omega z-2)\right)  }{2\omega-3}, \label{bd2.016}%
\end{align}
in which the new variables $\lambda$ and $\mu$ are defined as functions of $\phi$%
\begin{equation}
\lambda=\left(  \ln V\left(  \psi\right)  \right)  _{,\psi}~,~\mu=\phi\left(
\ln W\left(  \phi\right)  \right)  _{,\phi}~,~\tau=\ln a\left(  t\right) ,
\label{bd2.016a}%
\end{equation}
while the constraint equation (\ref{bd2.011}) becomes%
\begin{equation}
\Omega_{m}=x\left(  x\omega+\sqrt{6}\right)  +y^{2}+z+1, \label{bd2.018}%
\end{equation}
where $\Omega_{m}$ is the dimensionless density parameter of the dust fluid source defined as
$\Omega_{m}=\frac{\rho_{m}}{3\phi H^{2}},$ in which $\Omega_{m}\in\left[
0,1\right]  $. We also have an extra evolution equation%
\begin{equation}
\Omega_{m}^{\prime}=\Omega_{m} \left[  \frac{-6 x^{2} (\omega-1) \omega
+\sqrt{6} x (3-4 \omega)-6 y^{2} (\omega-1)-6 \mu z+6 \omega z-3}{2 \omega
-3}\right].
\end{equation}

As far as the variation of the new variables $\mu$ and $\lambda$ is concerned,
we find that they should satisfy the following two first-order ordinary
differential equations, namely,%
\begin{align}
\frac{d\mu}{d\tau}=\sqrt{6}x h(\lambda) ,\label{bd2.020}\\
\frac{d\lambda}{d\tau}=\sqrt{6}y f(\lambda) , \label{bd2.019}%
\end{align}
in which
\begin{align}
f(\lambda)= (\Gamma\left(  \lambda\right)  -1) \lambda^{2}, \quad h(\mu)=
\mu(\bar{\Gamma}\left(  \mu\right)  \mu-\mu+1), \quad\Gamma\left(
\lambda\right)  =\frac{V_{,\psi\psi}V}{\left(  V_{,\psi}\right)  ^{2}},
\quad\bar{\Gamma}\left(  \mu\right)  =\frac{W_{,\phi\phi}W}{\left(  W_{,\phi
}\right)  ^{2}}.
\end{align}

Consequently, the algebraic-differential system has dimension five, consisting
of the first-order differential equations (\ref{bd2.014}), (\ref{bd2.015}),
(\ref{bd2.016}), (\ref{bd2.019}) and (\ref{bd2.020}), while for specific forms
of the potentials $V\left(  \psi\right) ~$ and $W\left(  \phi\right)$, it
follows that, $\lambda = \mbox{const}$ and $\mu = \mbox{const}$, which mean that the dimension
of the dynamical system can be reduced by one or by two dimensions. In
addition, for the vacuum case, i.e., $\Omega_{m}=0$, the algebraic
equation (\ref{bd2.018}) can also be used to reduce the dimension of the
dynamical system by one dimension.

Any critical point describes a specific universe where the total equation of
state is $w_{tot}=-1-\frac{2}{3}\frac{\dot{H}}{H^{2}}$, which can be be expressed in terms of the dimensionless parameters as
\begin{equation}
w_{tot}=\frac{-2 x \omega\left(  3 x (\omega-1)+\sqrt{6}\right)  -6 (\omega-1)
y^{2}+6 z (\omega-\mu)-3}{6 \omega-9}. \label{bd2.021}%
\end{equation}
When $w_{tot}\neq-1$, the scale factor at the critical point is a power-law
function, while for $w_{tot}=-1$, the critical point describes a de Sitter universe.

\section{Cosmological evolution}
\label{sec4}

In the following, we set $V\left(  \psi\right)  =V_{0}e^{\lambda\phi}$ and
$W\left(  \phi\right)  =W_{0}\phi^{\mu}$, and we study the equilibrium points
and analyze the stability when $\Omega_{m}=0$ (vacuum case) and when $0<\Omega
_{m}\leq1$ (matter case). For the specific functions of the two potentials,
variables $\mu$ and $\lambda$ are constants, i.e. the right hand side of equations
(\ref{bd2.020}), (\ref{bd2.019}) are zero.

\subsection{Vacuum case}

In the vacuum case, namely for $\Omega_{m} = 0$, the algebraic equation
(\ref{bd2.018}) gives an additional restriction%
\begin{equation}
x\left(  x\omega+\sqrt{6}\right)  +y^{2}+z+1=0,
\end{equation}
that can be used to remove one phase-space variable, say $z$. Hence, we obtain
the reduced system:
\begin{align}
&  \frac{dx}{d\tau}=\frac{3x^{3}\omega(\mu-2\omega+1)+\sqrt{6}x^{2}(\mu
(\omega+3)-8\omega+3)+3x\left(  3\mu+y^{2}(\mu-2\omega+1)-2\omega-3\right)
+\sqrt{6}(\mu-2)\left(  y^{2}+1\right)  }{2\omega-3},\label{0bd2.014}\\
&  \frac{dy}{d\tau}=\frac{\left(  x\left(  x\omega+\sqrt{6}\right)
+y^{2}+1\right)  \left(  \sqrt{6}\lambda(2\omega-3)+6y(\mu-2\omega+1)\right)
}{4\omega-6}. \label{0bd2.015}%
\end{align}

The total equation of state parameter now becomes,
\begin{equation}
w_{tot}=\frac{6 \mu\left( x \left( x \omega+\sqrt{6}\right) +1\right) -2
\omega\left( x \left( 6 x \omega-3 x+4 \sqrt{6}\right) +3\right) +6 y^{2}
(\mu-2 \omega+1)-3}{6 \omega-9}. \label{bd2.021B}%
\end{equation}

For the choices $V\left(  \psi\right)  =V_{0}e^{\lambda\phi}$ and $W\left(
\phi\right)  =W_{0}\phi^{\mu}$, with $\mu$ and $\lambda$ being  constants, we
study the Equilibrium points and investigate the stability of the two-dimensional system
(\ref{0bd2.014}), (\ref{0bd2.015}).

The (lines of) equilibrium points are the following:

\begin{enumerate}
\item $A_{\pm}:$ The line
\begin{equation}
x_{c} \left( x_{c} \omega+\sqrt{6}\right) +y^{2}+1=0,
\end{equation}
corresponding to the massless fields, exists for $x_{c}<0, \omega<\frac{-\sqrt{6}
x_{c}-1}{x_{c}^{2}}$ or $x_{c}>0, \omega<\frac{-\sqrt{6} x_{c}-1}{x_{c}^{2}}$.
\newline The eigenvalues are: \newline{\small $\Big\{\frac{1}{2} \left(
\sqrt{6} (\mu+1) x_{c}-\sqrt{6} \sqrt{\lambda^{2} \left( -\left( x_{c} \left(
x_{c} \omega+\sqrt{6}\right) +1\right) \right) +(\mu+1) x_{c} \left( \mu
x_{c}+x_{c}+2 \sqrt{6}\right) +2 \lambda(\mu+1) x_{c} y+2 \sqrt{6} \lambda
y+6}+\sqrt{6} \lambda y+6\right) $,\newline$\frac{1}{2} \left( \sqrt{6}
(\mu+1) x_{c}+\sqrt{6} \sqrt{\lambda^{2} \left( -\left( x_{c} \left( x_{c}
\omega+\sqrt{6}\right) +1\right) \right) +(\mu+1) x_{c} \left( \mu x_{c}%
+x_{c}+2 \sqrt{6}\right) +2 \lambda(\mu+1) x_{c} y+2 \sqrt{6} \lambda
y+6}+\sqrt{6} \lambda y+6\right) \Big\}$, where $y=\epsilon\sqrt{-x_{c} \left(
x_{c} \omega+\sqrt{6}\right) -1}$, $\epsilon=\pm1$. } In Fig.  \ref{StCp} we show the stability conditions of $A_{+}$ for (a) $\omega = 1$, and for (b)
$\omega = 0$. For parameters in the red region (upper and lower left graphs of Fig. \ref{StCp}) $A_{+}$ is a source
and for parameters in the blue region (upper and lower right graphs of Fig.  \ref{StCp}) it is a sink.

Similarly, in Fig. \ref{StCn} we  present the stability conditions of $A_{-}$ for (a)
$\omega=1$, and for  (b) $\omega=0$. For parameters in the red region (upper and lower left graphs of Fig. \ref{StCn})
$A_{-}$ is a source and for parameters in the blue region (upper and lower right graphs of Fig. \ref{StCn}) it is
a sink.

The total EoS parameter and the fractional matter density at the equilibrium
points $A_{\pm}$ are the following: $w_{tot}(A_{\pm})=1+2\sqrt{\frac{2}{3}%
}x_{c}$. {They are accelerating solutions ($w_{tot}<-\frac{1}{3}$) for
$x_{c}<-\sqrt{\frac{2}{3}},~\omega<\frac{-\sqrt{6}x_{c}-1}{x_{c}^{2}}$.} For
$x_{c} =-\sqrt{\frac{3}{2}}$, it is a de Sitter solution ($w_{tot}(A_{\pm})=-1$).
It represents a decelerating solution for $-\sqrt{\frac{2}{3}}<x_{c}%
<0,\omega<\frac{-\sqrt{6}x_{c}-1}{x_{c}^{2}}$, or $x_{c}>0,~\omega<\frac
{-\sqrt{6}x_{c}-1}{x_{c}^{2}}$. In particular, it mimics a dust solution
($w_{tot}(A_{\pm})=0$) for $x_{c}=-\frac{1}{2}\sqrt{\frac{3}{2}}$.

\begin{figure}[ptb]
\textbf{ \includegraphics[width=0.8\textwidth]{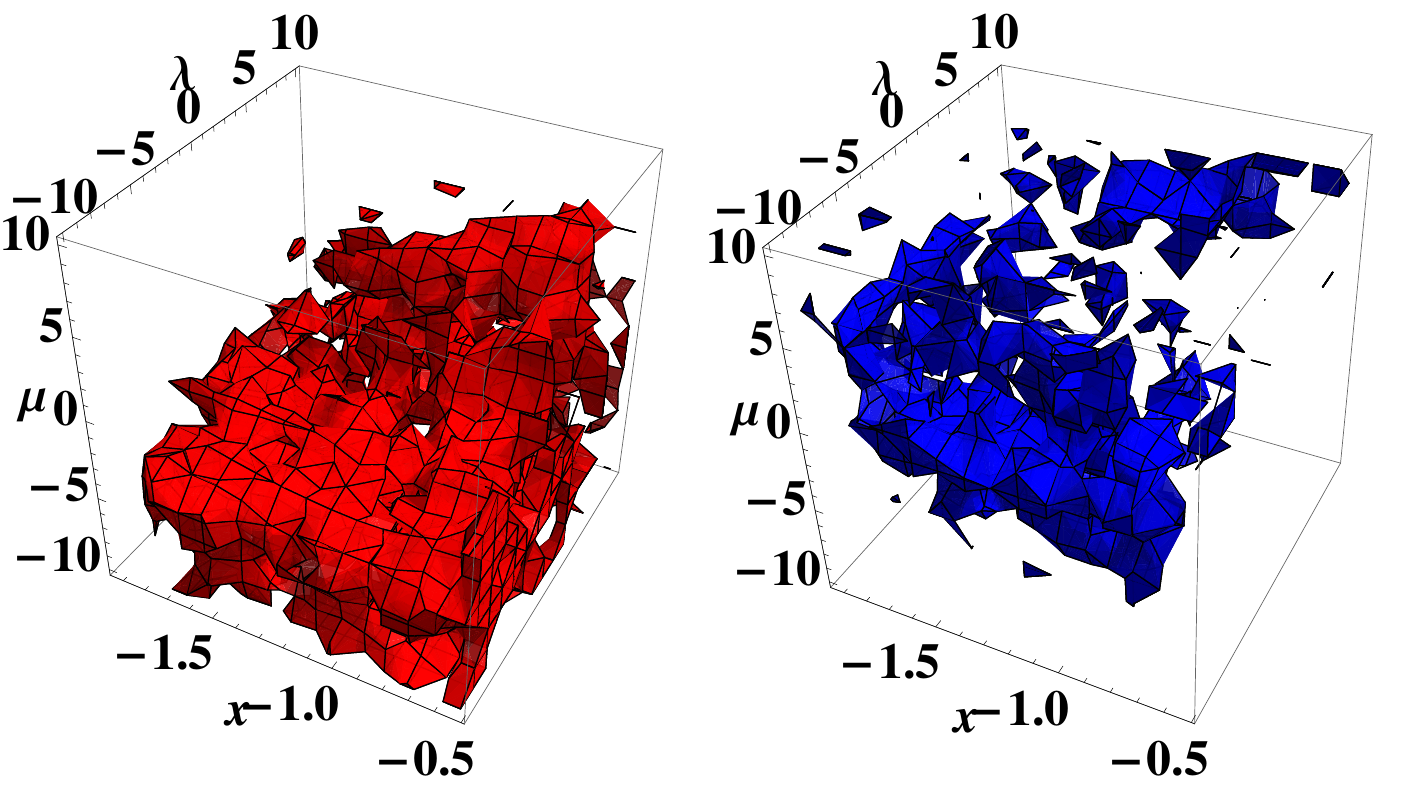} \newline}(a) \textbf{\includegraphics[width=0.8\textwidth]{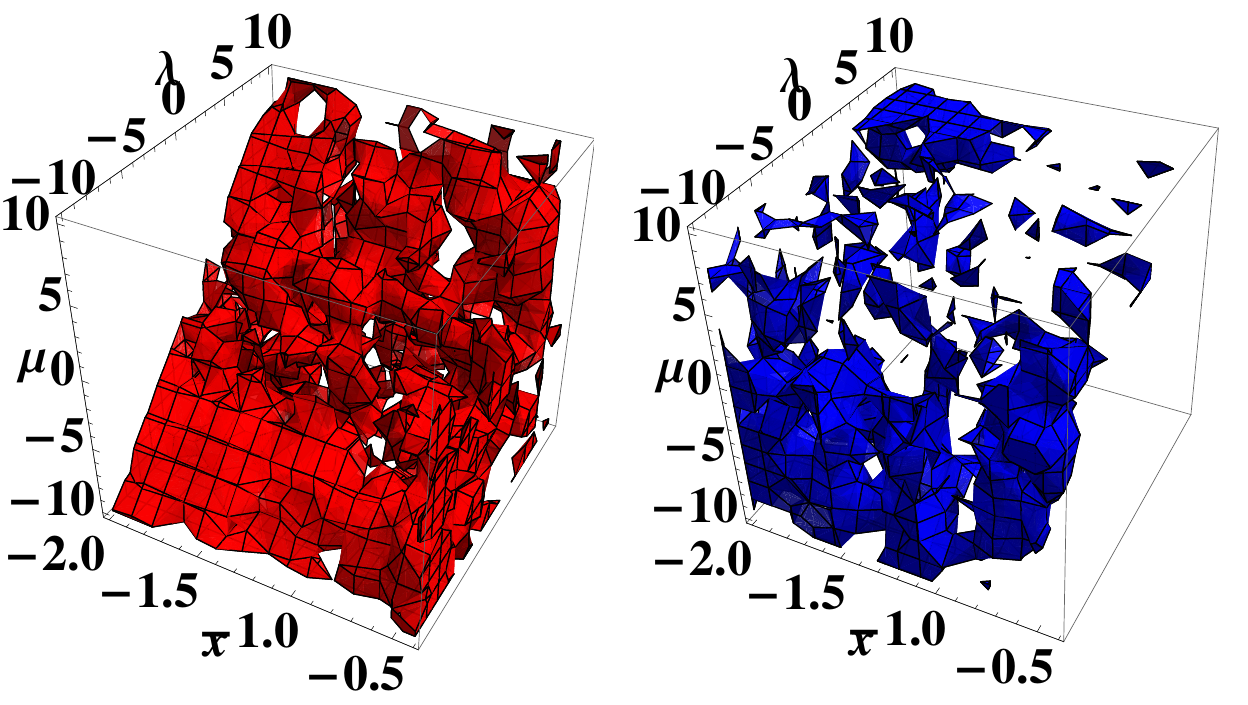} \newline} (b) \caption{Stability
of $A_{+}$ for (a) $\omega=1$, and for (b) $\omega=0$. For parameters in the red
region (left graph of each panel) $A_{+}$ is a source and for parameters in the blue region
(right graph of each panel) it is a sink. }%
\label{StCp}%
\end{figure}
\begin{figure}[ptb]
\textbf{ \includegraphics[width=0.8\textwidth]{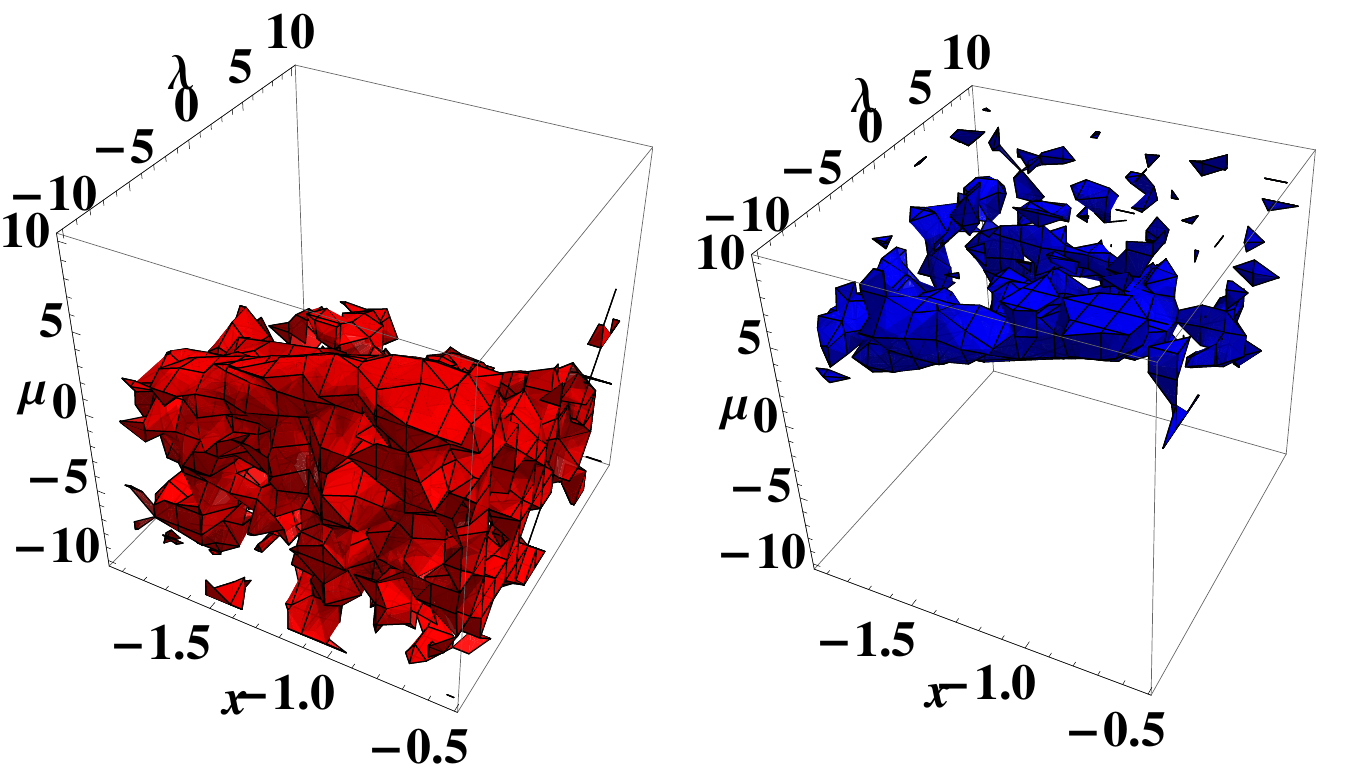} \newline}(a) \textbf{
\includegraphics[width=0.8\textwidth]{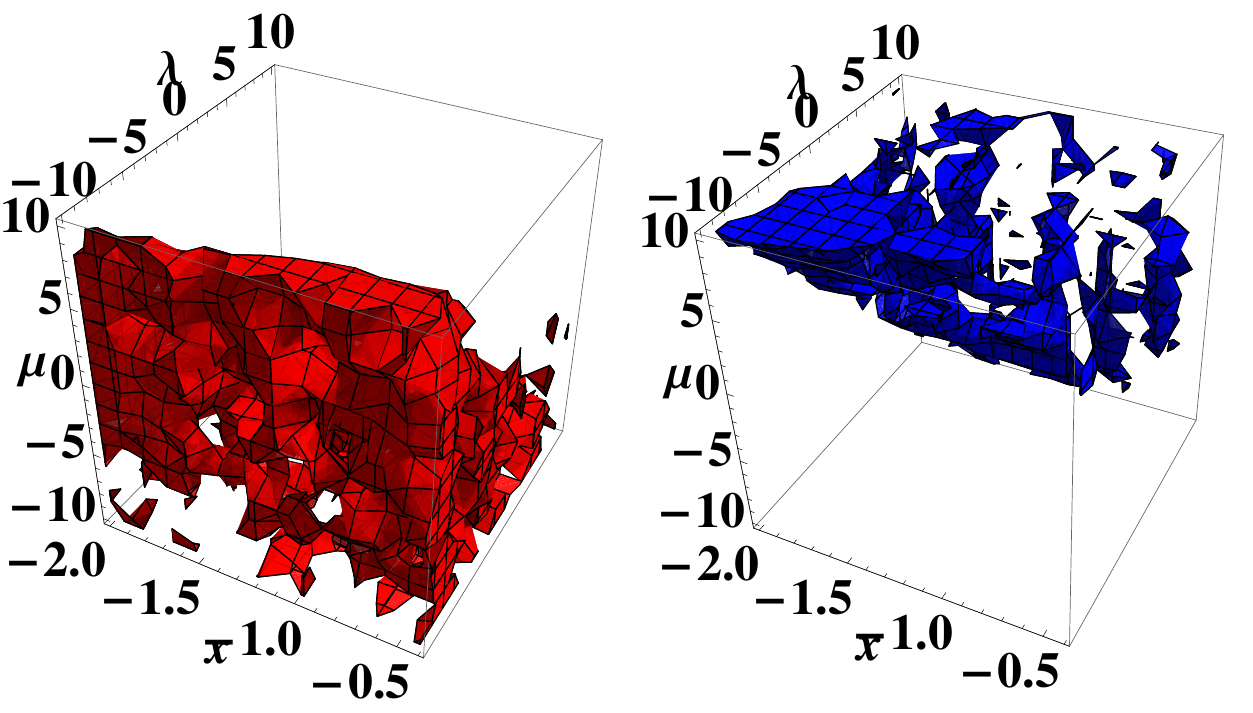} \newline}(b) \caption{Stability
of $A_{-}$ for (a) $\omega=1$, and for (b) $\omega=0$. For parameters in the red
region (left graph of each panel) $A_{-}$ is a source and for parameters in the blue region
(right graph of each panel) it is a sink. }%
\label{StCn}%
\end{figure}

\item $B:$ $(x,y)= \left( -\frac{\sqrt{\frac{2}{3}} (\mu-2)}{\mu-2 \omega
+1},\frac{\lambda(3-2 \omega)}{\sqrt{6} (\mu-2 \omega+1)}\right) $. \newline
Eigenvalues are: $\left\{ \frac{\lambda^{2} (2 \omega-3)+2 ((\mu-4) \mu+6
\omega-5)}{2 (\mu-2 \omega+1)},\frac{\lambda^{2} (2 \omega-3)+2 ((\mu-4) \mu+6
\omega-5)}{2 (\mu-2 \omega+1)}\right\} $. { The critical point is a }
stable node for
\begin{enumerate}
\item $2 \omega-3\neq0, \mu=2$, or

\item $\mu=-\frac{1}{2} \left( \lambda^{2}+2\right)  , \left( \lambda
^{2}+6\right)  \left( \lambda^{2}+4  \omega\right) \neq0$, or

\item $\lambda\in\mathbb{R}, -\frac{1}{2} \left( \lambda^{2}+2\right)  <\mu<2,
\omega>\frac{3 \lambda^{2}-2 \mu^{2}+8 \mu+10}{2 \left( \lambda^{2}+6\right)
}$, or

\item $\lambda\in\mathbb{R}, -\frac{1}{2} \left( \lambda^{2}+2\right) <\mu<2,
\omega<\frac{\mu+1}{2}$, or

\item $\lambda\in\mathbb{R}, \mu<-\frac{1}{2} \left( \lambda^{2}+2\right) ,
\omega>\frac{\mu+1}{2}$, or

\item $\mu>2, \omega>\frac{\mu+1}{2}$, or

\item $\lambda\in\mathbb{R}, \mu<-\frac{1}{2} \left( \lambda^{2}+2\right)  ,
\omega<\frac{3 \lambda^{2}-2 \mu^{2}+8 \mu+10}{2 \left( \lambda^{2}+6\right)
}$, or

\item $\lambda\in\mathbb{R}, \mu>2, \omega<\frac{3 \lambda^{2}-2 \mu^{2}+8
\mu+10}{2 \left( \lambda^{2}+6\right) }$.
\end{enumerate}

The critical point an unstable node for

\begin{enumerate}
\item $\lambda\in\mathbb{R}, \mu<-\frac{1}{2} \left( \lambda^{2}+2\right)  ,
\frac{3 \lambda^{2}-2 \mu^{2}+8 \mu+10}{2 \left( \lambda^{2}+6\right) }%
<\omega<\frac{\mu+1}{2}$, or

\item $\lambda\in\mathbb{R}, -\frac{1}{2} \left( \lambda^{2}+2\right)  <\mu<2,
\frac{\mu+1}{2}<\omega<\frac{3 \lambda^{2}-2 \mu^{2}+8 \mu+10}{2 \left(
\lambda^{2}+6\right) }$, or

\item $\lambda\in\mathbb{R}, \mu>2, \frac{3 \lambda^{2}-2 \mu^{2}+8 \mu+10}{2
\left( \lambda^{2}+6\right) }<\omega<\frac{\mu+1}{2}$.
\end{enumerate}

{Otherwise the critical point is nonhyperbolic.}

The total EoS parameter and the fractional matter density at the equilibrium
point $B$ are the following: $w_{tot}(B)=\frac{\lambda^{2}(2\omega-3)+\mu
(2\mu-9)+6\omega+1}{3(\mu-2\omega+1)}$. Therefore, they represent solutions
dominated by the scalar field and they are accelerating ($w_{tot}(B)<-\frac
{1}{3}$) for

\begin{enumerate}
\item $2 \omega-3\neq0, \mu=2$, or

\item $\mu=\frac{1}{2} \left( 2-\lambda^{2}\right) , \lambda^{4}+4 \lambda^{2}
\omega-2 \lambda^{2}+8 \omega-8\neq0$, or

\item $\lambda\in\mathbb{R}, \frac{1}{2} \left( 2-\lambda^{2}\right) <\mu<2,
\omega>\frac{3 \lambda^{2}-2 \mu^{2}+8 \mu-2}{2 \lambda^{2}+4}$, or

\item $\lambda\in\mathbb{R},  \frac{1}{2} \left( 2-\lambda^{2}\right) <\mu<2,
\omega<\frac{\mu+1}{2}$, or

\item $\lambda\in\mathbb{R}, \mu<\frac{1}{2} \left( 2-\lambda^{2}\right) ,
\omega>\frac{\mu+1}{2}$, or

\item $\mu>2,  \omega>\frac{\mu+1}{2}$, or

\item $\lambda\in\mathbb{R}, \mu<\frac{1}{2} \left( 2-\lambda^{2}\right) ,
\omega<\frac{3 \lambda^{2}-2 \mu^{2}+8 \mu-2}{2 \lambda^{2}+4}$, or

\item $\lambda\in\mathbb{R}, \mu>2,  \omega<\frac{3 \lambda^{2}-2 \mu^{2}+8
\mu-2}{2 \lambda^{2}+4}$.
\end{enumerate}

In particular, we have a de Sitter solution ($w_{tot}(B)=-1$) for

\begin{enumerate}
\item $\lambda=0, \mu=1, \omega\neq1$, or

\item $\lambda=0, \mu=2, 2 \omega\neq3$, or

\item $\omega=\frac{3}{2}-\frac{(\mu-2) (\mu-1)}{\lambda^{2}}, \lambda^{2}+2
\mu>2, \mu<2$, or

\item $\omega=\frac{3}{2}-\frac{(\mu-2) (\mu-1)}{\lambda^{2}}, \lambda^{2}+2
\mu<2$, or

\item $\omega=\frac{3}{2}-\frac{(\mu-2) (\mu-1)}{\lambda^{2}}, \mu>2$.
\end{enumerate}

The solution $B$ represents decelerating solutions for

\begin{enumerate}
\item $\mu<\frac{1}{2} \left( 2-\lambda^{2}\right) , \frac{3 \lambda^{2}-2
\mu^{2}+8 \mu-2}{2 \left( \lambda^{2}+2\right) }<\omega<\frac{\mu+1}{2}$, or

\item $\frac{1}{2} \left( 2-\lambda^{2}\right) <\mu<2, \frac{\mu+1}{2}%
<\omega<\frac{3 \lambda^{2}-2 \mu^{2}+8 \mu-2}{2 \left( \lambda^{2}+2\right)
}$, or

\item $\mu>2, \frac{3 \lambda^{2}-2 \mu^{2}+8 \mu-2}{2 \left( \lambda
^{2}+2\right) }<\omega<\frac{\mu+1}{2}$.
\end{enumerate}

In particular, for $\omega=\frac{3 \lambda^{2}+(9-2 \mu) \mu-1}{2 \left(
\lambda^{2}+3\right) }$, it mimics a dust fluid ($w_{tot}(B)=0$).

\item $C_{+}:$ $(x,y)=\left( -\frac{\sqrt{\frac{2}{3}} (\mu-2)}{\mu-2
\omega+1},\frac{\sqrt{\left( 1-\frac{2 \omega}{3}\right)  ((\mu-4) \mu+6
\omega-5)}}{\mu-2 \omega+1}\right) $.
\newline Eigenvalues are: $\left\{
0,\frac{10-2 (\mu-4) \mu-12 \omega+\sqrt{2} \lambda\sqrt{(3-2 \omega) ((\mu-4)
\mu+6 \omega-5)}}{\mu-2 \omega+1}\right\} $

{Nonhyperbolic critical point with a 1D stable manifold for:}

\begin{enumerate}
\item $\lambda\leq0, \mu<-1, -\frac{1}{6} (\mu-5) (\mu+1)<\omega<\frac{\mu
+1}{2}$, or

\item $\lambda\leq0, \mu>2, -\frac{1}{6} (\mu-5) (\mu+1)<\omega\leq\frac{3}%
{2}$, or

\item $\lambda>0, \mu<\frac{1}{2} \left( -\lambda^{2}-2\right) , \frac{3
\lambda^{2}-2 \mu^{2}+8 \mu+10}{2 \left( \lambda^{2}+6\right) }<\omega
<\frac{\mu+1}{2}$, or

\item $\lambda>0, \frac{1}{2} \left( -\lambda^{2}-2\right) <\mu\leq-1,
\frac{\mu+1}{2}<\omega<\frac{3 \lambda^{2}-2 \mu^{2}+8 \mu+10}{2 \left(
\lambda^{2}+6\right) }$, or

\item $\lambda>0, -1<\mu<2, -\frac{1}{6} (\mu-5) (\mu+1)<\omega<\frac{3
\lambda^{2}-2 \mu^{2}+8 \mu+10}{2 \left( \lambda^{2}+6\right) }$, or

\item $\lambda>0, \mu>2, \frac{3 \lambda^{2}-2 \mu^{2}+8 \mu+10}{2 \left(
\lambda^{2}+6\right) }<\omega\leq\frac{3}{2}$.
\end{enumerate}

{Nonhyperbolic critical point with a 1D unstable manifold for:}

\begin{enumerate}
\item $\lambda\leq0, \mu\leq-1, \frac{\mu+1}{2}<\omega\leq\frac{3}{2}$, or

\item $\lambda\leq0, -1<\mu<2, -\frac{1}{6} (\mu-5) (\mu+1)<\omega\leq\frac
{3}{2}$, or

\item $\lambda>0, \mu<\frac{1}{2} \left( -\lambda^{2}-2\right) , -\frac{1}{6}
(\mu-5) (\mu+1)<\omega<\frac{3 \lambda^{2}-2 \mu^{2}+8 \mu+10}{2 \left(
\lambda^{2}+6\right) }$, or

\item $\lambda>0, \mu<\frac{1}{2} \left( -\lambda^{2}-2\right) , \frac{\mu
+1}{2}<\omega\leq\frac{3}{2}$, or

\item $\mu=\frac{1}{2} \left( -\lambda^{2}-2\right) , -\frac{1}{6} (\mu-5)
(\mu+1)<\omega<\frac{3 \lambda^{2}-2 \mu^{2}+8 \mu+10}{2 \left( \lambda
^{2}+6\right) }$, or

\item $\mu=\frac{1}{2} \left( -\lambda^{2}-2\right) , \frac{3 \lambda^{2}-2
\mu^{2}+8 \mu+10}{2 \left( \lambda^{2}+6\right) }<\omega\leq\frac{3}{2}$, or

\item $\frac{1}{2} \left( -\lambda^{2}-2\right) <\mu\leq-1, -\frac{1}{6}
(\mu-5) (\mu+1)<\omega<\frac{\mu+1}{2}$, or

\item $\frac{1}{2} \left( -\lambda^{2}-2\right) <\mu\leq-1, \frac{3
\lambda^{2}-2 \mu^{2}+8 \mu+10}{2 \left( \lambda^{2}+6\right) }<\omega
\leq\frac{3}{2}$, or

\item $-1<\mu<2, \frac{3 \lambda^{2}-2 \mu^{2}+8 \mu+10}{2 \left( \lambda
^{2}+6\right) }<\omega\leq\frac{3}{2}$, or

\item $\mu>2, -\frac{1}{6} (\mu-5) (\mu+1)<\omega<\frac{3 \lambda^{2}-2
\mu^{2}+8 \mu+10}{2 \left( \lambda^{2}+6\right) }$.
\end{enumerate}

\item $C_{-}:$ $(x,y)=\left( -\frac{\sqrt{\frac{2}{3}} (\mu-2)}{\mu-2
\omega+1},-\frac{\sqrt{\left( 1-\frac{2 \omega}{3}\right)  ((\mu-4) \mu+6
\omega-5)}}{\mu-2 \omega+1}\right) $. \newline
Eigenvalues are: $\left\{
0,\frac{10-2 (\mu-4) \mu-12 \omega-\sqrt{2} \lambda\sqrt{(3-2 \omega) ((\mu-4)
\mu+6 \omega-5)}}{\mu-2 \omega+1}\right\} $. \newline

{Nonhyperbolic critical point with a 1D
stable manifold for:}

\begin{enumerate}
\item $\lambda<0, \mu<\frac{1}{2} \left( -\lambda^{2}-2\right) , \frac{3
\lambda^{2}-2 \mu^{2}+8 \mu+10}{2 \left( \lambda^{2}+6\right) }<\omega
<\frac{\mu+1}{2}$,

\item $\lambda<0, \frac{1}{2} \left( -\lambda^{2}-2\right) <\mu\leq-1,
\frac{\mu+1}{2}<\omega<\frac{3 \lambda^{2}-2 \mu^{2}+8 \mu+10}{2 \left(
\lambda^{2}+6\right) }$,

\item $\lambda<0, -1<\mu<2, -\frac{1}{6} (\mu-5) (\mu+1)<\omega<\frac{3
\lambda^{2}-2 \mu^{2}+8 \mu+10}{2 \left( \lambda^{2}+6\right) }$,

\item $ \lambda<0, \mu>2, \frac{3 \lambda^{2}-2 \mu^{2}+8 \mu+10}{2 \left(
\lambda^{2}+6\right) }<\omega\leq\frac{3}{2}$

\item $\lambda=0, \mu<-1, -\frac{1}{6} (\mu-5) (\mu+1)<\omega<\frac{\mu+1}{2}$,

\item $\lambda=0, \mu>2,  -\frac{1}{6} (\mu-5) (\mu+1)<\omega\leq\frac{3}{2}$

\item $\lambda>0, \mu<\frac{1}{2} \left( -\lambda^{2}-2\right) , -\frac{1}{6}
(\mu-5) (\mu+1)<\omega<\frac{\mu+1}{2}$,

\item $ \lambda>0, \mu=\frac{1}{2} \left( -\lambda^{2}-2\right) , -\frac{1}{6}
(\mu-5) (\mu+1)<\omega<\frac{3 \lambda^{2}-2 \mu^{2}+8 \mu+10}{2 \left(
\lambda^{2}+6\right) }$,

\item $\lambda>0, \frac{1}{2} \left( -\lambda^{2}-2\right) <\mu<-1,  -\frac
{1}{6} (\mu-5) (\mu+1)<\omega<\frac{\mu+1}{2}$,

\item $\lambda>0, \mu>2, -\frac{1}{6} (\mu-5) (\mu+1)<\omega\leq\frac{3}{2}$.
\end{enumerate}

Nonhyperbolic with a 1D unstable manifold for

\begin{enumerate}
\item $\lambda<0, \mu<\frac{1}{2} \left( -\lambda^{2}-2\right) , -\frac{1}{6}
(\mu-5) (\mu+1)<\omega<\frac{3 \lambda^{2}-2 \mu^{2}+8 \mu+10}{2 \left(
\lambda^{2}+6\right) }$,

\item $\lambda<0, \mu<\frac{1}{2} \left( -\lambda^{2}-2\right) , \frac{\mu
+1}{2}<\omega\leq\frac{3}{2}$,

\item $\lambda<0, \mu=\frac{1}{2} \left( -\lambda^{2}-2\right) , -\frac{1}{6}
(\mu-5)  (\mu+1)<\omega<\frac{3 \lambda^{2}-2 \mu^{2}+8 \mu+10}{2 \left(
\lambda^{2}+6\right) }$,

\item $\lambda<0, \mu=\frac{1}{2}  \left( -\lambda^{2}-2\right) , \frac{3
\lambda^{2}-2 \mu^{2}+8 \mu+10}{2 \left( \lambda^{2}+6\right) }<\omega
\leq\frac{3}{2}$,

\item $ \lambda<0, \frac{1}{2} \left( -\lambda^{2}-2\right) <\mu\leq-1,
-\frac{1}{6} (\mu-5) (\mu+1)<\omega<\frac{\mu+1}{2}$,

\item $\lambda<0, \frac{1}{2} \left( -\lambda^{2}-2\right) <\mu\leq-\frac{3
\lambda^{2}-2 \mu^{2}+8 \mu+10}{2 \left( \lambda^{2}+6\right) }<\omega
\leq\frac{3}{2}$,

\item $\lambda<0, -1<\mu<2, \frac{3 \lambda^{2}-2 \mu^{2}+8 \mu+10}{2 \left(
\lambda^{2}+6\right) }<\omega\leq\frac{3}{2}  $,

\item $\lambda<0, \mu>2, -\frac{1}{6} (\mu-5) (\mu+1)<\omega<\frac{3
\lambda^{2}-2 \mu^{2}+8 \mu+10}{2 \left( \lambda^{2}+6\right) }$,

\item $\lambda=0, \mu<-1, \frac{\mu+1}{2}<\omega\leq\frac{3}{2}$,

\item $\lambda=0, -1\leq\mu<2, -\frac{1}{6} (\mu-5) (\mu+1)<\omega\leq\frac
{3}{2}$,

\item $\lambda>0, \mu<\frac{1}{2} \left( -\lambda^{2}-2\right) , \frac{\mu
+1}{2}<\omega\leq\frac{3}{2}$,

\item $\lambda>0, \mu=\frac{1}{2} \left( -\lambda^{2}-2\right) , \frac{3
\lambda^{2}-2 \mu^{2}+8 \mu+10}{2 \left( \lambda^{2}+6\right) }<\omega
\leq\frac{3}{2}$,

\item $\lambda>0, \frac{1}{2} \left( -\lambda^{2}-2\right) <\mu\leq-1,
\frac{\mu+1}{2}<\omega\leq\frac{3}{2}$,

\item $\lambda>0, -1<\mu<2, -\frac{1}{6} (\mu-5) (\mu+1)<\omega\leq\frac{3}%
{2}$.
\end{enumerate}

The total EoS parameter and the fractional matter density at the equilibrium
point $C_{\pm}$ are the following: $w_{tot}(C_{\pm})=-\frac{\mu+6\omega
-11}{3(\mu-2\omega+1)}$.~They represent solutions dominated by the scalar
field and they are accelerating (i.e., $w_{tot}<-\frac{1}{3}$) for

\begin{enumerate}
\item $\mu<2, \frac{\mu+1}{2}<\omega<\frac{3}{2}$, or

\item $\mu>2, \frac{3}{2}<\omega<\frac{\mu+1}{2}$.
\end{enumerate}

In particular, for $\omega= \frac{\mu}{6}+\frac{7}{6}$, it represents a de
Sitter solution. It is is decelerating for

\begin{enumerate}

\item $\omega<\frac{3}{2}, \mu>2 \omega-1$, or

\item $\omega>\frac{3}{2}, \mu<2 \omega-1$.
\end{enumerate}

For $\omega= \frac{11}{6}-\frac{\mu}{6}$ it mimics a dust fluid.
\end{enumerate}

\subsubsection{Center manifold calculations}

Now, we proceed to the center manifold calculation of $C_{+}$, and $C_{-}$.

\paragraph{\textbf{The center manifold of $C_{+}$}:}

For the calculation of the center manifold of $C_{+}$, we introduce the new
variables
\begin{subequations}
\begin{align}
&  u=\frac{(\mu+1) (2 \omega-3) \left( \frac{\sqrt{\frac{2}{3}} (\mu-2)}{\mu-2
\omega+1}+x\right) }{\sqrt{2} \sqrt{(3-2 \omega) ((\mu-4) \mu+6 \omega-5)}},\\
&  v=\frac{-\sqrt{2} (\mu+1) x (2 \omega-3)+2 y \sqrt{(3-2 \omega) ((\mu-4)
\mu+6 \omega-5)}+2 \sqrt{3} (3-2 \omega)}{2 \sqrt{(3-2 \omega) ((\mu-4) \mu+6
\omega-5)}},
\end{align}
with inverse
\end{subequations}
\begin{subequations}
\label{00inverse}%
\begin{align}
&  x =\frac{\sqrt{2} \sqrt{(3-2 \omega) ((\mu-4) \mu+6 \omega-5)} \left(
u-\frac{(\mu-2) (\mu+1) (2 \omega-3)}{\sqrt{3} (\mu-2 \omega+1) \sqrt{(3-2
\omega) ((\mu-4) \mu+6 \omega-5)}}\right) }{(\mu+1) (2 \omega-3)},\\
& y = -\frac{\sqrt{3} \sqrt{(3-2 \omega) ((\mu-4) \mu+6 \omega-5)}+3 \mu u-6 u
\omega+3 u+3 \mu v-6 v \omega+3 v}{3 (-\mu+2 \omega-1)}.
\end{align}
Then, we obtain the dynamical system
\end{subequations}
\begin{subequations}
\label{00center1}%
\begin{align}
&  u^{\prime}=-\frac{9 u^{3} (\mu-2 \omega+1)^{3}}{(\mu+1)^{2} (3-2
\omega)^{2}}+\frac{6 u^{2} v (\mu-2 \omega+1)}{2 \omega-3}\nonumber\\
&  +u \left( \frac{3 v^{2} (\mu-2 \omega+1)}{2 \omega-3}+\frac{2 \sqrt{3} v
\sqrt{(3-2 \omega) ((\mu-4) \mu+6 \omega-5)}}{2 \omega-3}\right) ,\\
&  v^{\prime2 }\left( -\frac{3 (\mu-2 \omega+1)^{2} \left( \sqrt{6} \lambda(2
\omega-3)+2 \sqrt{3} \sqrt{(3-2 \omega) ((\mu-4) \mu+6 \omega-5)}\right) }{2
(\mu+1)^{2} (3-2 \omega)^{2}}-\frac{9 v (\mu-2 \omega+1)^{3}}{(\mu+1)^{2} (3-2
\omega)^{2}}\right) \nonumber\\
&  +u \left( \frac{6 v^{2} (\mu-2 \omega+1)}{2 \omega-3}+v \left( \sqrt{6}
\lambda+\frac{2 \sqrt{3} \sqrt{(3-2 \omega) ((\mu-4) \mu+6 \omega-5)}}{2
\omega-3}\right) \right) +\frac{3 v^{3} (\mu-2 \omega+1)}{2 \omega
-3}\nonumber\\
&  +v^{2} \left( \sqrt{\frac{3}{2}} \lambda+\frac{3 \sqrt{3} \sqrt{(3-2
\omega) ((\mu-4) \mu+6 \omega-5)}}{2 \omega-3}\right) \nonumber\\
&  +\frac{v \left( \sqrt{2} \lambda\sqrt{(3-2 \omega) ((\mu-4) \mu+6
\omega-5)}-2 (\mu-4) \mu-12 \omega+10\right) }{\mu-2 \omega+1}.
\end{align}

Hence, the center manifold is given by the graph $(u, v)=(u, h(u))$, with
$h(0)=0,~h^{\prime}(0)=0$, and satisfying the equation
\end{subequations}
\begin{align}
\label{00ODE1} &  u h^{\prime}(u) \left( \frac{h(u) \left( -3 (\mu+1) (h(u)+2
u)+6 \omega(h(u)+2 u)-2 \sqrt{3} \sqrt{-(2 \omega-3) ((\mu-4) \mu+6 \omega
-5)}\right) }{2 \omega-3}+\frac{9 u^{2} (\mu-2 \omega+1)^{3}}{(\mu+1)^{2} (3-2
\omega)^{2}}\right) \nonumber\\
&  -\frac{3 u^{2} (\mu-2 \omega+1)^{2} \left( 6 h(u) (\mu-2 \omega+1)+\sqrt{6}
\lambda(2 \omega-3)+2 \sqrt{3} \sqrt{-(2 \omega-3) ((\mu-4) \mu+6 \omega
-5)}\right) }{2 (\mu+1)^{2} (3-2 \omega)^{2}}\nonumber\\
&  +\frac{3 h(u)^{3} (\mu-2 \omega+1)}{2 \omega-3} +h(u)^{2} \left(
\sqrt{\frac{3}{2}} \lambda+\frac{3 \sqrt{3} \sqrt{(3-2 \omega) ((\mu-4) \mu+6
\omega-5)}}{2 \omega-3}\right) \nonumber\\
&  +\frac{h(u) \left( \sqrt{2} \lambda\sqrt{-(2 \omega-3) ((\mu-4) \mu+6
\omega-5)}-2 (\mu-4) \mu-12 \omega+10\right) }{\mu-2 \omega+1}\nonumber\\
&  +\frac{u h(u) \left( 6 h(u) (\mu-2 \omega+1)+\sqrt{6} \lambda(2 \omega-3)+2
\sqrt{3} \sqrt{-(2 \omega-3) ((\mu-4) \mu+6 \omega-5)}\right) }{2 \omega-3}=0.
\end{align}
Using the Taylor series expansion we propose the expansion
\begin{equation}
h(u)=a u^{2}+b u^{3}+c u^{4}+O\left( u^{5}\right) .
\end{equation}
Substituting this expansion into equation \eqref{00ODE1}, and equating the
coefficients of the same powers of $u$, we obtain
\begin{subequations}
\begin{align}
&  a=\frac{3 \sqrt{3} (\mu-2 \omega+1)^{3} \left( \sqrt{2} \lambda(2
\omega-3)+2 \sqrt{(3-2 \omega) ((\mu-4) \mu+6 \omega-5)}\right) }{2
(\mu+1)^{2} (3-2 \omega)^{2} \left( \sqrt{2} \lambda\sqrt{(3-2 \omega)
((\mu-4) \mu+6 \omega-5)}-2 (\mu-4) \mu-12 \omega+10\right) },\\
&  b=\frac{9 (\mu-2 \omega+1)^{4}}{2 (\mu+1)^{2} (3-2 \omega)^{2} ((\mu-4)
\mu+6 \omega-5)},\\
&  c= \frac{9 \sqrt{3} (\mu-2 \omega+1)^{5} \left( -4 (2 \mu+5) (\mu+1)
\omega+15 (\mu+1)^{2}+12 \omega^{2}\right)  \left( \sqrt{2} \lambda(2
\omega-3)+2 \sqrt{(3-2 \omega) ((\mu-4) \mu+6 \omega-5)}\right) }{8
(\mu+1)^{4} (3-2 \omega)^{4} ((\mu-4) \mu+6 \omega-5) \left( \sqrt{2}
\lambda\sqrt{(3-2 \omega) ((\mu-4) \mu+6 \omega-5)}-2 (\mu-4) \mu-12
\omega+10\right) }.
\end{align}
The evolution equation on the center manifold is reduced to $u^{\prime}=O(u^{6})$,
which means that $u$ is approximately constant on the center manifold.

\begin{figure}[t]
\textbf{ \includegraphics[width=0.6\textwidth]{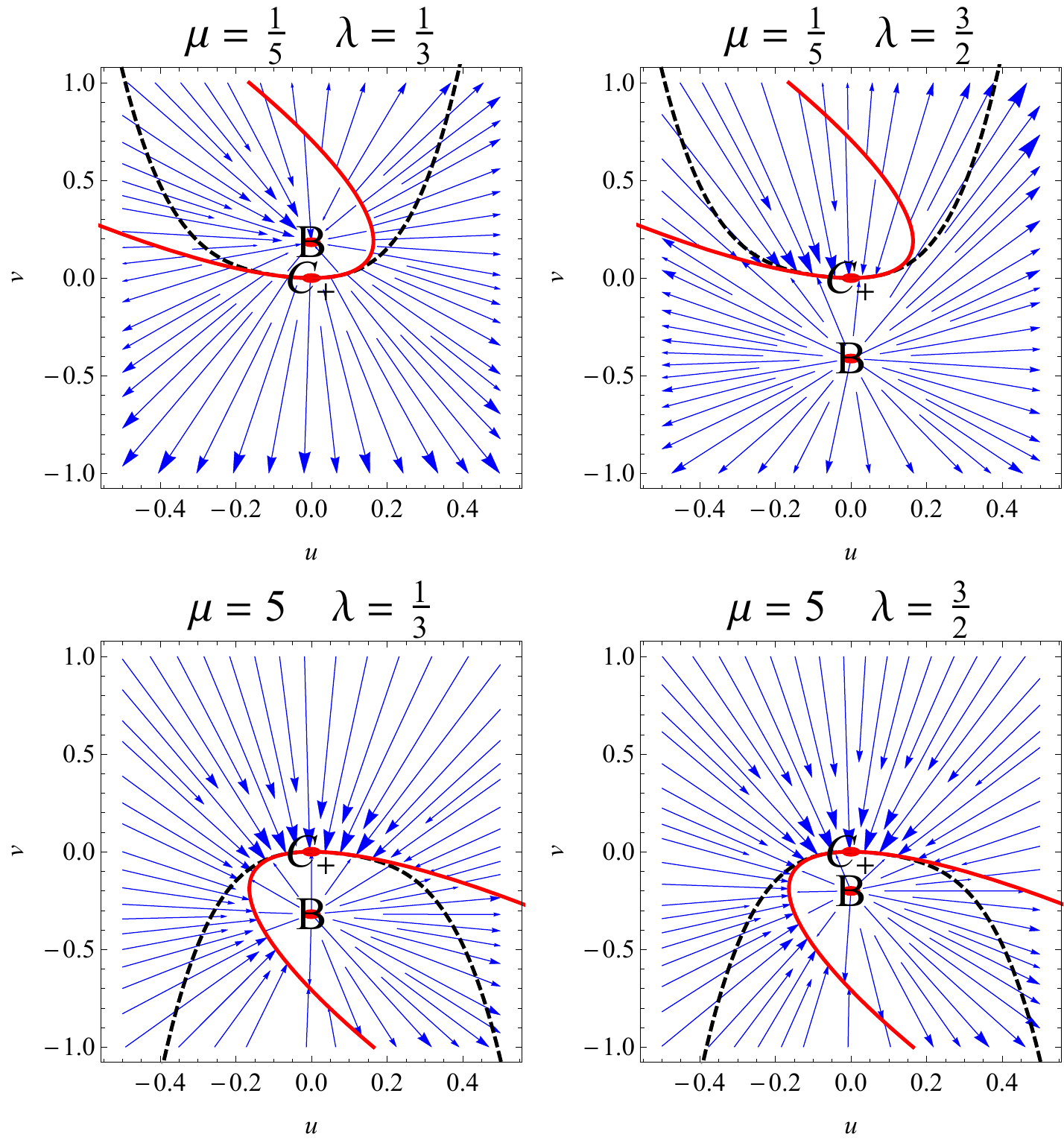} \newline}%
\caption{Qualitative evolution in the phase plane $(u,v)$ given by
\eqref{00center1} for $\omega=1$. The solid red lines denote the implicit
function $x(u) \left( x(u) \omega+\sqrt{6}\right) +y(u,v)^{2}+1=0$, with
$x(u)$ and $y(u,v)$ defined by \eqref{00inverse} for $\omega=1$, whereas the
dashed black lines represent the approximated center manifold of $C_{+}$.
Observe the accuracy for small $u$. }%
\label{Plot1}%
\end{figure}

In Fig. \ref{Plot1} we show the qualitative evolution in the phase plane $(u,v)$ given by \eqref{00center1} for $\omega = 1$. The solid red line in each plot denotes the implicit function $x(u) \left( x(u) +\sqrt{6}\right)
+y(u,v)^{2}+1=0$, with $x(u)$ and $y(u,v)$ defined by \eqref{inverse} for
$\omega = 1$, whereas the dashed black lines in each plot represent the approximated center
manifold of $C_{+}$. In Fig. \ref{Plot2} we show  the qualitative
evolution $(u,v)$ given by \eqref{00center1} for $\omega=0$. The solid red lines in each plot
denote the implicit function $\sqrt{6} x(u) +y(u,v)^{2}+1=0$, with $x(u)$ and
$y(u,v)$ defined by \eqref{00inverse} for $\omega=0$, whereas the dashed black
lines shown in each plot represent the approximated center manifold of $C_{+}$. Observe the
accuracy for small $u$.

The numerical solutions suggest that the exact center manifold of $C_{+}$ is
the line $x \left( x \omega+\sqrt{6}\right) +y^{2}+1=0$, where $x, y$ are
given by the inverse relations of \eqref{00inverse}.

Indeed, the two branches of this implicit equations are
\end{subequations}
\begin{subequations}
\begin{align}
&  h(u)\equiv v= -u-\frac{\sqrt{(3-2 \omega) ((\mu-4) \mu+6 \omega-5)}}%
{\sqrt{3} (\mu-2 \omega+1)}\nonumber\\
&  -\frac{\sqrt{-\frac{18}{\mu-2 \omega+1}+\frac{2 \omega\left( 3 u^{2} (\mu-2
\omega+1)^{2} ((\mu-4) \mu+6 \omega-5)-(\mu+1)^{2} (2 \omega-3) ((\mu-10)
\mu+12 \omega-2)\right) }{(\mu+1)^{2} (2 \omega-3) (\mu-2 \omega+1)^{2}}%
+\frac{2 \sqrt{3} u \sqrt{(3-2 \omega) ((\mu-4) \mu+6 \omega-5)}}{\mu-2
\omega+1}+3}}{\sqrt{3}}\\
&  h(u)\equiv v=-u -\frac{\sqrt{(3-2 \omega) ((\mu-4) \mu+6 \omega-5)}}%
{\sqrt{3} (\mu-2 \omega+1)}\nonumber\\
& +\frac{\sqrt{-\frac{18}{\mu-2 \omega+1}+\frac{2 \omega\left( 3 u^{2} (\mu-2
\omega+1)^{2} ((\mu-4) \mu+6 \omega-5)-(\mu+1)^{2} (2 \omega-3) ((\mu-10)
\mu+12 \omega-2)\right) }{(\mu+1)^{2} (2 \omega-3) (\mu-2 \omega+1)^{2}}%
+\frac{2 \sqrt{3} u \sqrt{(3-2 \omega) ((\mu-4) \mu+6 \omega-5)}}{\mu-2
\omega+1}+3}}{\sqrt{3}} .\label{47.b}%
\end{align}
Both are solutions of the equation \eqref{00ODE1}. But the only one that
satisfies the tangential conditions at the origin, say $h(0)=0, h^{\prime
}(0)=0$, is \eqref{47.b}. The exact equation that dictates the dynamics at the
center manifold is $u^{\prime}=0$. It means that $u$ is constant at the center manifold.

\begin{figure}[t]
\textbf{ \includegraphics[width=0.6\textwidth]{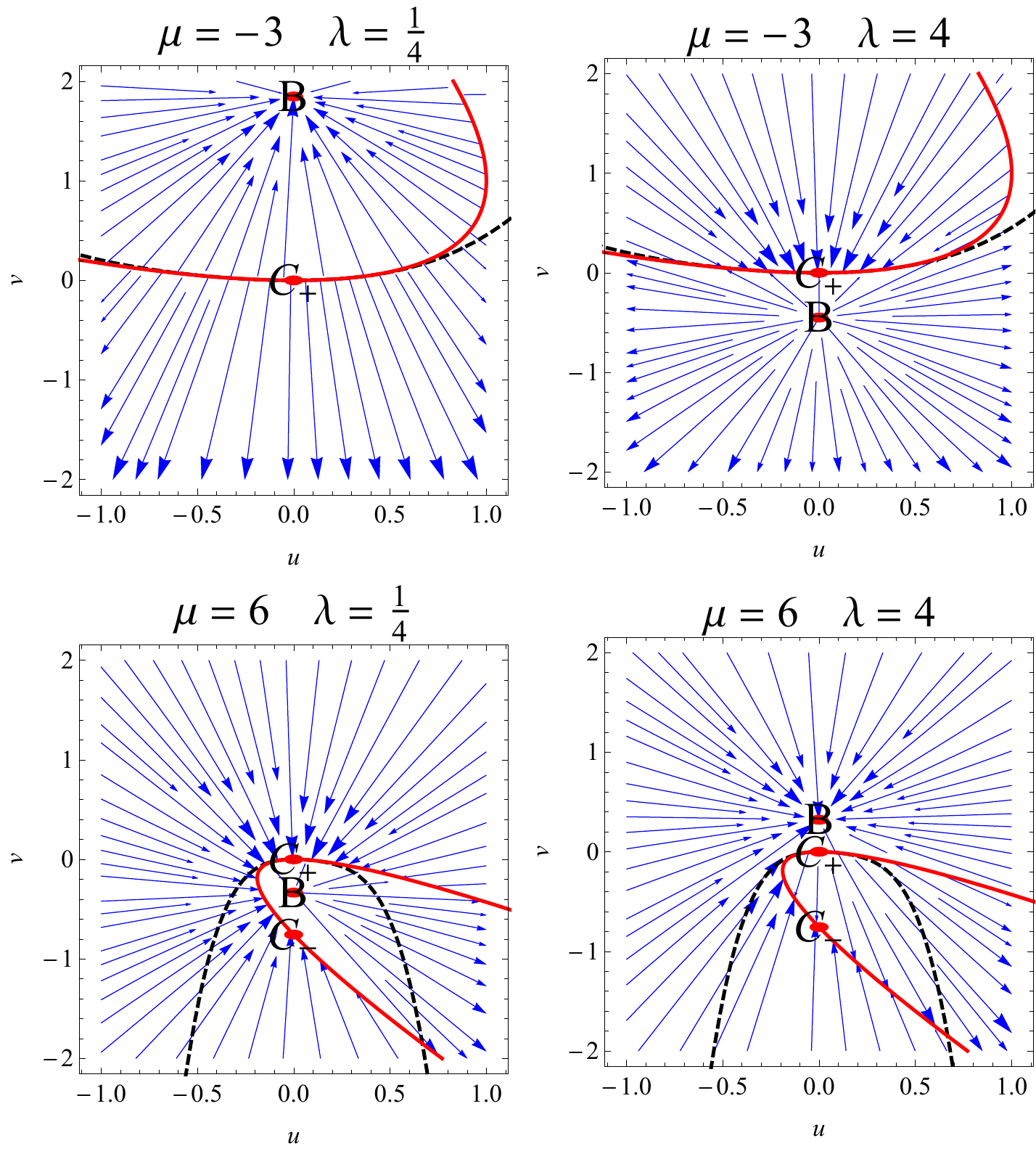} \newline}%
\caption{Qualitative evolution in the phase plane $(u,v)$ given by
\eqref{00center1} for $\omega=0$. The solid red lines denote the implicit
function $x(u) \left( x(u) \omega+\sqrt{6}\right) +y(u,v)^{2}+1=0$, with
$x(u)$ and $y(u,v)$ defined by \eqref{00inverse} for $\omega=0$, whereas the
dashed black lines represent the approximated center manifold of $C_{+}$.
Observe the accuracy for small $u$. }%
\label{Plot2}%
\end{figure}

\paragraph{\textbf{The center manifold of $C_{-}$}: }

For the calculation of the center manifold of $C_{-}$, we introduce the new
variables
\end{subequations}
\begin{subequations}
\begin{align}
& u=-\frac{(\mu+1) (2 \omega-3) \left( \frac{\sqrt{\frac{2}{3}} (\mu-2)}{\mu-2
\omega+1}+x\right) }{\sqrt{2} \sqrt{(3-2 \omega) ((\mu-4) \mu+6 \omega-5)}},\\
&  v=\frac{\sqrt{2} (\mu+1) x (2 \omega-3)+2 y \sqrt{(3-2 \omega) ((\mu-4)
\mu+6 \omega-5)}+2 \sqrt{3} (2 \omega-3)}{2 \sqrt{(3-2 \omega) ((\mu-4) \mu+6
\omega-5)}}.
\end{align}
with inverse
\end{subequations}
\begin{subequations}
\label{inverse01}%
\begin{align}
&  x= -\frac{\sqrt{2} \sqrt{(3-2 \omega) ((\mu-4) \mu+6 \omega-5)} \left(
\frac{(\mu-2) (\mu+1) (2 \omega-3)}{\sqrt{3} (\mu-2 \omega+1) \sqrt{(3-2
\omega) ((\mu-4) \mu+6 \omega-5)}}+u\right) }{(\mu+1) (2 \omega-3)},\\
&  y= -\frac{-\sqrt{3} \sqrt{(3-2 \omega) ((\mu-4) \mu+6 \omega-5)}+3 \mu u-6
u \omega+3 u+3 \mu v-6 v \omega+3 v}{3 (-\mu+2 \omega-1)}.
\end{align}

\begin{figure}[t]
\textbf{ \includegraphics[width=0.6\textwidth]{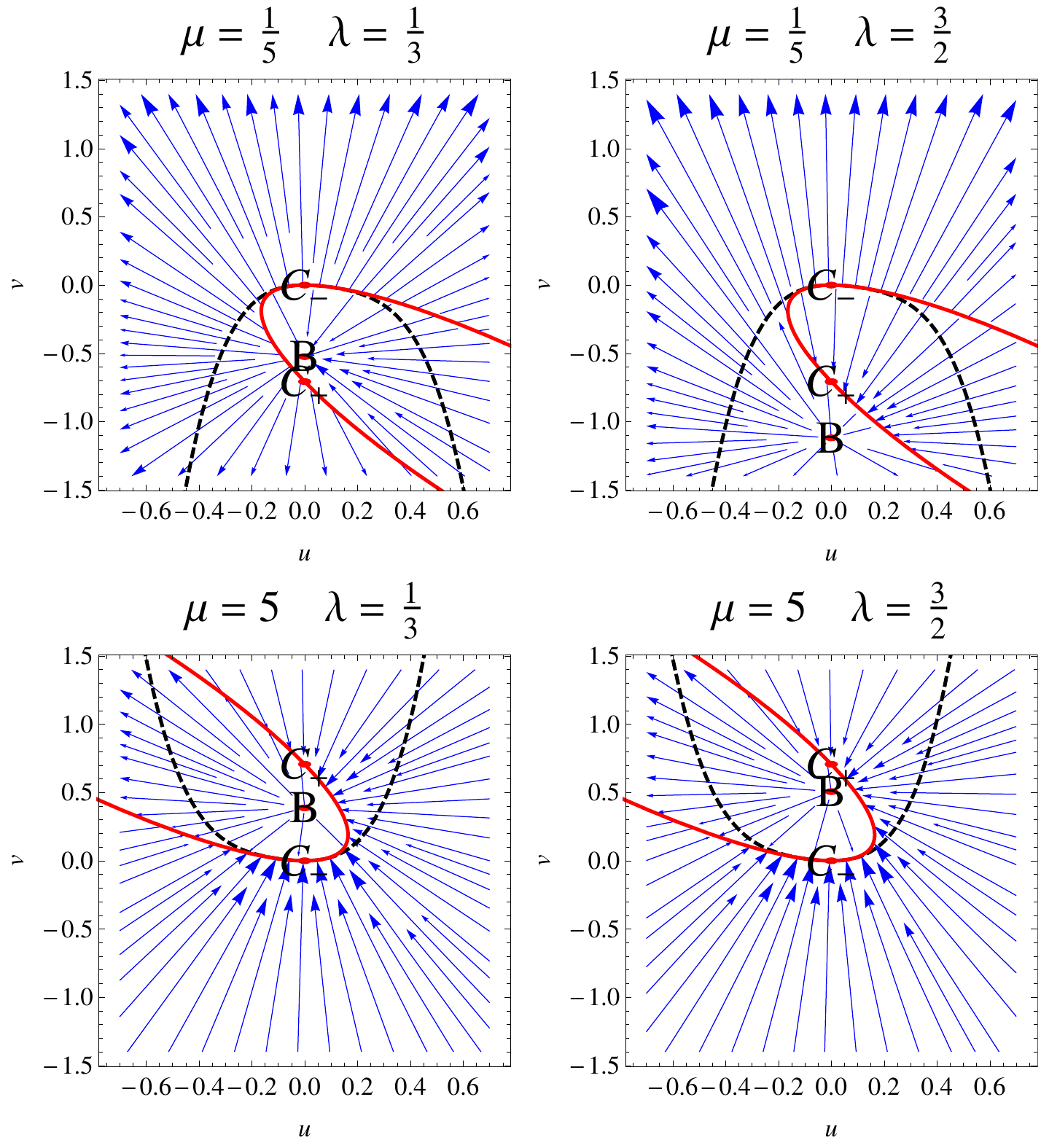} \newline}%
\caption{Qualitative evolution in the phase plane $(u,v)$ given by
\eqref{center01} for $\omega = 1$. The solid red line denotes the implicit
function $x(u) \left( x(u) \omega+\sqrt{6}\right) +y(u,v)^{2}+1=0$, with
$x(u)$ and $y(u,v)$ defined by \eqref{inverse01} for $\omega=1$, whereas the
dashed black lines represents the approximated center manifold of $C_{+}$.
Observe the accuracy for small $u$. }%
\label{Plot3}%
\end{figure}

Then, we obtain the dynamical system
\end{subequations}
\begin{subequations}
\label{center01}%
\begin{align}
&  u^{\prime}=-\frac{9 u^{3} (\mu-2 \omega+1)^{3}}{(\mu+1)^{2} (3-2
\omega)^{2}}+\frac{6 u^{2} v (\mu-2 \omega+1)}{2 \omega-3}+u \left( \frac{3
v^{2} (\mu-2 \omega+1)}{2 \omega-3}+\frac{2 \sqrt{3} v \sqrt{(3-2 \omega)
((\mu-4) \mu+6 \omega-5)}}{3-2 \omega}\right) ,\\
&  v^{\prime2 }\left( -\frac{3 \sqrt{3} (\mu-2 \omega+1)^{2} \left( \sqrt{2}
\lambda(2 \omega-3)-2 \sqrt{(3-2 \omega) ((\mu-4) \mu+6 \omega-5)}\right) }{2
(\mu+1)^{2} (3-2 \omega)^{2}}-\frac{9 v (\mu-2 \omega+1)^{3}}{(\mu+1)^{2} (3-2
\omega)^{2}}\right) \nonumber\\
&  +u \left( \frac{6 v^{2} (\mu-2 \omega+1)}{2 \omega-3}+v \left( \sqrt{6}
\lambda-\frac{2 \sqrt{3} \sqrt{(3-2 \omega) ((\mu-4) \mu+6 \omega-5)}}{2
\omega-3}\right) \right) +\frac{3 v^{3} (\mu-2 \omega+1)}{2 \omega
-3}\nonumber\\
&  +v^{2} \left( \sqrt{\frac{3}{2}} \lambda-\frac{3 \sqrt{3} \sqrt{(3-2
\omega) ((\mu-4) \mu+6 \omega-5)}}{2 \omega-3}\right) \nonumber\\
&  -\frac{v \left( \sqrt{2} \lambda\sqrt{(3-2 \omega) ((\mu-4) \mu+6
\omega-5)}+2 (\mu-4) \mu+12 \omega-10\right) }{\mu-2 \omega+1}.
\end{align}

Hence, the center manifold is given by the graph $(u, v)=(u, h(u))$, with
$h(0)=0, h^{\prime}(0)=0$, and satisfying the equation
\end{subequations}
\begin{align}
\label{ODE01} & u h^{\prime}(u) \left( \frac{h(u) \left( -3 (\mu+1) (h(u)+2
u)+6 \omega(h(u)+2 u)+2 \sqrt{3} \sqrt{-(2 \omega-3) ((\mu-4) \mu+6 \omega
-5)}\right) }{2 \omega-3}+\frac{9 u^{2} (\mu-2 \omega+1)^{3}}{(\mu+1)^{2} (3-2
\omega)^{2}}\right) \nonumber\\
&  -\frac{3 u^{2} (\mu-2 \omega+1)^{2} \left( 6 h(u) (\mu-2 \omega+1)+\sqrt{6}
\lambda(2 \omega-3)-2 \sqrt{3} \sqrt{-(2 \omega-3) ((\mu-4) \mu+6 \omega
-5)}\right) }{2 (\mu+1)^{2} (3-2 \omega)^{2}}\nonumber\\
& +\frac{3 h(u)^{3} (\mu-2 \omega+1)}{2 \omega-3}+h(u)^{2} \left( \sqrt
{\frac{3}{2}} \lambda-\frac{3 \sqrt{3} \sqrt{(3-2 \omega) ((\mu-4) \mu+6
\omega-5)}}{2 \omega-3}\right) \nonumber\\
&  -\frac{h(u) \left( \sqrt{2} \lambda\sqrt{-(2 \omega-3) ((\mu-4) \mu+6
\omega-5)}+2 (\mu-4) \mu+12 \omega-10\right) }{\mu-2 \omega+1}\nonumber\\
&  +\frac{u h(u) \left( 6 h(u) (\mu-2 \omega+1)+\sqrt{6} \lambda(2 \omega-3)-2
\sqrt{3} \sqrt{-(2 \omega-3) ((\mu-4) \mu+6 \omega-5)}\right) }{2 \omega-3}=0
\end{align}
Using the Taylor series we propose the expansion
\begin{equation}
h(u)=a u^{2}+b u^{3}+c u^{4}+O\left( u^{5}\right) .
\end{equation}
Substituting this expansion in the equation \eqref{ODE01}, and equating the
coefficients of the same powers of $u$, to obtain
\begin{subequations}
\begin{align}
&  a= -\frac{3 \sqrt{3} (\mu-2 \omega+1)^{3} \left( \sqrt{2} \lambda(2
\omega-3)-2 \sqrt{(3-2 \omega) ((\mu-4) \mu+6 \omega-5)}\right) }{2
(\mu+1)^{2} (3-2 \omega)^{2} \left( \sqrt{2} \lambda\sqrt{(3-2 \omega)
((\mu-4) \mu+6 \omega-5)}+2 (\mu-4) \mu+12 \omega-10\right) },\\
&  b= \frac{9 (\mu-2 \omega+1)^{4}}{2 (\mu+1)^{2} (3-2 \omega)^{2} ((\mu-4)
\mu+6 \omega-5)},\\
&  c= \frac{9 \sqrt{3} (\mu-2 \omega+1)^{5} \left( \sqrt{2} \lambda(2
\omega-3)-2 \sqrt{(3-2 \omega) ((\mu-4) \mu+6 \omega-5)}\right) }{2
(\mu+1)^{2} (2 \omega-3)^{3} ((\mu-4) \mu+6 \omega-5) \left( \sqrt{2}
\lambda\sqrt{(3-2 \omega) ((\mu-4) \mu+6 \omega-5)}+2 (\mu-4) \mu+12
\omega-10\right) }.
\end{align}
The evolution equation on the center manifold is reduced to
\end{subequations}
\begin{equation}
u^{\prime}=O\left( u^{5}\right) .
\end{equation}

\begin{figure}[ptb]
\textbf{ \includegraphics[width=0.6\textwidth]{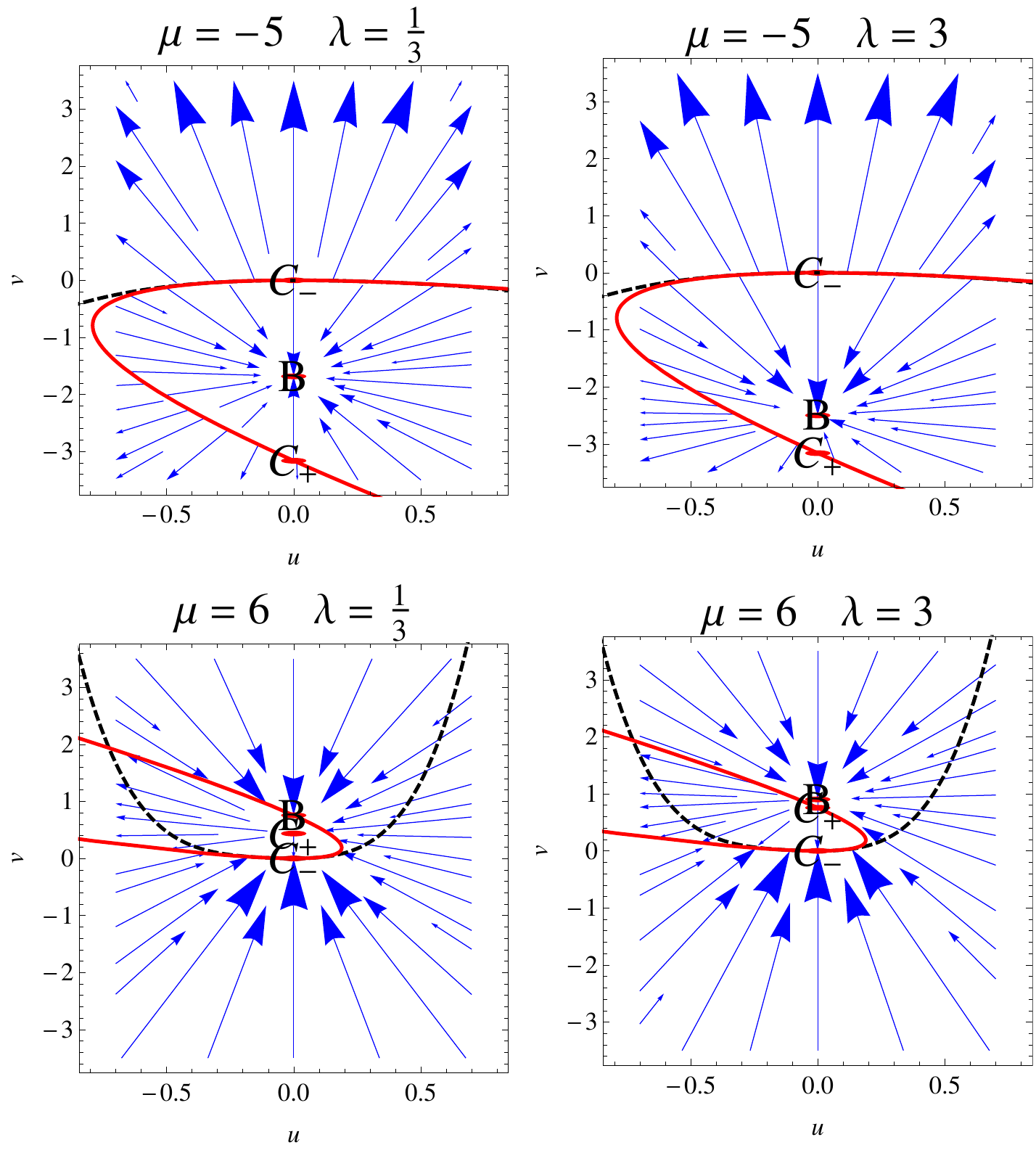} \newline}%
\caption{Qualitative evolution in the phase plane $(u,v)$ given by
\eqref{center01} for $\omega=0$. The solid red line denotes the implicit
function $x(u) \left( x(u) \omega+\sqrt{6}\right) +y(u,v)^{2}+1=0$, with
$x(u)$ and $y(u,v)$ defined by \eqref{inverse01} for $\omega=0$, whereas the
dashed black lines represents the approximated center manifold of $C_{+}$. Let us observe the accuracy for small $u$. }%
\label{Plot4}%
\end{figure}

In Fig. \ref{Plot3}, we show the qualitative evolution in the phase
plane $(u,v)$ given by \eqref{center01} for $\omega = 1$. The solid red lines
denote the implicit function $x(u) \left( x(u) +\sqrt{6}\right)
+y(u,v)^{2}+1=0$, with $x(u)$ and $y(u,v)$ defined by \eqref{inverse01} for
$\omega = 1$. The dashed black lines in Fig. \ref{Plot3} represent the approximated center
manifold of $C_{-}$. In Fig. \ref{Plot4}, we present the qualitative
evolution of $(u,v)$ given by \eqref{center01} for $\omega=0$. The solid red lines
denote the implicit function $\sqrt{6} x(u) +y(u,v)^{2}+1=0$, with $x(u)$ and
$y(u,v)$ defined by \eqref{inverse01} for $\omega=0$, whereas the dashed black
lines represent the approximated center manifold of $C_{-}$. Let us observe the accuracy for small $u$.

There are two branches of $C$:
\begin{subequations}
\begin{align}
&  v= h(u):=-u+\frac{\sqrt{(3-2 \omega) ((\mu-4) \mu+6 \omega-5)}}{\sqrt{3}
(\mu-2 \omega+1)}\nonumber\\
&  +\frac{\sqrt{-\frac{2 (3-2 \omega)^{2} \omega}{(\mu-2 \omega+1)^{2}}%
-\frac{2 (3-2 \omega)^{2}}{\mu-2 \omega+1}+\frac{36 u^{2} \omega^{2}}%
{(\mu+1)^{2} (2 \omega-3)}-\frac{36 u^{2} \omega}{(\mu+1) (2 \omega-3)}%
+\omega\left( \frac{6 u^{2}}{2 \omega-3}-2\right) -\frac{2 \sqrt{3} u
\sqrt{(3-2 \omega) ((\mu-4) \mu+6 \omega-5)}}{\mu-2 \omega+1}+3}}{\sqrt{3}},\\
&  v= h(u):= -u+\frac{\sqrt{(3-2 \omega) ((\mu-4) \mu+6 \omega-5)}}{\sqrt{3}
(\mu-2 \omega+1)}\nonumber\\
&  -\frac{\sqrt{-\frac{2 (3-2 \omega)^{2} \omega}{(\mu-2 \omega+1)^{2}}%
-\frac{2 (3-2 \omega)^{2}}{\mu-2 \omega+1}+\frac{36 u^{2} \omega^{2}}%
{(\mu+1)^{2} (2 \omega-3)}-\frac{36 u^{2} \omega}{(\mu+1) (2 \omega-3)}%
+\omega\left( \frac{6 u^{2}}{2 \omega-3}-2\right) -\frac{2 \sqrt{3} u
\sqrt{(3-2 \omega) ((\mu-4) \mu+6 \omega-5)}}{\mu-2 \omega+1}+3}}{\sqrt{3}%
},\label{40.b}%
\end{align}
Both are solutions of the equation \eqref{ODE01}. But the only one that
satisfies the tangential conditions at the origin, say $h(0)=0, h^{\prime
}(0)=0$, is \eqref{40.b}. In this case the exact equation that dictates the
dynamics over the center manifold of $C_{-}$ is $u^{\prime}=0$. It means that
$u$ is constant at the center manifold.

\subsection{Matter case}

In this section we determine the equilibrium points of the dimensionless
dynamical system (\ref{bd2.014}), (\ref{bd2.015}), (\ref{bd2.016}). For each
critical point $P$ we calculate the physical quantities of the solution, which
are the equation of state parameter for the effective fluid, i.e.
$w_{tot}\left(  P\right)$, the energy density of the dust fluid source,
$\Omega_{m}\left(  P\right)  $. As far as concerns the quantity $\Omega=x
\left(  x \omega+\sqrt{6}\right)  +y^{2}+z$, this describes the effective
energy density for the two scalar fields. Recall that there is an interaction
between the two scalar fields and the interaction in the dynamical term is
included in the variable $z$. Hence, there is no unique definition of
$\Omega_{\phi}$ and $\Omega_{\psi}$.

In order to avoid any selection of $\Omega_{\phi}$ and $\Omega_{\psi}$ which
can lead to false conclusion of the nature of the two scalar fields at the
equilibrium points, we choose to see the two-scalar fields as an effective
fluid. The only cases where a conclusion for the nature of the scalar field
can be done is, if at the critical point $P,$ one of the coordinates is zero,
that means, $x\left(  P\right)  =0\,,~$or $y\left(  P\right)  =0$, or $z\left(
P\right)  =0$. When $x\left(  P\right)  =0$, by definition $\phi
=\mbox{const}\,$\ which means that the Brans-Dicke field does not contribute to the
physical state of the solution. Moreover, when $y\left(  P\right)  =0$, it
follows $\psi= \mbox{const}\,$, where now the scalar field $\psi$ does not contribute
in the evolution of the universe. Finally, when $z\left(  P\right)  =0$, then only
the kinetic parts of the two scalar fields, namely, $\phi$ and $\psi$ contribute in
the cosmological evolution at the specific point.

Since $\Omega_{m}\geq0$, therefore, from the restriction
\end{subequations}
\begin{equation}
\Omega_{m}=x \left(  x \omega+\sqrt{6}\right)  +y^{2}+z+1
\end{equation}
we have that the phase space is given by
\begin{equation}
\left\{  (x,y,z)\in\mathbb{R}^{3}: 0\leq x \left(  x \omega+\sqrt{6}\right)
+y^{2}+z+1\leq1\right\}
\end{equation}

In this example we study the equilibrium points and their stability of the
dimensional system (\ref{bd2.014}), (\ref{bd2.015}), (\ref{bd2.016}).

\subsubsection{Equilibrium points for $\omega=1$}

\begin{enumerate}
\item $A_{1\pm}: (x,y,z)=\left(  x_{c}, \pm\sqrt{-x_{c} \left(  x_{c}+\sqrt
{6}\right)  -1}, 0\right)  $.

Eigenvalues: $\left\{  0,\sqrt{6} x_{c}+3,\sqrt{6} (\mu+1) x_{c}+\sqrt{6}
\epsilon\lambda\sqrt{-x_{c} \left(  x_{c}+\sqrt{6}\right)  -1} +6\right\} $,
$\epsilon=\pm1$.

$A_{1+}$ is nonhyperbolic with a 2D stable manifold for

\begin{enumerate}
\item $x_{c}=-\frac{1+\sqrt{3}}{\sqrt{2}}, \mu>\frac{\sqrt{3}-1}{1+\sqrt{3}}$, or

\item $\lambda\leq0,  -\frac{1+\sqrt{3}}{\sqrt{2}}<x_{c}<-\sqrt{\frac{3}{2}},
\mu>-\frac{x_{c} \sqrt{-\frac{\lambda^{2} \left( x_{c}^{2}+\sqrt{6}
x_{c}+1\right) }{x_{c}^{2}}}+x_{c}+\sqrt{6}}{x_{c}}$, or

\item $\lambda>0, -\frac{1+\sqrt{3}}{\sqrt{2}}<x_{c}<-\sqrt{\frac{3}{2}},
\mu>-\frac{x_{c} \left( -\sqrt{-\frac{\lambda^{2} \left( x_{c}^{2}+\sqrt{6}
x_{c}+1\right) }{x_{c}^{2}}}\right) +x_{c}+\sqrt{6}}{x_{c}}$.
\end{enumerate}

$A_{1+}$ is nonhyperbolic with a 2D unstable manifold when

\begin{enumerate}
\item $\lambda\leq0,  -\sqrt{\frac{3}{2}}<x_{c}<-\frac{\sqrt{3}-1}{\sqrt{2}},
\mu<-\frac{x_{c} \sqrt{-\frac{\lambda^{2} \left( x_{c}^{2}+\sqrt{6}
x_{c}+1\right) }{x_{c}^{2}}}+x_{c}+\sqrt{6}}{x_{c}}$, or

\item $\lambda>0, -\sqrt{\frac{3}{2}}<x_{c}<-\frac{\sqrt{3}-1}{\sqrt{2}},
\mu<-\frac{x_{c} \left( -\sqrt{-\frac{\lambda^{2} \left( x_{c}^{2}+\sqrt{6}
x_{c}+1\right) }{x_{c}^{2}}}\right) +x_{c}+\sqrt{6}}{x_{c}}$, or

\item $x_{c}=-\frac{\sqrt{3}-1}{\sqrt{2}}, \mu<\frac{1+\sqrt{3}}{\sqrt{3}-1}$.
\end{enumerate}

$A_{1-}$ is nonhyperbolic with a 2D stable manifold for

\begin{enumerate}
\item $x_{c}=-\frac{1+\sqrt{3}}{\sqrt{2}}, \mu>\frac{\sqrt{3}-1}{1+\sqrt{3}}$, or

\item $\lambda\leq0, -\frac{1+\sqrt{3}}{\sqrt{2}}<x_{c}<-\sqrt{\frac{3}{2}},
\mu>-\frac{x_{c} \left( -\sqrt{-\frac{\lambda^{2} \left( x_{c}^{2}+\sqrt{6}
x_{c}+1\right) }{x_{c}^{2}}}\right) +x_{c}+\sqrt{6}}{x_{c}}$, or

\item $\lambda>0, -\frac{1+\sqrt{3}}{\sqrt{2}}<x_{c}<-\sqrt{\frac{3}{2}},
\mu>-\frac{x_{c} \sqrt{-\frac{\lambda^{2} \left( x_{c}^{2}+\sqrt{6}
x_{c}+1\right) }{x_{c}^{2}}}+x_{c}+\sqrt{6}}{x_{c}}$.
\end{enumerate}

$A_{1-}$ is nonhyperbolic with a 2D unstable manifold for

\begin{enumerate}

\item $\lambda\leq0, -\sqrt{\frac{3}{2}}<x_{c}<-\frac{\sqrt{3}-1}{\sqrt{2}},
\mu<-\frac{x_{c} \left( -\sqrt{-\frac{\lambda^{2} \left( x_{c}^{2}+\sqrt{6}
x_{c}+1\right) }{x_{c}^{2}}}\right) +x_{c}+\sqrt{6}}{x_{c}}$, or

\item $\lambda>0,  -\sqrt{\frac{3}{2}}<x_{c}<-\frac{\sqrt{3}-1}{\sqrt{2}},
\mu<-\frac{x_{c} \sqrt{-\frac{\lambda^{2} \left( x_{c}^{2}+\sqrt{6}
x_{c}+1\right) }{x_{c}^{2}}}+x_{c}+\sqrt{6}}{x_{c}}$, or

\item $x_{c}=-\frac{\sqrt{3}-1}{\sqrt{2}}, \mu<\frac{1+\sqrt{3}}{\sqrt{3}-1}$.
\end{enumerate}

\item $A_{2}: (x,y,z)=\left(  -\frac{\sqrt{\frac{2}{3}} (\mu-2)}{\mu-1},
\frac{\lambda}{\sqrt{6} (\mu-1)}, -\frac{\lambda^{2}-2 ((\mu-4) \mu+1)}{6
(\mu-1)^{2}}\right)  $.

Eigenvalues $\left\{  -\frac{\lambda^{2}-2 \mu^{2}+8 \mu-2}{2 (\mu-1)}%
,-\frac{\lambda^{2}-2 \mu^{2}+8 \mu-2}{2 (\mu-1)},-\frac{\lambda^{2}-2 \mu
^{2}+7 \mu-3}{\mu-1}\right\}  $. It is a sink for

\begin{enumerate}
\item $\mu\leq-1, \lambda^{2}<(\mu-3) (2 \mu-1)$, or

\item $-1<\mu<2-\sqrt{3}, \lambda^{2}<2 \left( \mu^{2}-4 \mu+1\right) $, or

\item $1<\mu<3$, or

\item $\mu\geq3, \lambda^{2}>(\mu-3) (2 \mu-1)$.
\end{enumerate}

It is a source for

\begin{enumerate}
\item $\mu\leq-1, \lambda^{2}>2 \left( \mu^{2}-4 \mu+1\right) $, or

\item $-1<\mu\leq\frac{1}{2}, \lambda^{2}>(\mu-3) (2 \mu-1)$, or

\item $\frac{1}{2}<\mu<1, \lambda^{2}\geq0$, or

\item $\mu>2+\sqrt{3}, 0\leq\lambda^{2}<2 \left( \mu^{2}-4 \mu+1\right) $.
\end{enumerate}

It is a saddle for

\begin{enumerate}
\item $\mu<-1, (\mu-3) (2 \mu-1)<\lambda^{2}<2 \left( \mu^{2}-4 \mu+1\right)
$, or

\item $-1<\mu\leq2-\sqrt{3}, 2 \left( \mu^{2}-4 \mu+1\right) <\lambda^{2}%
<(\mu-3) (2 \mu-1)$, or

\item $(2-\sqrt{3}<\mu<\frac{1}{2}, 0\leq\lambda^{2}<(\mu-3) (2 \mu-1)$, or

\item $3<\mu<2+\sqrt{3}, 0\leq\lambda^{2}<(\mu-3) (2 \mu-1)$, or

\item $\mu=2+\sqrt{3}, 0<\lambda^{2}<\left( \sqrt{3}-1\right)  \left( 3+2
\sqrt{3}\right) $, or

\item $\mu>2+\sqrt{3}, 2 \left( \mu^{2}-4 \mu+1\right) <\lambda^{2}<(\mu-3) (2
\mu-1)$.
\end{enumerate}

\item $A_{3}: (x,y,z)=\left(  -\frac{\lambda^{2}+3 \mu-3}{\sqrt{6} (\mu-1)
\mu}, \frac{\lambda}{\sqrt{6} (\mu-1)}, \frac{\lambda^{2}-3 (\mu-1)^{2}}{6
(\mu-1)^{2} \mu}\right)  $.

Eigenvalues \newline$\Big\{\frac{\lambda^{2}-3 (\mu-1)^{2}}{2 (\mu-1) \mu},
\frac{\lambda^{2}-3 (\mu-1)^{2}-\sqrt{\left(  \lambda^{2}-3 (\mu
-1)^{2}\right)  \left(  \lambda^{2} (8 \mu-1)+\mu((59-16 \mu) \mu
-30)+3\right)  }}{4 (\mu-1) \mu}$,\newline$\frac{\lambda^{2}-3 (\mu
-1)^{2}+\sqrt{\left(  \lambda^{2}-3 (\mu-1)^{2}\right)  \left(  \lambda^{2} (8
\mu-1)+\mu((59-16 \mu) \mu-30)+3\right)  }}{4 (\mu-1) \mu}\Big\}$.

This point is physical, i.e., $0\leq\Omega_{m}=\frac{3 (\mu-1)^{2} \left( 2
\mu^{2}-7 \mu+3\right) +\lambda^{4}+\left( -5 \mu^{2}+13 \mu-6\right)
\lambda^{2}}{6 (\mu-1)^{2} \mu^{2}}\leq1$, and it stable, that is, all the
eigenvalues have negative real parts for

\begin{enumerate}
\item $0<\mu\leq0.0779056, 3 (\mu-1)^{2}<\lambda^{2}\leq\frac{16 \mu^{3}-59
\mu^{2}+30 \mu-3}{8 \mu-1}$, or

\item $0.0779056<\mu<1, 3 (\mu-1)^{2}<\lambda^{2}\leq\frac{1}{2} \left( 5
\mu^{2}+\sqrt{\mu^{2} \left( 25 \mu^{2}-46 \mu+25\right) }-13 \mu+6\right) $, or

\item $\mu=3.10266, \lambda=0$,

\item $\mu=3.10266, 0<\lambda^{2}<(\mu-3) (2 \mu-1)$, or

\item $\mu>3.10266, \frac{16 \mu^{3}-59 \mu^{2}+30 \mu-3}{8 \mu-1}<\lambda
^{2}<(\mu-3) (2 \mu-1)$, or

\item $\mu>3.10266, \lambda^{2}\leq\frac{16 \mu^{3}-59 \mu^{2}+30 \mu-3}{8
\mu-1}$, or

\item $0<\mu<0.0779056, \frac{16 \mu^{3}-59 \mu^{2}+30 \mu-3}{8 \mu-1}%
<\lambda^{2}\leq\frac{1}{2} \left( 5 \mu^{2}+\sqrt{\mu^{2} \left( 25 \mu
^{2}-46 \mu+25\right) }-13 \mu+6\right) $, or

\item $3<\mu<3.10266, \lambda^{2}<(\mu-3) (2 \mu-1)$.
\end{enumerate}

This point is physical, i.e., $0\leq\Omega_{m}=\frac{3 (\mu-1)^{2} \left( 2
\mu^{2}-7 \mu+3\right) +\lambda^{4}+\left( -5 \mu^{2}+13 \mu-6\right)
\lambda^{2}}{6 (\mu-1)^{2} \mu^{2}}\leq1$, and it unstable, that is, all the
eigenvalues have positive real parts for

\begin{enumerate}
\item $\mu=0.240719, \lambda^{2}=1.10824$, or

\item $\mu=0.240719, \frac{1}{2} \left( 5 \mu^{2}-13 \mu+6\right) -\frac{1}{2}
\sqrt{25 \mu^{4}-46 \mu^{3}+25 \mu^{2}}<\lambda^{2}<2 \mu^{2}-7 \mu+3$, or

\item $0.240719<\mu\leq\frac{3}{7}, \frac{1}{2} \left( 5 \mu^{2}-13
\mu+6\right) -\frac{1}{2} \sqrt{25 \mu^{4}-46 \mu^{3}+25 \mu^{2}}\leq
\lambda^{2}\leq\frac{16 \mu^{3}-59 \mu^{2}+30 \mu-3}{8 \mu-1}$, or

\item $\frac{3}{7}<\mu<0.450783, \lambda^{2}\leq\frac{16 \mu^{3}-59 \mu^{2}+30
\mu-3}{8 \mu-1}$, or

\item $\mu=0.450783, \lambda=0$, or

\item $\mu>1, 3 \mu^{2}-6 \mu+3<\lambda^{2}\leq\frac{1}{2} \left( 5 \mu^{2}-13
\mu+6\right) +\frac{1}{2} \sqrt{25 \mu^{4}-46 \mu^{3}+25 \mu^{2}}$, or

\item $0<\mu<0.240719, \frac{1}{2} \left( 5 \mu^{2}-13 \mu+6\right) -\frac
{1}{2} \sqrt{25 \mu^{4}-46 \mu^{3}+25 \mu^{2}}\leq\lambda^{2}<2 \mu^{2}-7
\mu+3$, or

\item $0.240719<\mu\leq0.450783, \frac{16 \mu^{3}-59 \mu^{2}+30 \mu-3}{8
\mu-1}<\lambda^{2}<2 \mu^{2}-7 \mu+3$, or

\item $0.450783<\mu<\frac{1}{2}, \lambda^{2}<2 \mu^{2}-7 \mu+3$.
\end{enumerate}
\end{enumerate}

{$A_{1\pm}$ are actually surfaces in the plane $\left\{  {x},y\right\}
$ ($z=0$)} which describe a universe with the cosmological observables $\Omega_{m}(A_{1\pm})=0, w_{tot}=1+2
\sqrt{\frac{2}{3}} x_{c}$.
$A_{1\pm}$ are
real if and only if $-\sqrt{2+\sqrt{3}}\leq{x}_{c}\leq-\sqrt{2-\sqrt{3}}$,
that is, the total equation of state parameter is bounded as $-1-\frac
{2\sqrt{3}}{3}\leq w_{tot}\left(  A_{1\pm}\right)  \leq-1+\frac{2\sqrt{3}}{3}%
$.
Therefore, they represent solutions dominated by
the scalar field and they are accelerating (i.e., $w_{tot}<-\frac{1}{3}$) for
$x_{c}<-\sqrt{\frac{2}{3}}, \omega<\frac{-\sqrt{6} x_{c}-1}{x_{c}^{2}}$. For
$x_{c} -\sqrt{\frac{3}{2}}$ it is a de Sitter solution ($w_{tot}(A_{\pm}%
)=-1$). It represents a decelerating solution for $-\sqrt{\frac{2}{3}}%
<x_{c}<0, \omega<\frac{-\sqrt{6} x_{c}-1}{x_{c}^{2}}$, or $x_{c}>0,
\omega<\frac{-\sqrt{6} x_{c}-1}{x_{c}^{2}}$. In particular, it mimics a dust
solution ($w_{tot}(A_{\pm})=0$) for $x_{c}=-\frac{1}{2}\sqrt{\frac{3}{2}}$. In
the limits where $\left\vert {x}_{c}+\frac{1}{2}\sqrt{6}\right\vert
=\frac{\sqrt{2}}{2}$, only the kinetic part of the Brans-Dicke field $\phi$
contributes in the evolution of the universe.

Point $A_{2}$ exists only when $\mu\neq1$ and it describes a universe with
$\Omega_{m}\left(  A_{2}\right)  =0$ and $w_{tot}\left(  A_{2}\right)
=-\frac{1}{3}\left(  7-2\mu+\frac{\lambda^{2}}{\mu-1}\right)  $, while for
$\lambda=0$, the scalar field $\psi$ does not contribute in the total matter
component of the solution. The exact solution at the critical point describes
a de Sitter universe when $\lambda^{2}=7-9\mu+2\mu^{2}$.

\begin{figure}[t]
\textbf{ \includegraphics[width=0.4\textwidth]{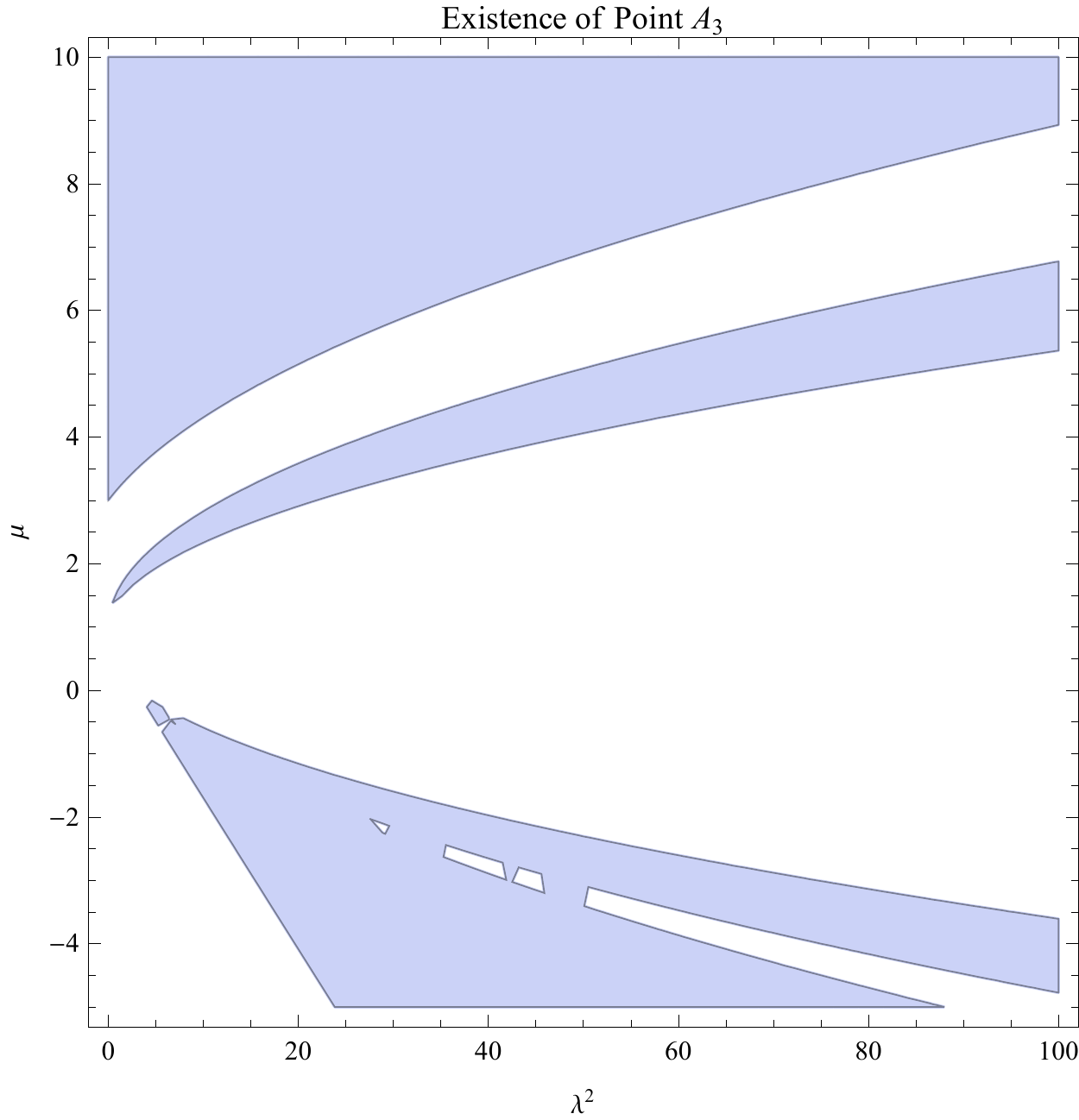} \newline}\caption{Region
plot in the space $\left\{  \lambda,\mu\right\}  $ where the critical point
$A_{3}$ is physically accepted.}%
\label{ff01}%
\end{figure}

Point $A_{3}$ exists only when $\mu\neq1$ and $\mu\neq0.$ In contrary to the
rest of the points, the energy density of the dust fluid is nonzero, that is,
$\Omega_{m}\left(  A_{3}\right)  =\frac{\left(  \lambda^{2}-3\left(
\mu-1\right)  ^{2}\right)  \left(  \lambda^{2}-3+\left(  7-2\mu\right)
\mu\right)  }{\mu^{2}\left(  \mu-1\right)  ^{2}}$. Hence, the point is
physically accepted when $0\leq\Omega_{m}\left(  A_{3}\right)  \leq1$, and from the
latter constraint it follows that for $\mu=-1,~$ parameter $\lambda$ is
constrained by%
\[
12-2\sqrt{6}\leq\lambda^{2}\leq12+2\sqrt{6}.
\]
When $\mu<-1$ and $\mu>0$ we find \textbf{ }%
\[
~\left(  5\mu-3\right)  \left(  \mu-2\right)  -\sqrt{\mu^{2}\left(
25+\mu\left(  25\mu-46\right)  \right)  }\leq2\lambda^{2}\leq6-2\left(
7-2\mu\right)  \mu,
\]

\[
6\left(  \mu-1\right)  ^{2}\leq2\lambda^{2}\leq\left(  5\mu-3\right)  \left(
\mu-2\right)  +\sqrt{\mu^{2}\left(  25+\mu\left(  25\mu-46\right)  \right)
},
\]
while when $-1<\mu<0$, the parameter $\lambda$ take values in the region
\[
\left(  5\mu-3\right)  \left(  \mu-2\right)  -\sqrt{\mu^{2}\left(
25+\mu\left(  25\mu-46\right)  \right)  }\leq2\lambda^{2}\leq6\left(
\mu-1\right)  ^{2},
\]%
\[
6-2\left(  7-2\mu\right)  \mu\leq2\lambda^{2}\leq\left(  5\mu-3\right)
\left(  \mu-2\right)  +\sqrt{\mu^{2}\left(  25+\mu\left(  25\mu-46\right)
\right)  }.
\]
The latter regions are presented in Fig. \ref{ff01}.

As far as the equation of state is concerned for the effective fluid
derived to be \newline$w_{tot}\left(  A_{3}\right)  =-\frac{1}{\mu}%
-\frac{\lambda^{2}}{3\mu\left(  \mu-1\right)  }$. From the latter it follows
that $-1\leq w_{tot}\left(  A_{3}\right)  <1$ when
\[
\mu<-1~,\left(  5\mu-3\right)  \left(  \mu-2\right)  -\sqrt{\mu^{2}\left(
25+\mu\left(  25\mu-46\right)  \right)  }<2\lambda^{2}\leq6-2\left(
7-2\mu\right)  \mu,\text{ and }\lambda^{2}=3\left(  \mu-1\right)  ^{2}%
\]
or%
\[
-1\leq\mu<0\text{ ,~}\left(  5\mu-3\right)  \left(  \mu-2\right)  -\sqrt
{\mu^{2}\left(  25+\mu\left(  25\mu-46\right)  \right)  }\leq2\lambda^{2}%
\leq6\left(  \mu-1\right)  ^{2}%
\]
\qquad or%
\[
0<\mu<1~,~6\left(  \mu-1\right)  ^{2}<2\lambda^{2}\leq\left(  5\mu-3\right)
\left(  \mu-2\right)  +\sqrt{\mu^{2}\left(  25+\mu\left(  25\mu-6\right)
\right)  }%
\]
or$~\left\{  1<\mu<3~,~\lambda^{2}=3\left(  \mu-1\right)  ^{2}\right\}
;~\left\{  \mu=3,\lambda^{2}=0,12\right\}  $ or $\left\{  \mu>3,~\lambda
^{2}\leq3-\left(  7-2\mu\right)  \mu,~\lambda^{2}=3\left(  \mu-1\right)
^{2}\right\}  $.

For large values of $\mu$ and for finite value of $\lambda^{2}$,
$w_{tot}\left(  A_{3}\right)  \rightarrow0$, which means that the effective
fluid mimics the dust fluid source. On the other hand, point $A_{3}$ describes
a de Sitter universe when $\lambda^{2}=3\left(  1-\mu^{2}\right)  \,\ $which
can not be true for any value of $\mu$ in order $\left\{  \lambda,\mu\right\}
$ belongs to the region of values where point~$A_{3}$ is physically accepted.

\begin{figure}[t]
\textbf{ \includegraphics[width=1\textwidth]{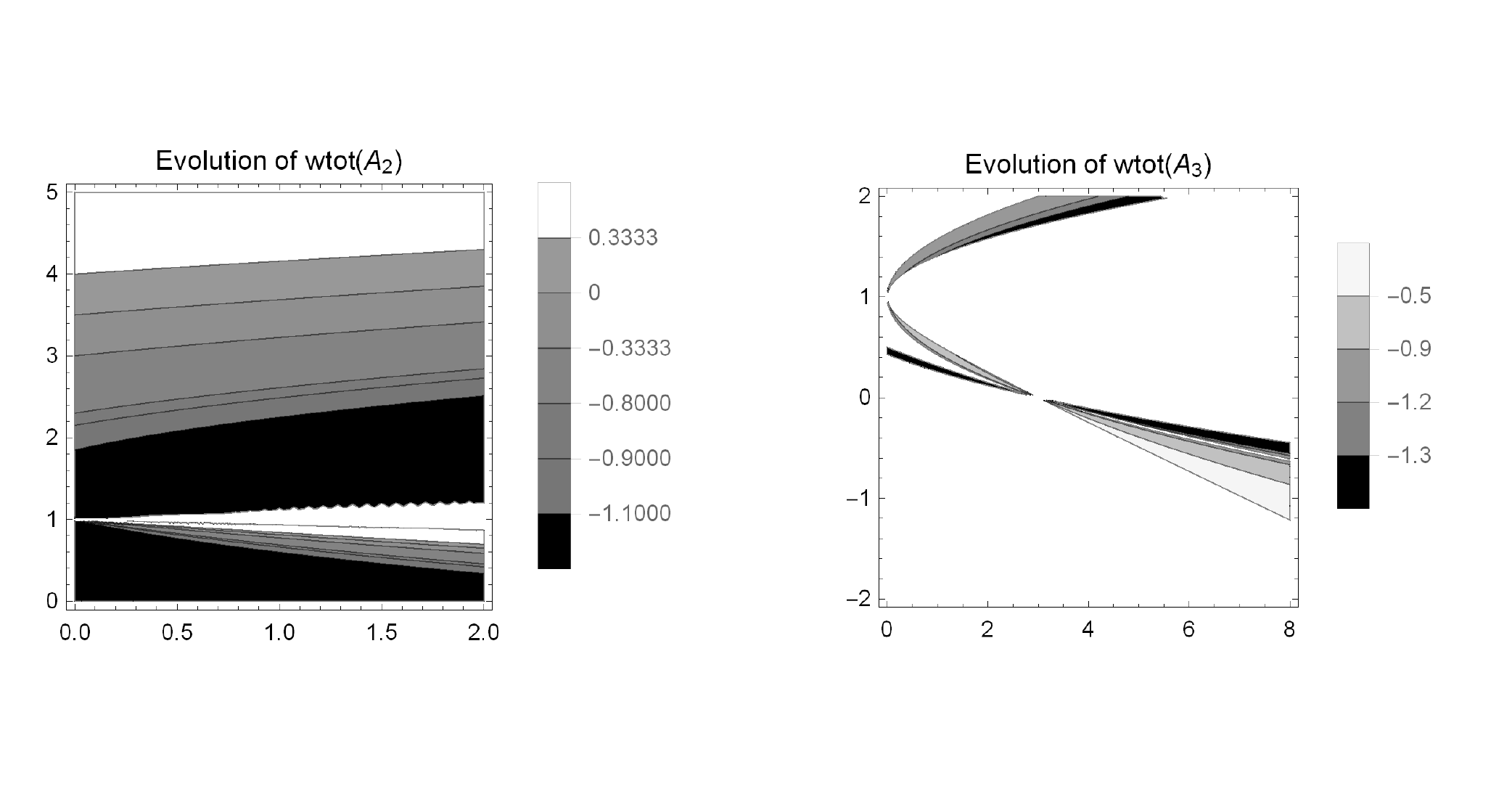} \newline}%
\caption{Qualitative evolution of the equation of state parameter $w_{tot}$ at
the equilibrium points $A_{2}~$ (left graph) and $A_{3}$ (right graph). }%
\label{ff02}%
\end{figure}

In Fig. \ref{ff02} we present the contour-plots for the effective equation of state
parameters $w_{tot}\left(  A_{2}\right)  $ and $w_{tot}\left(  A_{3}\right)  $
in the region space of the variables $\left\{  \lambda,\mu\right\}  $.

We continue our analysis by studying the stability conditions of the critical
points. As we shall see, one can easily conclude about the stability of the
equilibrium points $A_{2},~A_{3}$, while for points $A_{1\pm}$ the center
manifold theorem should be applied.

The eigenvalues of the linearized system around point $A_{2}$ are found to be%
\begin{equation}
e_{1}\left(  A_{2}\right)  =e_{2}\left(  A_{2}\right)  =-\frac{\lambda^{2}-2
\mu^{2}+8 \mu-2}{2 (\mu-1)}~,~e_{3}\left(  A_{3}\right)  =-\frac{\lambda^{2}-2
\mu^{2}+7 \mu-3}{\mu-1},
\end{equation}
from where we infer that the point is stable when \newline$\left\{  \mu\leq-1,
\lambda^{2}<(\mu-3) (2 \mu-1) \right\}  ,$ or \newline$\left\{  -1<\mu
<2-\sqrt{3},~ 0\leq\lambda^{2}<2 \left(  \mu^{2}-4 \mu+1\right)  \right\}  , ~
$ or \newline$\left\{  1<\mu<3, \lambda\in\mathbb{R}\right\}  ,$ or
\newline$\left\{  \mu\geq3,~ \lambda^{2}>(\mu-3) (2 \mu-1)\right\}  .$

As far as the critical point $A_{3}$ is concerned, the eigenvalues of the
linearized system are found to be%
\begin{equation}
e_{1}\left(  A_{3}\right)  =\frac{\lambda^{2}-3 (\mu-1)^{2}}{2 (\mu-1) \mu
}~,~e_{2\pm}\left(  A_{3}\right)  =\frac{\lambda^{2}-3 (\mu-1)^{2}\pm
\sqrt{f\left(  \lambda,\mu\right)  }}{4 (\mu-1) \mu},
\end{equation}
where
\begin{equation}
f\left(  \lambda,\mu\right)  =\left(  3 (\mu-1)^{2}-\lambda^{2}\right)
\left(  \lambda^{2} (8 \mu-1)+\mu((59-16 \mu) \mu-30)+3\right)  .
\end{equation}

\begin{figure}[t]
\textbf{ \includegraphics[width=1\textwidth]{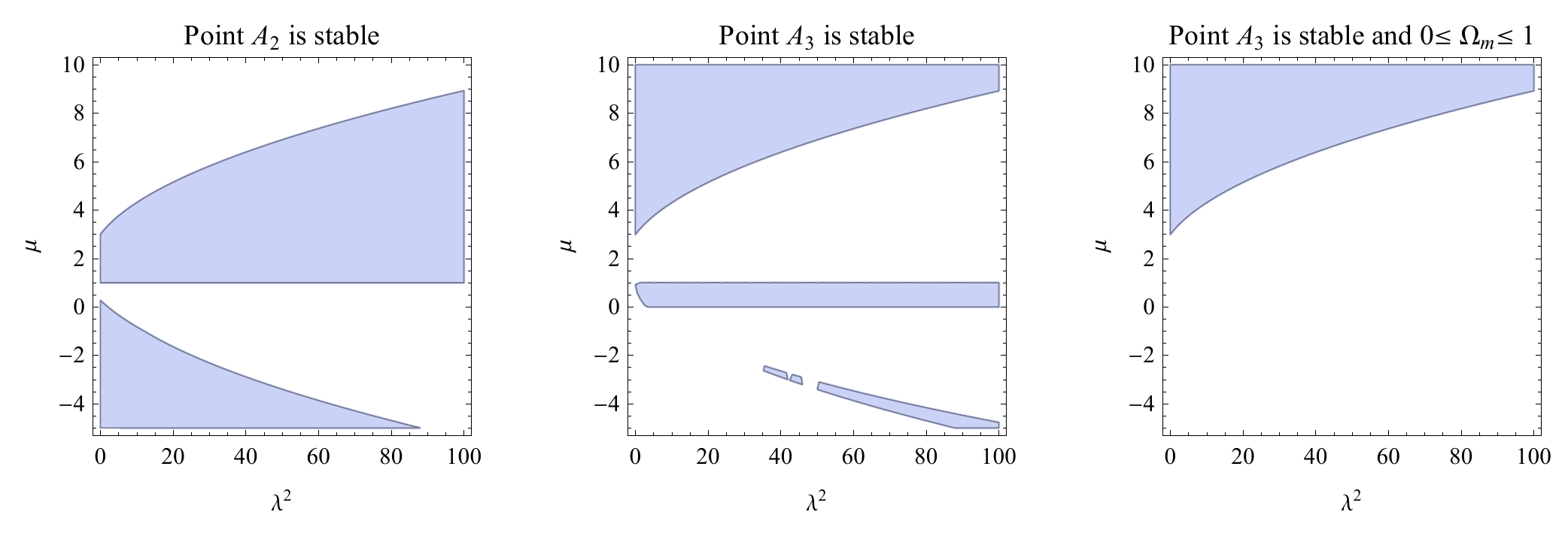} \newline}\caption{Region
plots in the space of variables $\left\{  \lambda,\mu\right\}  $ where the two
points $A_{2}$ (left plot) and $A_{3}$ (middle plot) are stable. In the
right plot it is superimposed to the stability region, the condition
$0\leq\Omega_{m}\leq1$, where the critical point $A_{3}$ is physically
acceptable.}%
\label{ff03}%
\end{figure}

Because of the nonlinear terms one can solve the stability conditions
$\operatorname{Re}\left(  e_{1}\right)  <0,$ $\operatorname{Re}\left(
e_{2\pm}\right)  <0$ numerically. The regions of values $\left\{  \lambda
,\mu\right\}  $ where the point is an attractor are presented in Fig.
\ref{ff03}. In addition in Fig. \ref{ff03} we show the regions in the space
$\left\{  \lambda,\mu\right\}  $ where point $A_{2}$ is stable. For point
$A_{3}$ we show also the regions where it is stable and satisfies the
condition $0\leq\Omega_{m}\leq1$ (physically acceptable). Combining both
conditions we obtain that $A_{3}$ ``exists'' ($0\leq\Omega_{m}\leq1$) and it
is stable for \newline$\left\{  0<\mu\leq0.0779056, 3 (\mu-1)^{2}<\lambda
^{2}\leq\frac{16 \mu^{3}-59 \mu^{2}+30 \mu-3}{8 \mu-1}\right\}  ,$ or
\newline$\left\{  0.0779056<\mu<1, 3 (\mu-1)^{2}<\lambda^{2}\leq\frac{1}{2}
\left(  5 \mu^{2}+\sqrt{\mu^{2} \left(  25 \mu^{2}-46 \mu+25\right)  }-13
\mu+6\right)  \right\}  ,$ or $\left\{  \mu=3.10266, \lambda=0\right\}  ,$ or
\newline$\left\{  \mu=3.10266, 0<\lambda^{2}<(\mu-3) (2 \mu-1)\right\}  ,$ or
$\left\{  \mu>3.10266, \frac{16 \mu^{3}-59 \mu^{2}+30 \mu-3}{8 \mu-1}%
<\lambda^{2}<(\mu-3) (2 \mu-1)\right\}  ,$ or \newline$\left\{  \mu>3.10266,
0\leq\lambda^{2}\leq\frac{16 \mu^{3}-59 \mu^{2}+30 \mu-3}{8 \mu-1}\right\}  ,$
or \newline$\left\{  0<\mu<0.0779056, \frac{16 \mu^{3}-59 \mu^{2}+30 \mu-3}{8
\mu-1}<\lambda^{2}\leq\frac{1}{2} \left(  5 \mu^{2}+\sqrt{\mu^{2} \left(  25
\mu^{2}-46 \mu+25\right)  }-13 \mu+6\right)  \right\}  ,$ or \newline$\left\{
3<\mu<3.10266, 0\leq\lambda^{2}<(\mu-3) (2 \mu-1)\right)  \}.$

\paragraph{Center manifold theorem for $A_{1\pm}$:}

The parametric equation for $A_{1\pm}$ can be derived from
\begin{equation}
\left( x+\sqrt{\frac{3}{2}}\right) ^{2}+y^{2}=\frac{1}{2},
\end{equation}
from which we can define
\begin{equation}
x=-\sqrt{\frac{3}{2}}+\frac{\sqrt{2}}{2}\cos\theta, \quad y= \frac{\sqrt{2}%
}{2}\sin\theta.
\end{equation}
The linearization matrix evaluated at $A_{1\pm}$ is then
\begin{equation}
J= \left(
\begin{array}
[c]{ccc}%
\sqrt{3} \cos(\theta) & \sqrt{3} \sin(\theta) & \frac{3 (\mu-1) \cos
(\theta)-\sqrt{3} \mu}{\sqrt{2}}\\
0 & 0 & \frac{3 (\mu-1) \sin(\theta)-\sqrt{3} \lambda}{\sqrt{2}}\\
0 & 0 & -3 \mu+\sqrt{3} (\mu+1) \cos(\theta)+\sqrt{3} \lambda\sin(\theta)+3\\
&  &
\end{array}
\right) ,
\end{equation}
with eigenvalues
\begin{equation}
\left\{ 0, \quad\sqrt{3} \cos(\theta), \quad\sqrt{3} \lambda\sin(\theta
)+\sqrt{3} (\mu+1) \cos(\theta)-3 \mu+3\right\}
\end{equation}
Due to the dynamics on the sets $A_{1\pm}$, which is essentially two-dimensional, to
study the dynamics over $A_{1\pm}$, we project the flow on the plane $(\theta,
z)$, where the dynamics is given by
\begin{align}
&  \theta^{\prime}=z \sec(\theta) \left( 3 (\mu-1) \sin(\theta)-\sqrt{3}
\lambda\right) ,\\
&  z^{\prime}=z \left( \sqrt{3} \lambda\sin(\theta)+\sqrt{3} (\mu+1)
\cos(\theta)+3 (\mu-1) (2 z-1)\right) .
\end{align}
For the calculation of the center manifold of $A_{1+}$ we introduce the new
variables
\begin{subequations}
\begin{align}
&  u=\theta-\theta_{c}+\frac{z \sec^{2}(\theta_{c}) \left( \sqrt{3} \lambda-3
(\mu-1) \sin(\theta_{c})\right) }{\sqrt{3} (\lambda\tan(\theta_{c})+\mu+1)-3
(\mu-1) \sec(\theta_{c})},\\
&  v=z,
\end{align}
with inverse
\end{subequations}
\begin{subequations}
\label{inverse}%
\begin{align}
&  \theta= \frac{\sqrt{3} (\theta_{c}+u) (\lambda\tan(\theta_{c})+\mu+1)-3
(\mu-1) \sec(\theta_{c}) (\theta_{c}+u-v \tan(\theta_{c}))-\sqrt{3} \lambda v
\sec^{2}(\theta_{c})}{\sqrt{3} (\lambda\tan(\theta_{c})+\mu+1)-3 (\mu-1)
\sec(\theta_{c})}.\\
& z =v,
\end{align}

\begin{figure}[t]
\textbf{ \includegraphics[width=0.6\textwidth]{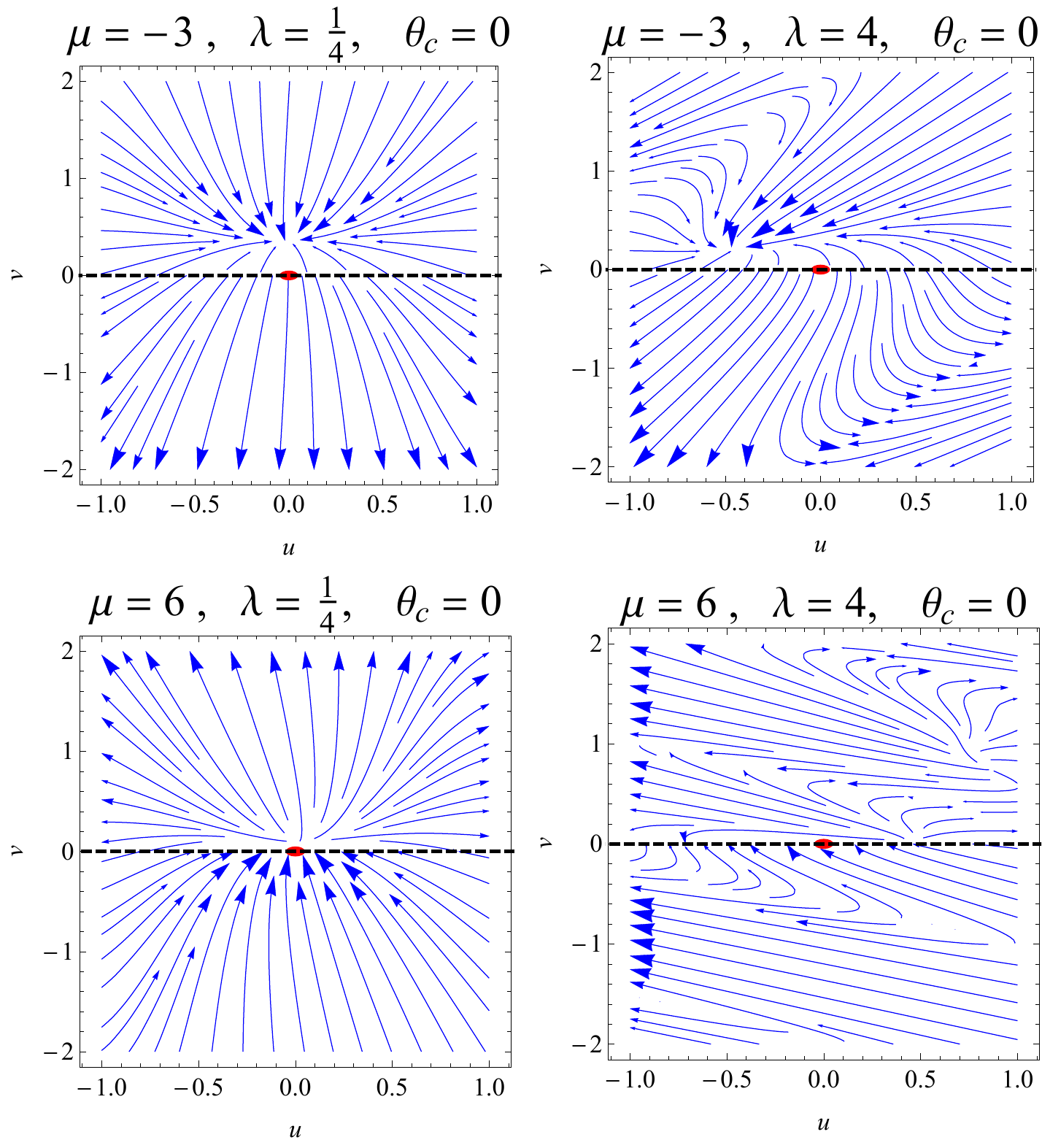} \newline}\caption{Phase
portrait of \eqref{center1} for different values of $\lambda, \mu$ ad $\theta_{c}%
$.}%
\label{F1}%
\end{figure}

where we have used $\theta=\theta_{c}$ to parametrize the system at an
specific point of $A_{1\pm}$ with coordinates
\end{subequations}
\begin{equation}
x=-\sqrt{\frac{3}{2}}+\frac{\sqrt{2}}{2}\cos\theta_{c}, \quad y= \frac
{\sqrt{2}}{2}\sin\theta_{c}%
\end{equation}
Then, we obtain the dynamical system
\begin{subequations}
\label{center1}%
\begin{align}
&  u^{\prime}=\frac{v \sec^{2}(\theta_{c}) \left( \sqrt{3} \lambda-3 (\mu-1)
\sin(\theta_{c})\right)  \left( \sqrt{3} \lambda\sin(\theta(u,v))+\sqrt{3}
(\mu+1) \cos(\theta(u,v))+3 (\mu-1) (2 v-1)\right) }{\sqrt{3} (\lambda
\tan(\theta_{c})+\mu+1)-3 (\mu-1) \sec(\theta_{c})}\nonumber\\
&  +v \sec(\theta(u,v)) \left( 3 (\mu-1) \sin(\theta(u,v))-\sqrt{3}
\lambda\right) ,\\
&  v^{\prime}= v \left( \sqrt{3} \lambda\sin(\theta(u,v))+\sqrt{3} (\mu+1)
\cos(\theta(u,v))+3 (\mu-1) (2 v-1)\right) ,
\end{align}
where
\end{subequations}
\begin{equation}
\theta(u,v)=\frac{\sqrt{3} (\theta_{c}+u) (\lambda\tan(\theta_{c})+\mu+1)-3
(\mu-1) \sec(\theta_{c}) (\theta_{c}+u-v \tan(\theta_{c}))-\sqrt{3} \lambda v
\sec^{2}(\theta_{c})}{\sqrt{3} (\lambda\tan(\theta_{c})+\mu+1)-3 (\mu-1)
\sec(\theta_{c})}.
\end{equation}
In the coordinates $u, v$, the eigensystem of the linearization is - as
expected -
\[
\left(
\begin{array}
[c]{cc}%
0 & -3 \mu+\sqrt{3} (\mu+1) \cos(\theta_{c})+\sqrt{3} \lambda\sin(\theta
_{c})+3\\
\{1,0\} & \{0,1\}\\
&
\end{array}
\right) ,
\]
Hence, the center manifold is given by the graph $(u, v)=(u, h(u))$, with
$h(0)=0, h^{\prime}(0)=0$, and satisfying the equation {\small
\begin{align}
\label{A1ODE1} & \Big(-h^{\prime}(u) \Big(\frac{\sec^{2}(\theta_{c}) \left(
\sqrt{3} \lambda-3 (\mu-1) \sin(\theta_{c})\right)  \left( \sqrt{3}
\lambda\sin(\theta(u,h(u)))+\sqrt{3} (\mu+1) \cos(\theta(u,h(u)))+6 (\mu-1)
h(u)-3 \mu+3\right) }{\sqrt{3} (\lambda\tan(\theta_{c})+\mu+1)-3 (\mu-1)
\sec(\theta_{c})}\nonumber\\
&  +\sec(\theta(u,h(u))) \left( 3 (\mu-1) \sin(\theta(u,h(u)))-\sqrt{3}
\lambda\right) \Big)\nonumber\\
&  +\sqrt{3} \lambda\sin(\theta(u,h(u)))+\sqrt{3} (\mu+1) \cos(\theta
(u,h(u)))+6 (\mu-1) h(u)-3 \mu+3\Big) h(u) =0.
\end{align}
} For which we have the trivial solution. Now, imposing the compatibility
conditions
\begin{subequations}
\begin{align}
& \mu=\frac{2}{\sqrt{3} \cos(\theta_{c})-1}+1,\\
&  \lambda=-\frac{6 \sin(\theta_{c})}{\sqrt{3}-3 \cos(\theta_{c})},
\end{align}
we can seek for another, non-trivial solution using Taylor series
\end{subequations}
\begin{equation}
h(u)=a u^{2}+O\left( u^{3}\right) .
\end{equation}
with results on $a=\frac{1}{4}$. In the two cases the evolution equation on
the center manifold is reduced to
\begin{equation}
u^{\prime}=O\left( u^{3}\right) ,
\end{equation}
that means that $u$ is constant at the invariant manifold.

\begin{figure}[t]
\textbf{ \includegraphics[width=0.6\textwidth]{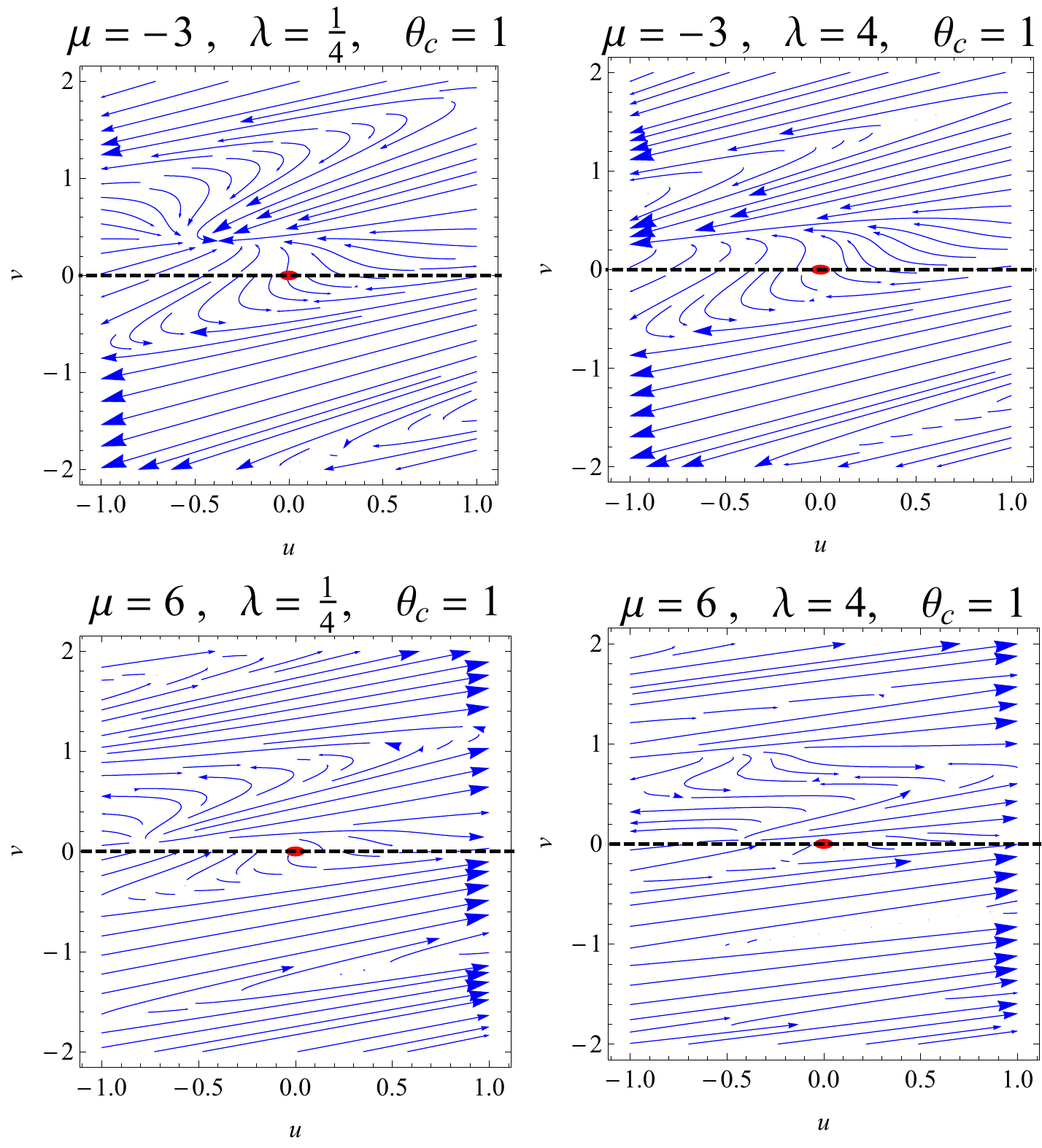} \newline}\caption{Phase
portrait of \eqref{center1} for different values of $\lambda, \mu$ ad $\theta_{c}%
$.}%
\label{F2}%
\end{figure}
In Figs. \ref{F1} and \ref{F2} we  present some orbits of the
phase portraits of \eqref{center1} for different values of $\lambda, \mu$ ad
$\theta_{c}$.

\subsubsection{Equilibrium points for $\omega=0$}

The case $\omega=0$ is of special interest because the Brans-Dicke field
reduces to that of O'Hanlon theory \cite{Hanlon}. The latter theory is
dynamically equivalent with the fourth-order theory known as $f\left(  R\right)
$-gravity \cite{s01}. The dynamically equivalent scenario can be found easily with the use of
a Lagrange multiplier, for more details we refer the reader to \cite{s01} and
references therein.

For $\omega=0$, we evaluated the cosmological observables
\begin{align}
& \Omega_{m}= \sqrt{6} x+y^{2}+z+1,\\
& w_{tot}=\frac{1}{9} \left( -6 y^{2}+6 \mu z+3\right) ,
\end{align}
and discuss the stability conditions of the equilibrium points of the
dimensionless dynamical system (\ref{bd2.014}), (\ref{bd2.015}),
(\ref{bd2.016}), as follows.

\begin{enumerate}
\item $B_{1}: (x,y,z)= \left( -\frac{y_{c}^{2}+1}{\sqrt{6}},y_{c},0\right) $.

Eigenvalues: $\left\{ 0,2-y_{c}^{2},-(\mu+1) y_{c}^{2}+\sqrt{6} \lambda
y_{c}-\mu+5\right\} $. Nonhyperbolic with a 2D stable manifold for

\begin{enumerate}
\item $\lambda\in\mathbb{R}, y_{c}<-\sqrt{2}, \mu>\frac{-y_{c}^{2}+\sqrt{6}
\lambda y_{c}+5}{y_{c}^{2}+1}$, or

\item $\lambda\in\mathbb{R}, y_{c}>\sqrt{2}, \mu>\frac{-y_{c}^{2}+\sqrt{6}
\lambda y_{c}+5}{y_{c}^{2}+1}$.
\end{enumerate}

Nonhyperbolic with a 2D unstable manifold for $\lambda\in\mathbb{R}, -\sqrt
{2}<y_{c}<\sqrt{2}, \mu<\frac{-y_{c}^{2}+\sqrt{6} \lambda y_{c}+5}{y_{c}%
^{2}+1}$.

The cosmological observables are $\Omega_{m}(B_{1})=0$, and $w_{tot}%
(B_{1})=\frac{1}{3} \left( 1-2 y_{c}^{2}\right) $. That means the solution
corresponds to a scalar field dominated universe, that is accelerating for
$y_{c}^{2}>1$. In particular, it is a de Sitter solution for $y_{c}=\pm
\sqrt{2}$. It is decelerating for $y_{c}^{2}<1$. In particular, it mimics a
dust solution for $y_{c}=\pm\frac{\sqrt{2}}{2}$.

\item $B_{2}: (x,y,z)= \left( -\frac{\sqrt{\frac{2}{3}} (\mu-2)}{\mu+1}%
,\frac{\sqrt{\frac{3}{2}} \lambda}{\mu+1},1-\frac{3 \left( \lambda^{2}+4
\mu+4\right) }{2 (\mu+1)^{2}}\right) $.

Eigenvalues: $\left\{ -\frac{3 \lambda^{2}}{2 (\mu+1)}+\mu-5,-\frac{3
\lambda^{2}}{2 (\mu+1)}+\mu-5,2 \mu-\frac{3 \left( \lambda^{2}+3 \mu+1\right)
}{\mu+1}\right\} $.  It is a sink for

\begin{enumerate}

\item $\mu=\frac{1}{4} \left( 7+\sqrt{73}\right) , \lambda\neq0$, or

\item $\mu=\frac{1}{4} \left( 7-\sqrt{73}\right) , \lambda\neq0$, or

\item $-7<\mu<-1, \lambda^{2}<\frac{2}{3} (\mu-5) (\mu+1)$, or

\item $\frac{1}{4} \left( 7-\sqrt{73}\right) <\mu<\frac{1}{4} \left(
7+\sqrt{73}\right) $, or

\item $\mu\leq-7, \lambda^{2}<\frac{1}{3} \left( 2 \mu^{2}-7 \mu-3\right) $,
or

\item $-1<\mu<\frac{1}{4} \left( 7-\sqrt{73}\right) , \lambda^{2}>\frac{1}{3}
\left( 2 \mu^{2}-7 \mu-3\right) $, or

\item $\mu>\frac{1}{4} \left( 7+\sqrt{73}\right) , \lambda^{2}>\frac{1}{3}
\left( 2 \mu^{2}-7 \mu-3\right) $.
\end{enumerate}

It is a source for

\begin{enumerate}

\item $\mu\leq-7, \lambda^{2}>\frac{2}{3} (\mu-5) (\mu+1)$, or

\item $-7<\mu<-1, \lambda^{2}>\frac{1}{3} \left( 2 \mu^{2}-7 \mu-3\right) $,
or

\item $\mu>5, \lambda^{2}<\frac{2}{3} (\mu-5) (\mu+1)$.
\end{enumerate}

The cosmological observables are $\Omega_{m}(B_{2})=0$, and $w_{tot}%
(B_{2})=\frac{-3 \lambda^{2}+2 \mu^{2}-9 \mu+1}{3 \mu+3}$. That is, the
solution corresponds to a scalar field dominated universe, that is
accelerating for

\begin{enumerate}
\item $\mu<-1, -\frac{\sqrt{2 \mu^{2}-8 \mu+2}}{\sqrt{3}}<\lambda<\frac
{\sqrt{2 \mu^{2}-8 \mu+2}}{\sqrt{3}}$, or

\item $-1<\mu<2-\sqrt{3}, \lambda<-\frac{\sqrt{2 \mu^{2}-8 \mu+2}}{\sqrt{3}}$, or

\item $-1<\mu<2-\sqrt{3}, \lambda>\frac{\sqrt{2 \mu^{2}-8 \mu+2}}{\sqrt{3}}$, or

\item $\mu=2-\sqrt{3}, \lambda<0$, or

\item $\mu=2-\sqrt{3}, \lambda>0$, or

\item $2-\sqrt{3}<\mu<2+\sqrt{3}$, or

\item $\mu=2+\sqrt{3}, \lambda<0$, or

\item $\mu=2+\sqrt{3}, \lambda>0$, or

\item $\mu>2+\sqrt{3}, \lambda<-\frac{\sqrt{2 \mu^{2}-8 \mu+2}}{\sqrt{3}}$, or

\item $\mu>2+\sqrt{3}, \lambda>\frac{\sqrt{2 \mu^{2}-8 \mu+2}}{\sqrt{3}}$.
\end{enumerate}

In particular, it is a de Sitter solution for

\begin{enumerate}
\item $\mu\neq-1, \lambda=-\sqrt{\frac{2}{3}} \sqrt{\mu^{2}-3 \mu+2}$, or

\item $\mu\neq-1, \lambda=\sqrt{\frac{2}{3}} \sqrt{\mu^{2}-3 \mu+2}$.
\end{enumerate}

Is is decelerating for

\begin{enumerate}
\item $\mu<-1, \lambda<-\frac{\sqrt{2 \mu^{2}-8 \mu+2}}{\sqrt{3}}$, or

\item $\mu<-1, \lambda>\frac{\sqrt{2 \mu^{2}-8 \mu+2}}{\sqrt{3}}$, or

\item $-1<\mu<2-\sqrt{3}, -\frac{\sqrt{2 \mu^{2}-8 \mu+2}}{\sqrt{3}}%
<\lambda<\frac{\sqrt{2 \mu^{2}-8 \mu+2}}{\sqrt{3}}$, or

\item $\mu>2+\sqrt{3}, -\frac{\sqrt{2 \mu^{2}-8 \mu+2}}{\sqrt{3}}%
<\lambda<\frac{\sqrt{2 \mu^{2}-8 \mu+2}}{\sqrt{3}}$.
\end{enumerate}

In particular, it mimics a dust solution for

\begin{enumerate}
\item $\mu\neq-1, \lambda=-\frac{\sqrt{2 \mu^{2}-9 \mu+1}}{\sqrt{3}}$, or

\item $\mu\neq-1, \lambda=\frac{\sqrt{2 \mu^{2}-9 \mu+1}}{\sqrt{3}}$.
\end{enumerate}

\item $B_{3}: (x,y,z)= \left( \frac{\lambda^{2}+3 \mu-3}{\sqrt{6} \left(
\lambda^{2}-\mu^{2}+\mu\right) },-\frac{\lambda(\mu+3)}{\sqrt{6} \left(
\lambda^{2}-\mu^{2}+\mu\right) },\frac{\left( 4 \lambda^{2}-3 (\mu
-1)^{2}\right)  (\mu+3)}{6 \left( \lambda^{2}-\mu^{2}+\mu\right) ^{2}}\right)
$.

Eigenvalues: $\left\{ \frac{3 (\mu-1)^{2}-4 \lambda^{2}}{2 \left( \lambda
^{2}-\mu^{2}+\mu\right) },-\frac{12 \lambda^{2}-9 (\mu-1)^{2}+\delta_{1}}{12
\left( \lambda^{2}-\mu^{2}+\mu\right) },\frac{-12 \lambda^{2}+9 (\mu
-1)^{2}+\delta_{1}}{12 \left( \lambda^{2}-\mu^{2}+\mu\right) }\right\}
$,\newline where $\delta_{1}=\sqrt{3} \sqrt{-\left( 4 \lambda^{2}-3
(\mu-1)^{2}\right)  \left( 12 \lambda^{2} (2 \mu+5)+\mu((17-16 \mu)
\mu+174)+81\right) }$.

It is a sink for

\begin{enumerate}

\item $\mu<-7, \frac{16 \mu^{3}-17 \mu^{2}-174 \mu-81}{24 \mu+60}\leq
\lambda^{2}<\frac{1}{4} \left( 3 \mu^{2}-6 \mu+3\right) $, or

\item $-3<\mu<-\frac{5}{2}, \frac{1}{4} \left( 3 \mu^{2}-6 \mu+3\right)
<\lambda^{2}\leq\frac{16 \mu^{3}-17 \mu^{2}-174 \mu-81}{24 \mu+60}$, or

\item $-\frac{5}{2}\leq\mu<-2.48926, \lambda^{2}>\frac{1}{4} \left( 3 \mu
^{2}-6 \mu+3\right) $, or

\item $\lambda^{2}>9.13121, \mu=-2.48926$, or

\item $-2.48926<\mu<-0.501725, \lambda^{2}\leq\frac{16 \mu^{3}-17 \mu^{2}-174
\mu-81}{24 \mu+60}$, or

\item $-2.48926<\mu<-0.501725, \lambda^{2}>\frac{1}{4} \left( 3 \mu^{2}-6
\mu+3\right) $, or

\item $\lambda^{2}>1.69138, \mu=-0.501725$, or

\item $-0.501725<\mu\leq1, \lambda^{2}>\frac{1}{4} \left( 3  \mu^{2}-6
\mu+3\right) $, or

\item $1<\mu<4.05349, \lambda^{2}>\mu^{2}-\mu$, or

\item $\lambda^{2}>12.3773, \mu=4.05349$, or

\item $\mu>4.05349, \lambda^{2}\leq\frac{16 \mu^{3}-17 \mu^{2}-174 \mu-81}{24
\mu+60}$, or

\item $\mu>4.05349, \lambda^{2}>\mu^{2}-\mu$, or

\item $\mu<-7, \frac{1}{3} \left( 2 \mu^{2}-7 \mu-3\right) <\lambda^{2}%
<\frac{16 \mu^{3}-17 \mu^{2}-174 \mu-81}{24 \mu+60}$, or

\item $-3<\mu<-\frac{5}{2}, 0\leq\lambda^{2}<\mu^{2}-\mu$, or

\item $-3<\mu<-\frac{5}{2}, \lambda^{2}>\frac{16 \mu^{3}-17 \mu^{2}-174
\mu-81}{24 \mu+60}$, or

\item $-\frac{5}{2}\leq\mu<-2.48926, 0\leq\lambda^{2}<\mu^{2}-\mu$, or

\item $0<\lambda^{2}<8.68569, \mu=-2.48926$, or

\item $-2.48926<\mu\leq-1, \frac{16 \mu^{3}-17 \mu^{2}-174 \mu-81}{24 \mu
+60}<\lambda^{2}<\mu^{2}-\mu$, or

\item $-1<\mu<-0.501725, \frac{16 \mu^{3}-17 \mu^{2}-174 \mu-81}{24 \mu
+60}<\lambda^{2}<\frac{1}{3} \left( 2 \mu^{2}-7 \mu-3\right) $, or

\item $0<\lambda^{2}<0.33851, \mu=-0.501725$, or

\item $-0.501725<\mu<\frac{1}{4} \left( 7-\sqrt{73}\right) , \lambda^{2}%
<\frac{1}{3} \left( 2 \mu^{2}-7 \mu-3\right) $, or

\item $\frac{1}{4} \left( 7+\sqrt{73}\right) <\mu<4.05349, 0\leq\lambda
^{2}<\frac{1}{3} \left( 2 \mu^{2}-7 \mu-3\right) $, or

\item $0<\lambda^{2}<0.495701, \mu=4.05349$, or

\item $\mu>4.05349, \frac{16 \mu^{3}-17 \mu^{2}-174 \mu-81}{24 \mu+60}%
<\lambda^{2}<\frac{1}{3} \left( 2  \mu^{2}-7 \mu-3\right) $, or

\item $\lambda=0, \mu\in\left\{ -2.48926, -0.501725, 4.05349\right\} $.
\end{enumerate}

It is a source for

\begin{enumerate}

\item $-7<\mu<-3, \frac{1}{4} \left( 3 \mu^{2}-6 \mu+3\right) <\lambda
^{2}<\frac{1}{3} \left( 2 \mu^{2}-7 \mu-3\right) $, or

\item $-3<\mu<-1, \mu^{2}-\mu<\lambda^{2}<\frac{1}{3} \left( 2 \mu^{2}-7
\mu-3\right) $, or

\item $\mu>1, \frac{1}{4} \left( 3 \mu^{2}-6 \mu+3\right) <\lambda^{2}<\mu
^{2}-\mu$.
\end{enumerate}

The cosmological observables are $\Omega_{m}(B_{3})=\frac{\left( 4 \lambda
^{2}-3 (\mu-1)^{2}\right)  \left( 3 \lambda^{2}+(7-2 \mu) \mu+3\right) }{6
\left( \lambda^{2}-\mu^{2}+\mu\right) ^{2}}$, and $w_{tot}(B_{3}%
)=\frac{\lambda^{2}+3 \mu-3}{3 \left( \lambda^{2}-\mu^{2}+\mu\right) }$. That
means, the solution is accelerating for

\begin{enumerate}
\item $\lambda<-2 \sqrt{3}, 2-\sqrt{2 \lambda^{2}+1}<\mu<\frac{1}{2}-\frac
{1}{2} \sqrt{4 \lambda^{2}+1}$, or

\item $\lambda<-2 \sqrt{3}, \frac{1}{2} \sqrt{4 \lambda^{2}+1}+\frac{1}{2}%
<\mu<\sqrt{2 \lambda^{2}+1}+2$, or

\item $-2 \sqrt{3}\leq\lambda\leq2 \sqrt{3}, \frac{1}{2}-\frac{1}{2} \sqrt{4
\lambda^{2}+1}<\mu<2-\sqrt{2 \lambda^{2}+1}$, or

\item $-2 \sqrt{3}\leq\lambda\leq2 \sqrt{3}, \frac{1}{2} \sqrt{4 \lambda
^{2}+1}+\frac{1}{2}<\mu<\sqrt{2 \lambda^{2}+1}+2$, or

\item $\lambda>2 \sqrt{3}, 2-\sqrt{2 \lambda^{2}+1}<\mu<\frac{1}{2}-\frac
{1}{2} \sqrt{4 \lambda^{2}+1}$, or

\item $\lambda>2 \sqrt{3}, \frac{1}{2}  \sqrt{4 \lambda^{2}+1}+\frac{1}{2}%
<\mu<\sqrt{2 \lambda^{2}+1}+2.$
\end{enumerate}

In particular, it is a de Sitter solution for

\begin{enumerate}
\item $\lambda^{2}+3 \mu-3\neq0, \mu=\frac{1}{3} \left( 3-2 \sqrt{3}
\lambda\right) $, or

\item $\lambda^{2}+3 \mu-3\neq0, \mu=\frac{1}{3} \left( 2 \sqrt{3}
\lambda+3\right) $.
\end{enumerate}

It is decelerating for

\begin{enumerate}
\item $\lambda<-2 \sqrt{3}, \mu<2-\sqrt{2 \lambda^{2}+1}$, or

\item $\lambda<-2 \sqrt{3}, \frac{1}{2}-\frac{1}{2} \sqrt{4 \lambda^{2}+1}%
<\mu<\frac{1}{2} \sqrt{4 \lambda^{2}+1}+\frac{1}{2}$, or

\item $\lambda<-2 \sqrt{3}, \mu>\sqrt{2 \lambda^{2}+1}+2$, or

\item $-2 \sqrt{3}\leq\lambda\leq2 \sqrt{3}, \mu<\frac{1}{2}-\frac{1}{2}
\sqrt{4 \lambda^{2}+1}$, or

\item $-2 \sqrt{3}\leq\lambda\leq2 \sqrt{3}, 2-\sqrt{2 \lambda^{2}+1}%
<\mu<\frac{1}{2} \sqrt{4 \lambda^{2}+1}+\frac{1}{2}$, or

\item $-2 \sqrt{3}\leq\lambda\leq2 \sqrt{3}, \mu>\sqrt{2 \lambda^{2}+1}+2$, or

\item $\lambda>2 \sqrt{3}, \mu<2-\sqrt{2 \lambda^{2}+1}$, or

\item $\lambda>2 \sqrt{3}, \frac{1}{2}-\frac{1}{2} \sqrt{4 \lambda^{2}+1}%
<\mu<\frac{1}{2} \sqrt{4 \lambda^{2}+1}+\frac{1}{2}$, or

\item $\lambda>2 \sqrt{3}, \mu>\sqrt{2 \lambda^{2}+1}+2$.
\end{enumerate}

In particular, it mimics a dust solution for $\mu=\frac{1}{3} \left(
3-\lambda^{2}\right) , \lambda\left( \lambda^{2}-12\right) \neq0$.

\item $B_{4}: (x,y,z)= \left( \frac{1}{\sqrt{6}},0,0\right) $. Eigenvalues:
$\{-2,-2,\mu+3\}$. Nonhyperbolic for $\mu=-3$, sink for $\mu<-3$, saddle for
$\mu>-3$. The cosmological observables are $\Omega_{m}(B_{4})=2$, and
$w_{tot}(B_{4})=\frac{1}{3}$. That means, it mimics a radiation dominated
universe. But this point has no physical meaning since it satisfies
$\Omega_{m}>1$.
\end{enumerate}

Points $B_{1}: (x,y,z)= \left( -\frac{y_{c}^{2}+1}{\sqrt{6}},y_{c},0\right) $
can also be written as $B_{1\pm}=\left( {x}_{c}, \pm\sqrt{-1-\sqrt{6}{x}_{c}%
},0\right) $, equivalently, points $A_{1\pm}$ for the case $\omega=0$,
describe surfaces a family of points in the surface $y^{2}=-\left( 1+\sqrt
{6}{x}_{c}\right)  $ in the plane $\left\{  x,y\right\}  $. The points are
real, when $\sqrt{6}{x}_{c}\leq-1$. {The energy density for the dust fluid is
found to be $\Omega_{m}=0$, while the equation of state for the effective
fluid is $w_{tot}\left(  B_{1\pm}\right)  =1+\frac{2}{3}\sqrt{6}{x}_{c}$,
similar expression with that of $A_{1\pm}$.} In the case where $\sqrt
{6}{x}_{c}=-1$ where only the field $\phi$ contributes to the evolution of the
universe, we find that $w_{tot}\left(  B_{1\pm}\right)  |_{\sqrt{6}{x}%
_{c}\rightarrow-1}=\frac{1}{3}$, where the critical point in which the
$f\left(  R\right)  $-gravity describes the radiation era is recovered.

\subsection{Center manifold of $B_{1}$}

It is more convenient to use the parametrization $B_{1}: (x,y,z)= \left(
-\frac{y_{c}^{2}+1}{\sqrt{6}},y_{c},0\right) $, which is solved globally for
$x$, rather than considering two branches $B_{1\pm}$, which provide with only a
local description of the model.

We choose $y_{c}^{2} \neq2$, in which the center manifold is one
dimensional. Next, we introduce the new variables
\begin{align}
&  u=-\frac{\sqrt{6} x {y_{c}}}{{y_{c}}^{2}-2}+y \left( -\frac{4}{{y_{c}}%
^{2}-2}-1\right) +\frac{{y_{c}}}{{y_{c}}^{2}-2}\nonumber\\
&  -\frac{\sqrt{6} {y_{c}} z \left( \frac{\mu-(\mu+1) {y_{c}}^{2}-5}{\sqrt{6}
\left( \mu+(\mu+1) {y_{c}}^{2}-5\right) -6 \lambda{y_{c}}}+\frac{\left(
{y_{c}}^{2}+2\right)  \left( -\sqrt{6} \lambda\left( \mu+(3 \mu+2) {y_{c}}%
^{2}-3\right) +2 (\mu+1) {y_{c}} \left( \mu+\mu{y_{c}}^{2}-3\right) +6
\lambda^{2} {y_{c}}\right) }{2 {y_{c}} \left( \sqrt{6} (\mu-5) (\mu
-3)+\sqrt{6} \mu(\mu+1) {y_{c}}^{4}-6 {y_{c}}^{3} (2 \lambda\mu+\lambda
)+\sqrt{6} {y_{c}}^{2} \left( 6 \lambda^{2}+\mu(2 \mu-7)-3\right) -12
\lambda(\mu-4) {y_{c}}\right) }\right) }{{y_{c}}^{2}-2},\\
&  v= \frac{\sqrt{6} x {y_{c}}}{{y_{c}}^{2}-2}+y \left( \frac{4}{{y_{c}}%
^{2}-2}+2\right) +\frac{{y_{c}} z}{{y_{c}}^{2}-2}+\frac{{y_{c}}-{y_{c}}^{3}%
}{{y_{c}}^{2}-2},\\
&  w= z,
\end{align}
with inverse
\begin{align}
&  x= -\sqrt{\frac{2}{3}} u {y_{c}}-\frac{v \left( {y_{c}}^{2}+2\right)
}{\sqrt{6} {y_{c}}}+\frac{w \left( -\mu+(\mu+1) {y_{c}}^{2}+5\right) }%
{\sqrt{6} \left( \mu+(\mu+1) {y_{c}}^{2}-5\right) -6 \lambda{y_{c}}}%
-\frac{{y_{c}}^{2}+1}{\sqrt{6}},\\
&  y= u+v+w +\frac{\left( {y_{c}}^{2}+2\right)  {y_{c}}}{2-{y_{c}}^{2}}%
+\frac{2 {y_{c}}^{3}}{{y_{c}}^{2}-2}+ \Big[\frac{6 \lambda{y_{c}}^{2}-2
\sqrt{6} (\mu+1) {y_{c}}^{3}}{\left( {y_{c}}^{2}-2\right)  \left( \sqrt{6}
\left( \mu+(\mu+1) {y_{c}}^{2}-5\right) -6 \lambda{y_{c}}\right) }\nonumber\\
&  +\frac{\left( {y_{c}}^{2}+2\right)  \left( -3 \lambda(\mu-3)+\sqrt{6}
\mu(\mu+1) {y_{c}}^{3}-3 \lambda(3 \mu+2) {y_{c}}^{2}+\sqrt{6} {y_{c}} \left(
3 \lambda^{2}+(\mu-3) (\mu+1)\right) \right) }{\left( {y_{c}}^{2}-2\right)
\left( \sqrt{6} (\mu-5) (\mu-3)+\sqrt{6} \mu(\mu+1) {y_{c}}^{4}-6 {y_{c}}^{3}
(2 \lambda\mu+\lambda)+\sqrt{6} {y_{c}}^{2} \left( 6 \lambda^{2}+\mu(2
\mu-7)-3\right) -12 \lambda(\mu-4) {y_{c}}\right) }\Big],\\
&  z= w.
\end{align}
The resulting system, which can be written symbolically as
\begin{align}
&  u^{\prime}=0. u + f_{1}(u,v,w),\\
&  v^{\prime}= (2-{y_{c}}^{2}) v +f_{2}(u,v,w),\\
&  w^{\prime}= (-(\mu+1) {y_{c}}^{2}+\sqrt{6} \lambda{y_{c}}-\mu+5) w
+f_{3}(u,v,w),
\end{align}
{where $f_{i}$'s ($i = 1, 2, 3$) are at least of second order in $(u,v,w)$. Hence the latter system has the following
eigensystem of the linearization} \newline$\left(
\begin{array}
[c]{ccc}%
0 & 2-{y_{c}}^{2} & -(\mu+1) {y_{c}}^{2}+\sqrt{6} \lambda{y_{c}}-\mu+5\\
\{1,0,0\} & \{0,1,0\} & \{0,0,1\}\\
&  &
\end{array}
\right) $.

That means, the center manifold of the origin is given locally by the
graph $(u,v,w)=(u, h_{1}(u), h_{2}(u))$, where
\begin{align}
&  f_{2}(u, h_{1}(u), h_{2}(u))-h_{1}^{\prime}(u) f_{1}(u, h_{1}(u),
h_{2}(u))=0, h_{1}(0)=0 , h_{1}^{\prime}(0)=0,\\
&  f_{3}(u, h_{1}(u), h_{2}(u))-h_{2}^{\prime}(u) f_{1}(u, h_{1}(u),
h_{2}(u))=0, h_{2}(0)=0 , h_{2}^{\prime}(0)=0.
\end{align}
Using the Taylor series we propose the expansion
\begin{equation}
\label{ODEx2}h_{1}(u)=a_{1} u^{2}+b_{1} u^{3}+c_{1} u^{4}+d_{1} u^{5}+O\left(
u^{6}\right) , \quad h_{2}(u)=a_{2} u^{2}+b_{2} u^{3}+c_{2} u^{4}+d_{2}
u^{5}+O\left( u^{6}\right) .
\end{equation}
\begin{figure}[t]
\textbf{ \includegraphics[width=0.6\textwidth]{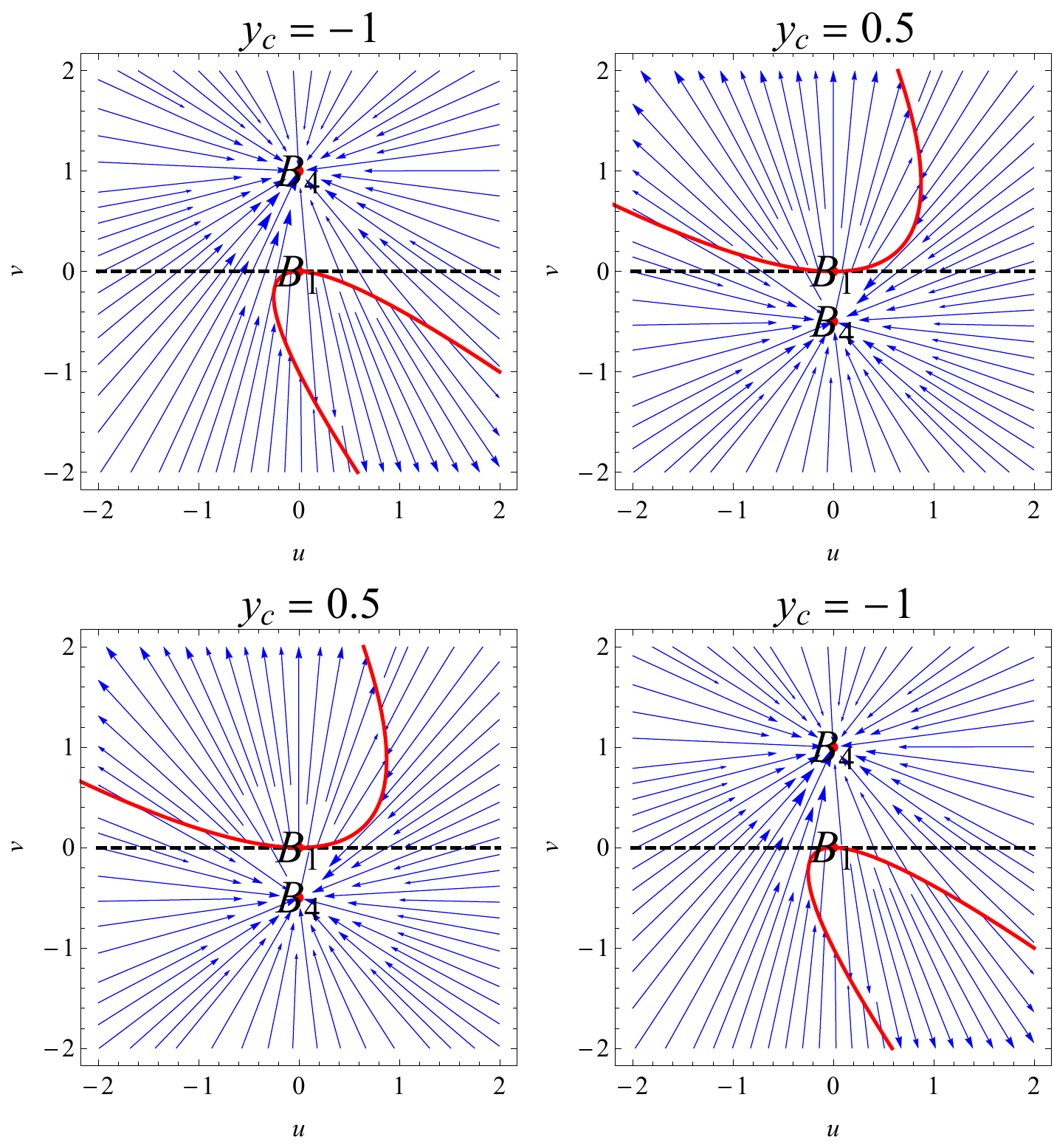} \newline}%
\caption{Qualitative evolution in the phase plane $(u,v)$ given by
\eqref{2ndcenter1} for $\omega=0$. The solid red lines denote the lines
$-v\pm\frac{\sqrt{-v \text{yc} \left( \text{yc}^{2}-2\right) }}{\text{yc}}$,
whereas the dashed black lines represent the approximated center manifold of
$B_{1}$. Observe the accuracy for small $u$. The solutions typically converge
to the unphysical point $B_{4}$.}%
\label{F3}%
\end{figure}Substituting this expansion in the above equations, and equating
the coefficients of the same powers of $u$, we obtain
\begin{align}
&  a_{1}= \frac{y_{c}}{2-y_{c}^{2}},\quad b_{1}= \frac{2 y_{c}^{2}}{\left(
y_{c}^{2}-2\right) ^{2}},\quad c_{1}= -\frac{5 y_{c}^{3}}{\left( y_{c}%
^{2}-2\right) ^{3}},\quad d_{1}= \frac{14 y_{c}^{4}}{\left( y_{c}^{2}-2\right)
^{4}},\\
&  a_{2}= 0, \quad b_{2}= 0,\quad c_{2}=0,\quad d_{2}=0.
\end{align}
The evolution at the center manifold is $u^{\prime 7})$. That is, $u$ is
constant at the center manifold. In the invariant set $w=0$, the evolution
equations in a neighborhood of a point of $B_{1}$ parametrized by $y_{c}$ can
be simplified to
\begin{align}
\label{2ndcenter1} & u^{\prime}= u \left( -(u+v)^{2}-v y_{c}+\frac{2 v}{y_{c}%
}\right) ,\nonumber\\
& v^{\prime}=-\frac{(v+y_{c}) \left( y_{c} (u+v)^{2}+v y_{c}^{2}-2 v\right)
}{y_{c}}.
\end{align}

From the Fig. \ref{F3}, we see that
\begin{equation}
v= h_{1}(u):= \frac{\sqrt{y_{c}^{2}-2} \sqrt{4 u y_{c}+y_{c}^{2}-2}-y_{c} (2
u+y_{c})+2}{2 y_{c}}, w=h_{2}(u):=0,
\end{equation}
is the exact solution for the center manifold, which has the expansion up to
fifth order given before (and they satisfy the equations and the initial
conditions in \eqref{ODEx2}). The equation on the center manifold is
$u^{\prime}=0$. As before, $u$ is constant at the center manifold.

\subsection{Comments on arbitrary $\omega$}

\label{Sect_3} For arbitrary $\omega$, we evaluated the cosmological
observables
\begin{align}
& \Omega_{m}= x \left( x \omega+\sqrt{6}\right) +y^{2}+z+1,\\
& w_{tot}=\frac{-2 x \omega\left( 3 x (\omega-1)+\sqrt{6}\right) -6 (\omega-1)
y^{2}+6 z (\omega-\mu)-3}{6 \omega-9},
\end{align}
and discuss the stability conditions of the equilibrium points of the
dimensionless dynamical system (\ref{bd2.014}), (\ref{bd2.015}),
(\ref{bd2.016}), as follows.

\begin{enumerate}
\item $A_{1+}:(x, y, z)=(x_{c}, \sqrt{-x_{c} \left( x_{c} \omega+\sqrt
{6}\right) -1}, 0)$, with eigenvalues \newline$\left\{ 0,\sqrt{6}
x_{c}+3,\sqrt{6} \lambda\sqrt{-x_{c} \left( \omega x_{c}+\sqrt{6}\right)
-1}+\sqrt{6} (\mu+1) x_{c}+6\right\} $. \newline$A_{1+}$ nonhyperbolic with a
2D stable manifold for

\begin{enumerate}
\item $\lambda\leq0, x_{c}<-\sqrt{\frac{3}{2}}, \omega\leq\frac{-\sqrt{6}
x_{c}-1}{x_{c}^{2}}, \mu>\frac{-\sqrt{6} x_{c}-6}{\sqrt{6} x_{c}}-\sqrt
{\frac{-\omega\lambda^{2} x_{c}^{2}-\sqrt{6} \lambda^{2} x_{c}-\lambda^{2}%
}{x_{c}^{2}}}$, or

\item $\lambda>0,  x_{c}<-\sqrt{\frac{3}{2}}, \omega\leq\frac{-\sqrt{6}
x_{c}-1}{x_{c}^{2}}, \mu>\sqrt{\frac{-\omega\lambda^{2} x_{c}^{2}-\sqrt{6}
\lambda^{2} x_{c}-\lambda^{2}}{x_{c}^{2}}}+\frac{-\sqrt{6} x_{c}-6}{\sqrt{6}
x_{c}}$.
\end{enumerate}

It is nonhyperbolic with a 2D unstable manifold for

\begin{enumerate}
\item $\lambda\leq0, -\sqrt{\frac{3}{2}}<x_{c}<0, \omega\leq\frac{-\sqrt{6}
x_{c}-1}{x_{c}^{2}}, \mu<\frac{-\sqrt{6} x_{c}-6}{\sqrt{6} x_{c}}-\sqrt
{\frac{-\omega\lambda^{2} x_{c}^{2}-\sqrt{6} \lambda^{2} x_{c}-\lambda^{2}%
}{x_{c}^{2}}}$, or

\item $\lambda\leq0, x_{c}>0, \omega\leq\frac{-\sqrt{6} x_{c}-1}{x_{c}^{2}},
\mu>\sqrt{\frac{-\omega\lambda^{2} x_{c}^{2}-\sqrt{6} \lambda^{2}
x_{c}-\lambda^{2}}{x_{c}^{2}}}+\frac{-\sqrt{6} x_{c}-6}{\sqrt{6} x_{c}}$, or

\item $\lambda>0, -\sqrt{\frac{3}{2}}<x_{c}<0, \omega\leq\frac{-\sqrt{6}
x_{c}-1}{x_{c}^{2}}, \mu<\sqrt{\frac{-\omega\lambda^{2} x_{c}^{2}-\sqrt{6}
\lambda^{2} x_{c}-\lambda^{2}}{x_{c}^{2}}}+\frac{-\sqrt{6} x_{c}-6}{\sqrt{6}
x_{c}}$, or

\item $\lambda>0, x_{c}>0, \omega\leq\frac{-\sqrt{6} x_{c}-1}{x_{c}^{2}},
\mu>\frac{-\sqrt{6} x_{c}-6}{\sqrt{6} x_{c}}-\sqrt{\frac{-\omega\lambda^{2}
x_{c}^{2}-\sqrt{6} \lambda^{2} x_{c}-\lambda^{2}}{x_{c}^{2}}}$.
\end{enumerate}

\item $A_{1-}:(x, y, z)=(x_{c}, -\sqrt{-x_{c} \left( x_{c} \omega+\sqrt
{6}\right) -1}, 0)$, with eigenvalues \newline$\left\{ 0,\sqrt{6}
x_{c}+3,\sqrt{6} (\mu+1) x_{c}+6-\sqrt{6} \lambda\sqrt{-x_{c} \left( \omega
x_{c}+\sqrt{6}\right) -1}\right\} $.\newline$A_{1-}$ nonhyperbolic with a 2D
stable manifold for

\begin{enumerate}
\item $\lambda\leq0, x_{c}<-\sqrt{\frac{3}{2}}, \omega\leq\frac{-\sqrt{6}
x_{c}-1}{x_{c}^{2}}, \mu>\sqrt{\frac{-\omega\lambda^{2} x_{c}^{2}-\sqrt{6}
\lambda^{2} x_{c}-\lambda^{2}}{x_{c}^{2}}}+\frac{-\sqrt{6} x_{c}-6}{\sqrt{6}
x_{c}}$, or

\item $\lambda>0, x_{c}<-\sqrt{\frac{3}{2}}, \omega\leq\frac{-\sqrt{6}
x_{c}-1}{x_{c}^{2}}, \mu>\frac{-\sqrt{6} x_{c}-6}{\sqrt{6} x_{c}}-\sqrt
{\frac{-\omega\lambda^{2} x_{c}^{2}-\sqrt{6} \lambda^{2} x_{c}-\lambda^{2}%
}{x_{c}^{2}}}$.
\end{enumerate}

It is nonhyperbolic with a 2D unstable manifold for

\begin{enumerate}
\item $\lambda\leq0, -\sqrt{\frac{3}{2}}<x_{c}<0, \omega\leq\frac{-\sqrt{6}
x_{c}-1}{x_{c}^{2}}, \mu<\sqrt{\frac{\lambda^{2} (-\omega) x_{c}^{2}-\sqrt{6}
\lambda^{2} x_{c}-\lambda^{2}}{x_{c}^{2}}}+\frac{-x_{c}-\sqrt{6}}{x_{c}}$, or

\item $\lambda\leq0, x_{c}>0,  \omega\leq\frac{-\sqrt{6} x_{c}-1}{x_{c}^{2}},
\mu>\frac{-x_{c}-\sqrt{6}}{x_{c}}-\sqrt{\frac{-\omega\lambda^{2} x_{c}%
^{2}-\sqrt{6} \lambda^{2} x_{c}-\lambda^{2}}{x_{c}^{2}}}$, or

\item $\lambda>0, -\sqrt{\frac{3}{2}}<x_{c}<0, \omega\leq\frac{-\sqrt{6}
x_{c}-1}{x_{c}^{2}}, \mu<\frac{-x_{c}-\sqrt{6}}{x_{c}}-\sqrt{\frac
{-\omega\lambda^{2} x_{c}^{2}-\sqrt{6} \lambda^{2} x_{c}-\lambda^{2}}%
{x_{c}^{2}}}$, or

\item $\lambda>0, x_{c}>0, \omega\leq\frac{-\sqrt{6} x_{c}-1}{x_{c}^{2}},
\mu>\sqrt{\frac{-\omega\lambda^{2} x_{c}^{2}-\sqrt{6} \lambda^{2}
x_{c}-\lambda^{2}}{x_{c}^{2}}}+\frac{-x_{c}-\sqrt{6}}{x_{c}}$.
\end{enumerate}

The cosmological observables are $\Omega_{m}(A_{1\pm})=0, w_{tot}=1+2
\sqrt{\frac{2}{3}} x_{c}$. Therefore, they represent solutions dominated by
the scalar field and they are accelerating (i.e., $w_{tot}<-\frac{1}{3}$) for
$x_{c}<-\sqrt{\frac{2}{3}}, \omega<\frac{-\sqrt{6} x_{c}-1}{x_{c}^{2}}$. For
$x_{c} -\sqrt{\frac{3}{2}}$, it is a de Sitter solution ($w_{tot}(A_{\pm}%
)=-1$). It represents a decelerating solution for $-\sqrt{\frac{2}{3}}%
<x_{c}<0, \omega<\frac{-\sqrt{6} x_{c}-1}{x_{c}^{2}}$, or $x_{c}>0,
\omega<\frac{-\sqrt{6} x_{c}-1}{x_{c}^{2}}$. In particular, it mimics a dust
solution ($w_{tot}(A_{\pm})=0$) for $x_{c}=-\frac{1}{2}\sqrt{\frac{3}{2}}$.

\item $A_{2}: (x,y,z)=\left( -\frac{\sqrt{\frac{2}{3}} (\mu-2)}{\mu-2
\omega+1}, \frac{\lambda(3-2 \omega)}{\sqrt{6} (\mu-2 \omega+1)},  -\frac{(2
\omega-3) \left( (2 \omega-3) \lambda^{2}+2 ((\mu-4) \mu+6 \omega-5)\right)
}{6 (\mu-2 \omega+1)^{2}}\right) $, \newline with eigenvalues $\left\{
\frac{(2 \omega-3) \lambda^{2}+\mu(2 \mu-7)+6 \omega-3}{\mu-2 \omega+1}%
,\frac{(2 \omega-3) \lambda^{2}+2 ((\mu-4) \mu+6 \omega-5)}{2 (\mu-2
\omega+1)},\frac{(2 \omega-3) \lambda^{2}+2 ((\mu-4) \mu+6 \omega-5)}{2 (\mu-2
\omega+1)}\right\} $.

It is a source for

\begin{enumerate}
\item $\mu<\frac{1}{2} \left( -\lambda^{2}-2\right) , \frac{3 \lambda^{2}-2
\mu^{2}+8 \mu+10}{2 \left( \lambda^{2}+6\right) }<\omega<\frac{\mu+1}{2}$, or

\item $-\frac{\lambda^{2}}{2}<\mu<\frac{1}{6} \left( 6-\lambda^{2}\right) ,
\frac{\mu+1}{2}<\omega<\frac{3 \lambda^{2}-2 \mu^{2}+7 \mu+3}{2 \left(
\lambda^{2}+3\right) }$, or

\item $\frac{1}{6} \left( 6-\lambda^{2}\right) \leq\mu<2, \frac{\mu+1}%
{2}<\omega<\frac{3 \lambda^{2}-2 \mu^{2}+8 \mu+10}{2 \left( \lambda
^{2}+6\right) }$, or

\item $\mu>2, \frac{3 \lambda^{2}-2 \mu^{2}+8 \mu+10}{2 \left( \lambda
^{2}+6\right) }<\omega<\frac{\mu+1}{2}$.
\end{enumerate}

It is a sink for

\begin{enumerate}
\item $\mu<\frac{1}{2} \left( -\lambda^{2}-2\right) , \omega>\frac{\mu+1}{2}$, or

\item $\frac{1}{2}\left( -\lambda^{2}-2\right) \leq\mu\leq-\frac{\lambda^{2}%
}{2}, \omega<\frac{3 \lambda^{2}-2 \mu^{2}+7 \mu+3}{2 \left( \lambda
^{2}+3\right) }$, or

\item $\frac{1}{2} \left( -\lambda^{2}-2\right) \leq\mu\leq-\frac{\lambda^{2}%
}{2}, \omega>\frac{3 \lambda^{2}-2 \mu^{2}+8 \mu+10}{2 \left( \lambda
^{2}+6\right) }$, or

\item $-\frac{\lambda^{2}}{2}<\mu\leq\frac{1}{6} \left( 6-\lambda^{2}\right) ,
\omega<\frac{\mu+1}{2}$, or

\item $-\frac{\lambda^{2}}{2}<\mu\leq\frac{1}{6} \left( 6-\lambda^{2}\right) ,
\omega>\frac{3 \lambda^{2}-2 \mu^{2}+8 \mu+10}{2 \left( \lambda^{2}+6\right)
}$, or

\item $\frac{1}{6} \left( 6-\lambda^{2}\right) <\mu<2, \omega<\frac{\mu+1}{2}%
$, or

\item $\frac{1}{6} \left( 6-\lambda^{2}\right) <\mu<2, \omega>\frac{3
\lambda^{2}-2 \mu^{2}+7 \mu+3}{2 \left( \lambda^{2}+3\right) }$, or

\item $\mu=2, \omega\neq\frac{3}{2}$, or

\item $\mu>2, \omega<\frac{3 \lambda^{2}-2 \mu^{2}+7 \mu+3}{2 \left(
\lambda^{2}+3\right) }$, or

\item $\mu>2, \omega>\frac{\mu+1}{2}$.
\end{enumerate}

The cosmological observables are $\Omega_{m}(A_{2})=0, w_{tot}(A_{2}%
)=\frac{\lambda^{2} (2 \omega-3)+\mu(2 \mu-9)+6 \omega+1}{3 (\mu-2 \omega+1)}%
$. It corresponds to a scalar field dominated universe. It is an accelerating
solution for

\begin{enumerate}
\item $\lambda\in\mathbb{R}, \mu<1, \omega\geq\frac{3}{2}$, or

\item $\lambda\in\mathbb{R}, \mu=2, \omega>\frac{3}{2}$, or

\item $\lambda\in\mathbb{R}, \mu>2, \omega>\frac{\mu+1}{2}$, or

\item $\lambda\in\mathbb{R}, \mu>2, \omega<\frac{1}{2} \left( -\mu^{2}+4
\mu-1\right) $, or

\item $\lambda\in\mathbb{R}, \mu<1, \omega<\frac{1}{2} \left( -\mu^{2}+4
\mu-1\right) $, or

\item $\lambda\in\mathbb{R}, \mu=2, \omega<\frac{3}{2}$, or

\item $\mu>2, \omega=\frac{1}{2} \left( -\mu^{2}+4 \mu-1\right) , \lambda>0$, or

\item $\mu<1, \omega=\frac{1}{2} \left( -\mu^{2}+4 \mu-1\right) , \lambda>0$, or

\item $\lambda\in\mathbb{R}, 1\leq\mu<2, \omega\geq\frac{3}{2}$, or

\item $\mu>2, \omega=\frac{1}{2} \left( -\mu^{2}+4 \mu-1\right) , \lambda<0$, or

\item $\mu<1, \omega=\frac{1}{2} \left( -\mu^{2}+4 \mu-1\right) , \lambda<0$, or

\item $\lambda\in\mathbb{R}, 1\leq\mu<2, \omega<\frac{\mu+1}{2}$, or

\item $\mu>2, \frac{1}{2} \left( -\mu^{2}+4 \mu-1\right) <\omega<\frac{3}{2},
\lambda>\sqrt{2} \sqrt{\frac{-\mu^{2}+4 \mu-2 \omega-1}{2 \omega-3}}$, or

\item $\mu<1, \frac{1}{2} \left( -\mu^{2}+4 \mu-1\right) <\omega<\frac{\mu
+1}{2}, \lambda>\sqrt{2} \sqrt{\frac{-\mu^{2}+4 \mu-2 \omega-1}{2 \omega-3}}$, or

\item $\mu>2, \frac{1}{2} \left( -\mu^{2}+4 \mu-1\right) <\omega<\frac{3}{2},
\lambda<-\sqrt{2}  \sqrt{\frac{-\mu^{2}+4 \mu-2 \omega-1}{2 \omega-3}}$, or

\item $\mu<1, \frac{1}{2} \left( -\mu^{2}+4 \mu-1\right) <\omega<\frac{\mu
+1}{2}, \lambda<-\sqrt{2} \sqrt{\frac{-\mu^{2}+4 \mu-2 \omega-1}{2 \omega-3}}%
$, or

\item $\mu<1, \frac{\mu+1}{2}<\omega<\frac{3}{2}, -\sqrt{2} \sqrt{\frac
{-\mu^{2}+4 \mu-2 \omega-1}{2 \omega-3}}<\lambda<\sqrt{2} \sqrt{\frac{-\mu
^{2}+4 \mu-2 \omega-1}{2 \omega-3}}$, or

\item $1\leq\mu<2, \frac{1}{2} \left( -\mu^{2}+4 \mu-1\right) <\omega<\frac
{3}{2}, -\sqrt{2} \sqrt{\frac{-\mu^{2}+4 \mu-2 \omega-1}{2 \omega-3}}%
<\lambda<\sqrt{2} \sqrt{\frac{-\mu^{2}+4 \mu-2 \omega-1}{2 \omega-3}}$.
\end{enumerate}

In particular, it mimics a de Sitter solution for

\begin{enumerate}
\item $\lambda=-\frac{\sqrt{2} \sqrt{(2-\mu) (\mu-1)} \sqrt{2 \omega-3}}{3-2
\omega}$, or

\item $\lambda= \frac{\sqrt{2} \sqrt{(2-\mu) (\mu-1)} \sqrt{2 \omega-3}}{3-2
\omega}$, or

\item $\mu= 1, \omega= \frac{3}{2}$.
\end{enumerate}

It is a decelerating solution for

\begin{enumerate}
\item $\lambda\in\mathbb{R}, \mu>2, \frac{3}{2}\leq\omega<\frac{\mu+1}{2}$, or

\item $\mu\leq1, \frac{\mu+1}{2}<\omega<\frac{3}{2}, \lambda>\sqrt{2}
\sqrt{\frac{-\mu^{2}+4 \mu-2 \omega-1}{2 \omega-3}}$, or

\item $1<\mu<2, \omega=\frac{1}{2} \left( -\mu^{2}+4 \mu-1\right) , \lambda
>0$, or

\item $\mu\leq1, \frac{\mu+1}{2}<\omega<\frac{3}{2}, \lambda<-\sqrt{2}
\sqrt{\frac{-\mu^{2}+4 \mu-2 \omega-1}{2 \omega-3}}$, or

\item $\lambda\in\mathbb{R}, 1<\mu<2, \frac{\mu+1}{2}<\omega<\frac{1}{2}
\left( -\mu^{2}+4 \mu-1\right) $, or

\item $1<\mu<2, \omega=\frac{1}{2} \left( -\mu^{2}+4 \mu-1\right) , \lambda
<0$, or

\item $1<\mu<2, \frac{1}{2} \left( -\mu^{2}+4 \mu-1\right) <\omega<\frac{3}%
{2}, \lambda>\sqrt{2} \sqrt{\frac{-\mu^{2}+4 \mu-2 \omega-1}{2 \omega-3}}$, or

\item $\mu>2,  \frac{1}{2} \left( -\mu^{2}+4 \mu-1\right) <\omega<\frac{3}{2},
-\sqrt{2} \sqrt{\frac{-\mu^{2}+4 \mu-2 \omega-1}{2 \omega-3}}<\lambda<\sqrt{2}
\sqrt{\frac{-\mu^{2}+4 \mu-2 \omega-1}{2 \omega-3}}$, or

\item $\mu<1, \frac{1}{2} \left( -\mu^{2}+4 \mu-1\right) <\omega<\frac{\mu
+1}{2}, -\sqrt{2} \sqrt{\frac{-\mu^{2}+4 \mu-2 \omega-1}{2 \omega-3}}%
<\lambda<\sqrt{2} \sqrt{\frac{-\mu^{2}+4 \mu-2 \omega-1}{2 \omega-3}}$, or

\item $1<\mu<2, \frac{1}{2} \left( -\mu^{2}+4 \mu-1\right) <\omega<\frac{3}%
{2}, \lambda<-\sqrt{2} \sqrt{\frac{-\mu^{2}+4 \mu-2 \omega-1}{2 \omega-3}}$.
\end{enumerate}

In particular, it mimics a dust solution for

\begin{enumerate}
\item $\omega=\frac{3}{2}, \mu=\frac{5}{2}$, or

\item $\mu+1\neq2 \omega, 2 \omega\neq3, \lambda=-\frac{\sqrt{-2 \mu^{2}+9
\mu-6 \omega-1}}{\sqrt{2 \omega-3}}$, or

\item $\mu+1\neq2 \omega, 2 \omega\neq3, \lambda=\frac{\sqrt{-2 \mu^{2}+9
\mu-6 \omega-1}}{\sqrt{2 \omega-3}}$.
\end{enumerate}

\item $A_{3}: (x, y, z)= \left( \frac{-\lambda^{2}-3 \mu+3}{\sqrt{6} \left(
(\omega-1) \lambda^{2}+(\mu-1) \mu\right) }, \frac{\lambda(\mu-3 \omega
+3)}{\sqrt{6} \left( (\omega-1) \lambda^{2}+(\mu-1) \mu\right) }, -\frac
{(\mu-3 \omega+3) \left( (3 \omega-4) \lambda^{2}+3 (\mu-1)^{2}\right) }{6
\left( (\omega-1) \lambda^{2}+(\mu-1) \mu\right) ^{2}}\right) $ \newline with
eigenvalues \newline$\Big\{\frac{1}{2} \left( \frac{\lambda^{2}+3 \mu
-3}{(\omega-1) \lambda^{2}+(\mu-1) \mu}-3\right) $,\newline$\frac{1}{4} \left(
\frac{\lambda^{2}+3 \mu-3}{(\omega-1) \lambda^{2}+(\mu-1) \mu}-3-\frac
{\sqrt{(8 \mu-21 \omega+20) (2 \omega-3) \lambda^{2}-144 \omega^{2}+\mu(\mu(16
\mu-17)-174)+6 (34-7 \mu) \mu\omega+222 \omega-81} \sqrt{(3 \omega-4)
\lambda^{2}+3 (\mu-1)^{2}}}{\left( (\omega-1) \lambda^{2}+(\mu-1) \mu\right)
\sqrt{2 \omega-3}}\right) $,\newline$\frac{1}{4} \left( \frac{\lambda^{2}+3
\mu-3}{(\omega-1) \lambda^{2}+(\mu-1) \mu}-3+\frac{\sqrt{(8 \mu-21 \omega+20)
(2 \omega-3) \lambda^{2}-144 \omega^{2}+\mu(\mu(16 \mu-17)-174)+6 (34-7 \mu)
\mu\omega+222 \omega-81} \sqrt{(3 \omega-4) \lambda^{2}+3 (\mu-1)^{2}}}{\left(
(\omega-1) \lambda^{2}+(\mu-1) \mu\right)  \sqrt{2 \omega-3}}\right) \Big\}$.
In Fig. \ref{StP} we show a portion of the parameters space with
stability conditions of $P$ when it is a sink (left panel) and when it is a source (right panel).

\begin{figure}[ptb]
\textbf{ \includegraphics[width=0.4\textwidth]{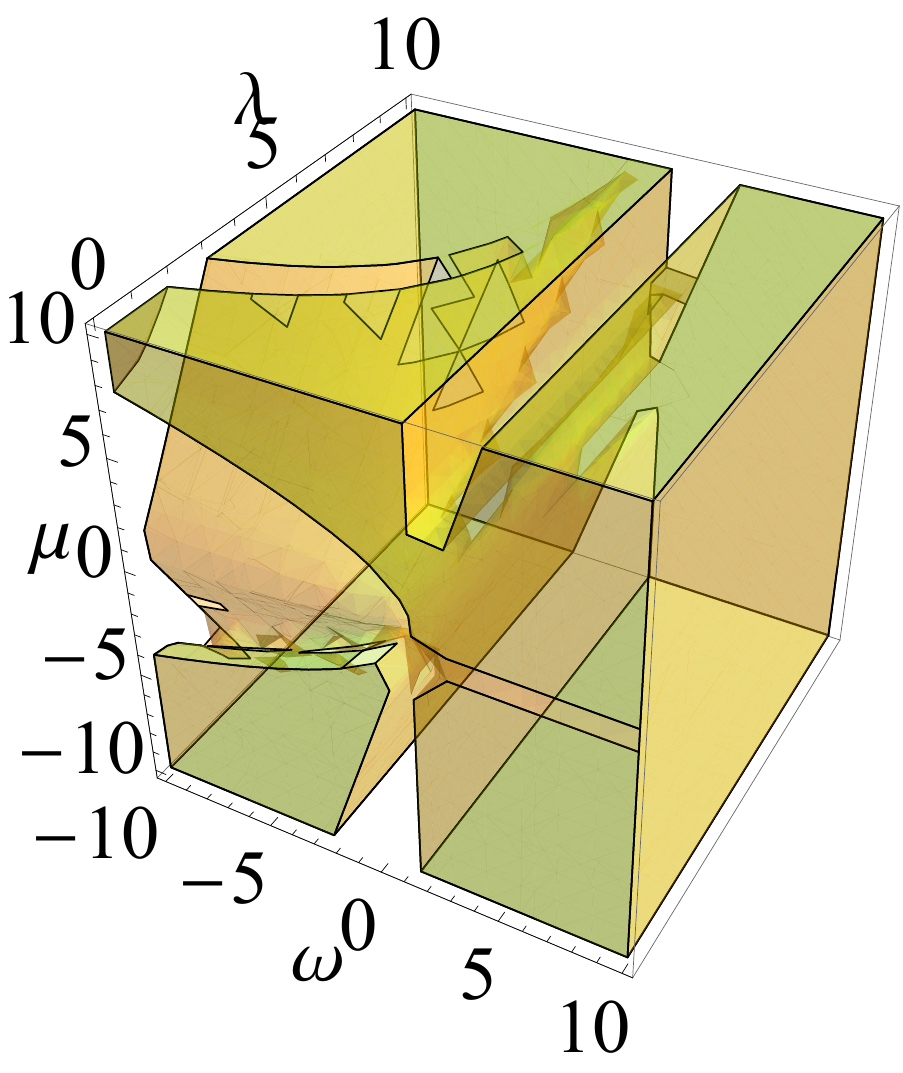}}\textbf{
\includegraphics[width=0.4\textwidth]{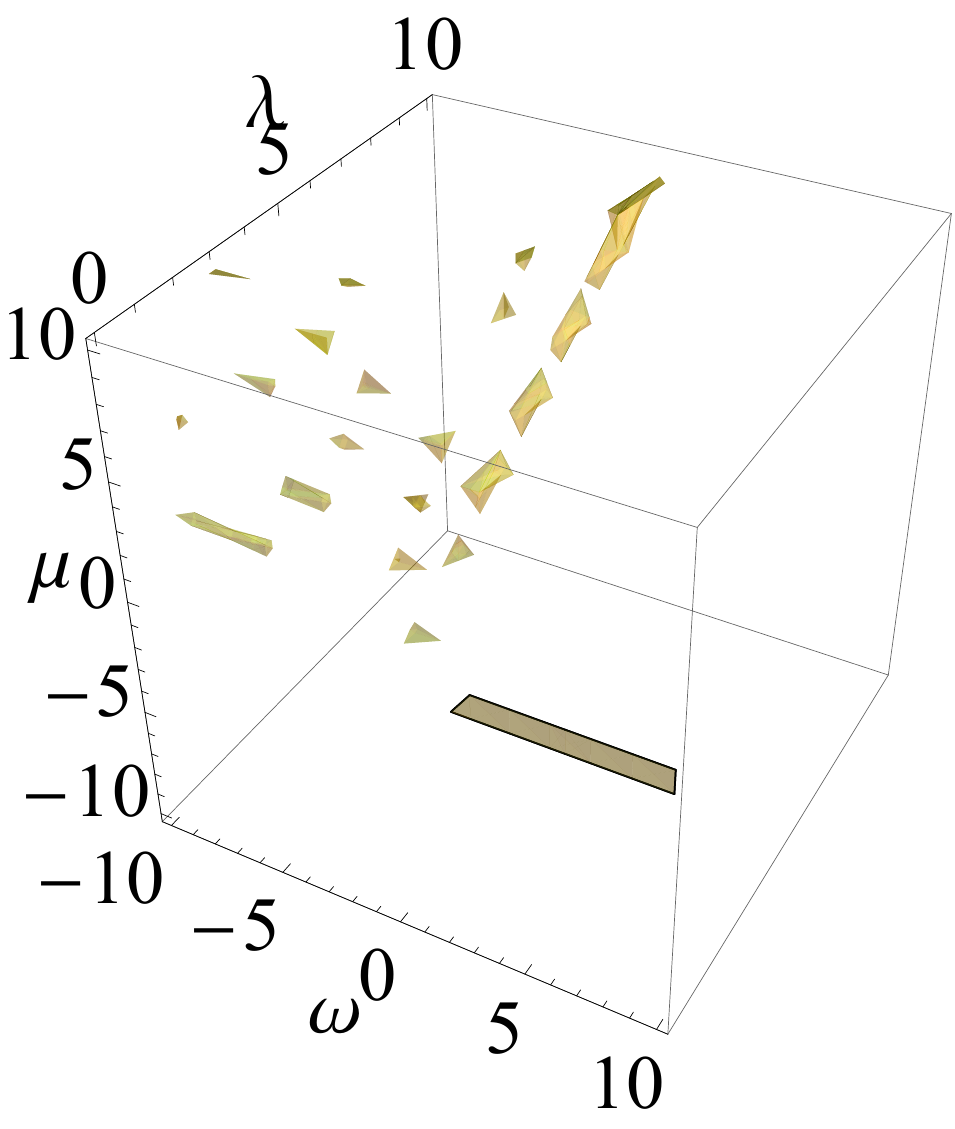}}\caption{Stability of $A_{3}$ when
it is a sink (left panel) and when it is a source (right panel). }%
\label{StP}%
\end{figure}

The cosmological observables are \newline$\Omega_{m}(A_{3})=\frac{\left(
\lambda^{2} (2 \omega-3)+\mu(2 \mu-7)+6 \omega-3\right)  \left( \lambda^{2} (3
\omega-4)+3 (\mu-1)^{2}\right) }{6 \left( \lambda^{2} (\omega-1)+(\mu-1)
\mu\right) ^{2}}$,  $w_{tot}(A_{3})=-\frac{\lambda^{2}+3 \mu-3}{3 \left(
\lambda^{2} (\omega-1)+(\mu-1) \mu\right) }$.

It represents an accelerating universe for

\begin{enumerate}
\item $(\omega\in\mathbb{R}, \lambda=0, 0<\mu<1$, or

\item $\omega\in\mathbb{R}, \lambda=0, 1<\mu<3$, or

\item $\lambda>\sqrt{3}, \frac{1}{3} \left( 3-\lambda^{2}\right) <\mu\leq0,
\frac{\lambda^{2}-\mu^{2}+\mu}{\lambda^{2}}<\omega<\frac{2 \lambda^{2}-\mu
^{2}+4 \mu-3}{\lambda^{2}}$, or

\item $\lambda>0, \mu\geq3, \frac{\lambda^{2}-\mu^{2}+\mu}{\lambda^{2}}%
<\omega<\frac{2 \lambda^{2}-\mu^{2}+4 \mu-3}{\lambda^{2}}$, or

\item $\lambda>0, \mu>1, \frac{\lambda^{2}-\mu^{2}+\mu}{\lambda^{2}}%
<\omega<\frac{2 \lambda^{2}-\mu^{2}+4 \mu-3}{\lambda^{2}}$, or

\item $\lambda>0, \mu=1, 1<\omega<2$, or

\item $\lambda<0,  \mu\geq3, \frac{\lambda^{2}-\mu^{2}+\mu}{\lambda^{2}%
}<\omega<\frac{2 \lambda^{2}-\mu^{2}+4 \mu-3}{\lambda^{2}}$, or

\item $\lambda<-\sqrt{3}, \frac{1}{3} \left( 3-\lambda^{2}\right) <\mu\leq0,
\frac{\lambda^{2}-\mu^{2}+\mu}{\lambda^{2}}<\omega<\frac{2 \lambda^{2}-\mu
^{2}+4 \mu-3}{\lambda^{2}}$, or

\item $\lambda<0, \mu>1, \frac{\lambda^{2}-\mu^{2}+\mu}{\lambda^{2}}%
<\omega<\frac{2 \lambda^{2}-\mu^{2}+4 \mu-3}{\lambda^{2}}$, or

\item $\lambda<0, \mu=1, 1<\omega<2$, or

\item $0<\lambda\leq\sqrt{3}, \frac{1}{3} \left( 3-\lambda^{2}\right) <\mu<1,
\frac{\lambda^{2}-\mu^{2}+\mu}{\lambda^{2}}<\omega<\frac{2 \lambda^{2}-\mu
^{2}+4 \mu-3}{\lambda^{2}}$, or

\item $\lambda>\sqrt{3}, 0<\mu<1, \frac{\lambda^{2}-\mu^{2}+\mu}{\lambda^{2}%
}<\omega<\frac{2 \lambda^{2}-\mu^{2}+4 \mu-3}{\lambda^{2}}$, or

\item $\lambda\leq-\sqrt{3}, \mu<\frac{1}{3} \left( 3-\lambda^{2}\right) ,
\frac{2 \lambda^{2}-\mu^{2}+4 \mu-3}{\lambda^{2}}<\omega<\frac{\lambda^{2}%
-\mu^{2}+\mu}{\lambda^{2}}$, or

\item $-\sqrt{3}<\lambda<0, \mu\leq0, \frac{2 \lambda^{2}-\mu^{2}+4 \mu
-3}{\lambda^{2}}<\omega<\frac{\lambda^{2}-\mu^{2}+\mu}{\lambda^{2}}$, or

\item $0<\lambda<\sqrt{3}, \mu\leq0, \frac{2 \lambda^{2}-\mu^{2}+4 \mu
-3}{\lambda^{2}}<\omega<\frac{\lambda^{2}-\mu^{2}+\mu}{\lambda^{2}}$, or

\item $\lambda=\sqrt{3}, \mu<0, \frac{1}{3} \left( -\mu^{2}+4 \mu+3\right)
<\omega<\frac{1}{3} \left( -\mu^{2}+\mu+3\right) $, or

\item $\lambda>\sqrt{3}, \mu<\frac{1}{3} \left( 3-\lambda^{2}\right) , \frac{2
\lambda^{2}-\mu^{2}+4 \mu-3}{\lambda^{2}}<\omega<\frac{\lambda^{2}-\mu^{2}%
+\mu}{\lambda^{2}}$, or

\item $\lambda\leq-\sqrt{3}, 0<\mu<1, \frac{\lambda^{2}-\mu^{2}+\mu}%
{\lambda^{2}}<\omega<\frac{2 \lambda^{2}-\mu^{2}+4 \mu-3}{\lambda^{2}}$, or

\item $-\sqrt{3}<\lambda<0, \frac{1}{3} \left( 3-\lambda^{2}\right) <\mu<1,
\frac{\lambda^{2}-\mu^{2}+\mu}{\lambda^{2}}<\omega<\frac{2 \lambda^{2}-\mu
^{2}+4 \mu-3}{\lambda^{2}}$, or

\item $-\sqrt{3}<\lambda<0, 0<\mu<\frac{1}{3} \left( 3-\lambda^{2}\right) ,
\frac{2 \lambda^{2}-\mu^{2}+4 \mu-3}{\lambda^{2}}<\omega<\frac{\lambda^{2}%
-\mu^{2}+\mu}{\lambda^{2}}$, or

\item $0<\lambda<\sqrt{3}, 0<\mu<\frac{1}{3} \left( 3-\lambda^{2}\right) ,
\frac{2 \lambda^{2}-\mu^{2}+4 \mu-3}{\lambda^{2}}<\omega<\frac{\lambda^{2}%
-\mu^{2}+\mu}{\lambda^{2}}$.
\end{enumerate}

It represents a de Sitter solution for

\begin{enumerate}
\item $\omega=\frac{4}{3}, \mu=1, \lambda\neq0$, or

\item $3 \omega-4\neq0, (\mu-1) (\mu-3 \omega+3)\neq0, \lambda=-\frac{\sqrt{3}
\sqrt{-(\mu-1)^{2}}}{\sqrt{3 \omega-4}}$, or

\item $3 \omega-4\neq0, (\mu-1) (\mu-3 \omega+3)\neq0, \lambda=\frac{\sqrt{3}
\sqrt{-(\mu-1)^{2}}}{\sqrt{3 \omega-4}}$.
\end{enumerate}

It represents a decelerating universe for

\begin{enumerate}
\item $(\omega\in\mathbb{R}, \lambda=0, \mu>3$, or

\item $\omega\in\mathbb{R}, \lambda=0, \mu<0$, or

\item $\lambda>\sqrt{3}, \frac{1}{3} \left( 3-\lambda^{2}\right) <\mu<0,
\omega>\frac{2 \lambda^{2}-\mu^{2}+4 \mu-3}{\lambda^{2}}$, or

\item $\lambda>0, \mu>3, \omega>\frac{2 \lambda^{2}-\mu^{2}+4 \mu-3}%
{\lambda^{2}}$, or

\item $\lambda<0, \mu>3, \omega>\frac{2 \lambda^{2}-\mu^{2}+4 \mu-3}%
{\lambda^{2}}$, or

\item $\lambda>0, \mu\geq1, \omega>\frac{2 \lambda^{2}-\mu^{2}+4 \mu
-3}{\lambda^{2}}$, or

\item $\lambda<-\sqrt{3}, \frac{1}{3} \left( 3-\lambda^{2}\right) <\mu<0,
\omega>\frac{2 \lambda^{2}-\mu^{2}+4 \mu-3}{\lambda^{2}}$, or

\item $\lambda<0, \mu\geq1, \omega>\frac{2 \lambda^{2}-\mu^{2}+4 \mu
-3}{\lambda^{2}}$, or

\item $\lambda<-\sqrt{3}, \mu=\frac{1}{3} \left( 3-\lambda^{2}\right) ,
\omega>\frac{\lambda^{2}-\mu^{2}+\mu}{\lambda^{2}}$, or

\item $\lambda>\sqrt{3}, \mu=\frac{1}{3} \left( 3-\lambda^{2}\right) ,
\omega>\frac{\lambda^{2}-\mu^{2}+\mu}{\lambda^{2}}$, or

\item $\lambda>\sqrt{3}, \frac{1}{3} \left( 3-\lambda^{2}\right) <\mu<0,
\omega<\frac{\lambda^{2}-\mu^{2}+\mu}{\lambda^{2}}$, or

\item $\lambda>0, \mu>3, \omega<\frac{\lambda^{2}-\mu^{2}+\mu}{\lambda^{2}}$, or

\item $\lambda<0, \mu>3, \omega<\frac{\lambda^{2}-\mu^{2}+\mu}{\lambda^{2}}$, or

\item $\lambda>0, \mu\geq1, \omega<\frac{\lambda^{2}-\mu^{2}+\mu}{\lambda^{2}%
}$, or

\item $\lambda<-\sqrt{3},  \frac{1}{3} \left( 3-\lambda^{2}\right) <\mu<0,
\omega<\frac{\lambda^{2}-\mu^{2}+\mu}{\lambda^{2}}$, or

\item $\lambda<-\sqrt{3}, \mu=\frac{1}{3} \left( 3-\lambda^{2}\right) ,
\omega<\frac{\lambda^{2}-\mu^{2}+\mu}{\lambda^{2}}$, or

\item $\lambda>\sqrt{3}, \mu=\frac{1}{3} \left( 3-\lambda^{2}\right) ,
\omega<\frac{\lambda^{2}-\mu^{2}+\mu}{\lambda^{2}}$, or

\item $\lambda<0, \mu\geq1, \omega<\frac{\lambda^{2}-\mu^{2}+\mu}{\lambda^{2}%
}$, or

\item $0<\lambda\leq\sqrt{3}, \frac{1}{3} \left( 3-\lambda^{2}\right) <\mu<1,
\omega>\frac{2 \lambda^{2}-\mu^{2}+4 \mu-3}{\lambda^{2}}$, or

\item $\lambda>\sqrt{3}, 0\leq\mu<1, \omega>\frac{2 \lambda^{2}-\mu^{2}+4
\mu-3}{\lambda^{2}}$, or

\item $\lambda<-\sqrt{3}, 0\leq\mu<1, \omega>\frac{2 \lambda^{2}-\mu^{2}+4
\mu-3}{\lambda^{2}}$, or

\item $\lambda=-\sqrt{3},  0<\mu<1, \omega>\frac{1}{3} \left( -\mu^{2}+4
\mu+3\right) $, or

\item $-\sqrt{3}<\lambda<0, \frac{1}{3} \left( 3-\lambda^{2}\right) <\mu<1,
\omega>\frac{2 \lambda^{2}-\mu^{2}+4 \mu-3}{\lambda^{2}}$, or

\item $\lambda\leq-\sqrt{3}, \mu<\frac{1}{3} \left( 3-\lambda^{2}\right) ,
\omega>\frac{\lambda^{2}-\mu^{2}+\mu}{\lambda^{2}}$, or

\item $-\sqrt{3}<\lambda<0, \mu<0, \omega>\frac{\lambda^{2}-\mu^{2}+\mu
}{\lambda^{2}}$, or

\item $0<\lambda\leq\sqrt{3}, \mu<0, \omega>\frac{\lambda^{2}-\mu^{2}+\mu
}{\lambda^{2}}$, or

\item $\lambda>\sqrt{3}, \mu<\frac{1}{3} \left( 3-\lambda^{2}\right) ,
\omega>\frac{\lambda^{2}-\mu^{2}+\mu}{\lambda^{2}}$, or

\item $-\sqrt{3}\leq\lambda<0, \mu=\frac{1}{3} \left( 3-\lambda^{2}\right) ,
\omega>\frac{\lambda^{2}-\mu^{2}+\mu}{\lambda^{2}}$, or

\item $0<\lambda\leq\sqrt{3}, \mu=\frac{1}{3} \left( 3-\lambda^{2}\right) ,
\omega>\frac{\lambda^{2}-\mu^{2}+\mu}{\lambda^{2}}$, or

\item $0<\lambda\leq\sqrt{3}, \frac{1}{3} \left( 3-\lambda^{2}\right) <\mu<1,
\omega<\frac{\lambda^{2}-\mu^{2}+\mu}{\lambda^{2}}$, or

\item $\lambda>\sqrt{3}, 0\leq\mu<1, \omega<\frac{\lambda^{2}-\mu^{2}+\mu
}{\lambda^{2}}$, or

\item $\lambda\leq-\sqrt{3}, \mu<\frac{1}{3}  \left( 3-\lambda^{2}\right) ,
\omega<\frac{2 \lambda^{2}-\mu^{2}+4 \mu-3}{\lambda^{2}}$, or

\item $-\sqrt{3}<\lambda<0, \mu<0, \omega<\frac{2 \lambda^{2}-\mu^{2}+4 \mu
-3}{\lambda^{2}}$, or

\item $0<\lambda\leq\sqrt{3}, \mu<0, \omega<\frac{2 \lambda^{2}-\mu^{2}+4
\mu-3}{\lambda^{2}}$, or

\item $\lambda>\sqrt{3}, \mu<\frac{1}{3} \left( 3-\lambda^{2}\right) ,
\omega<\frac{2 \lambda^{2}-\mu^{2}+4 \mu-3}{\lambda^{2}}$, or

\item $\lambda<-\sqrt{3}, 0\leq\mu<1, \omega<\frac{\lambda^{2}-\mu^{2}+\mu
}{\lambda^{2}}$, or

\item $\lambda=-\sqrt{3}, 0<\mu<1, \omega<\frac{1}{3} \left( -\mu^{2}%
+\mu+3\right) $, or

\item $-\sqrt{3}<\lambda<0, \frac{1}{3} \left( 3-\lambda^{2}\right) <\mu<1,
\omega<\frac{\lambda^{2}-\mu^{2}+\mu}{\lambda^{2}}$, or

\item $-\sqrt{3}\leq\lambda<0, \mu=\frac{1}{3} \left( 3-\lambda^{2}\right) ,
\omega<\frac{\lambda^{2}-\mu^{2}+\mu}{\lambda^{2}}$, or

\item $0<\lambda\leq\sqrt{3}, \mu=\frac{1}{3} \left( 3-\lambda^{2}\right) ,
\omega<\frac{\lambda^{2}-\mu^{2}+\mu}{\lambda^{2}}$, or

\item $-\sqrt{3}<\lambda<0, 0\leq\mu<\frac{1}{3} \left( 3-\lambda^{2}\right) ,
\omega>\frac{\lambda^{2}-\mu^{2}+\mu}{\lambda^{2}}$, or

\item $0<\lambda<\sqrt{3}, 0\leq\mu<\frac{1}{3} \left( 3-\lambda^{2}\right) ,
\omega>\frac{\lambda^{2}-\mu^{2}+\mu}{\lambda^{2}}$, or

\item $-\sqrt{3}<\lambda<0, 0\leq\mu<\frac{1}{3} \left( 3-\lambda^{2}\right) ,
\omega<\frac{2 \lambda^{2}-\mu^{2}+4 \mu-3}{\lambda^{2}}$, or

\item $0<\lambda<\sqrt{3}, 0\leq\mu<\frac{1}{3} \left( 3-\lambda^{2}\right) ,
\omega<\frac{2 \lambda^{2}-\mu^{2}+4 \mu-3}{\lambda^{2}}$.
\end{enumerate}

It represents a dust solution for

\begin{enumerate}

\item $(\mu-1) (\mu-3 \omega+3)\neq0, \lambda=-\sqrt{3} \sqrt{1-\mu}$, or

\item $(\mu-1) (\mu-3 \omega+3)\neq0, \lambda=\sqrt{3} \sqrt{1-\mu}$.
\end{enumerate}

\item $A_{4}: (x,y,z)=\left( \frac{1}{\sqrt{6} (\omega-1)}, 0, 0\right) $,
with eigenvalues $\left\{ \frac{1}{2} \left( \frac{1}{\omega-1}-3\right)
,\frac{1}{2}  \left( \frac{1}{\omega-1}-3\right) ,3-\frac{\mu}{\omega
-1}\right\} $. It is a sink for

\begin{enumerate}

\item $\omega<1, \mu<3 (\omega-1)$, or

\item $\omega>\frac{4}{3}, \mu>3 (\omega-1)$.
\end{enumerate}

It is a source for $1<\omega<\frac{4}{3}, \mu<3 (\omega-1)$.

The cosmological observables are $\Omega_{m}(A_{4})=\frac{(2 \omega-3) (3
\omega-4)}{6 (\omega-1)^{2}}$,  $w_{tot}(A_{4})=-\frac{1}{3 (\omega-1)}$. It
is associated with an accelerating solution for $1<\omega<2$. It represents a de
Sitter solution for $\omega=\frac{4}{3}$. Is corresponds to a decelerated
solution for $\omega<1$ por $\omega>2$. For $\omega=0$, it corresponds to a
radiation dominated solution.

\item For $\omega=\frac{3 \lambda^{2}+(5-2 \mu) \mu-3}{2 \lambda^{2}}$, we
have the equilibrium point $A_{5}: (x, y, z)=  \left( -\sqrt{\frac{2}{3}},
\frac{2 \mu-3}{\sqrt{6} \lambda}, \frac{3-2 \mu}{6 \lambda^{2}}\right) $, with
eigenvalues  \newline$\Big\{-\frac{1}{2}, -\frac{1}{4} \left( 1+\frac
{\sqrt{(17 \mu-25) \lambda^{2}+24 (\mu-1) (2 \mu-3)}}{\lambda\sqrt{\mu-1}%
}\right) , -\frac{1}{4} \left( 1-\frac{\sqrt{(17 \mu-25) \lambda^{2}+24
(\mu-1) (2 \mu-3)}}{\lambda\sqrt{\mu-1}}\right) \Big\}$. It is a sink for

\begin{enumerate}
\item $\lambda<0, \frac{1}{96} \left( -17 \lambda^{2}-\sqrt{289 \lambda
^{4}+720 \lambda^{2}+576}+120\right) \leq\mu<\frac{1}{3} \left( 3-\lambda
^{2}\right) $, or

\item $\lambda<0, \frac{1}{96} \left( -17 \lambda^{2}+\sqrt{289 \lambda
^{4}+720 \lambda^{2}+576}+120\right) \leq\mu<\frac{3}{2}$, or \

\item $\lambda>0, \frac{1}{96} \left( -17  \lambda^{2}-\sqrt{289 \lambda
^{4}+720 \lambda^{2}+576}+120\right) \leq\mu<\frac{1}{3} \left( 3-\lambda
^{2}\right) $, or

\item $\lambda>0, \frac{1}{96} \left( -17 \lambda^{2}+\sqrt{289 \lambda
^{4}+720 \lambda^{2}+576}+120\right) \leq\mu<\frac{3}{2}$.
\end{enumerate}

It is a saddle point for

\begin{enumerate}
\item $\lambda<0, \frac{1}{3} \left( 3-\lambda^{2}\right) <\mu<1$, or

\item $\lambda<0, \mu>\frac{3}{2}$, or

\item $\lambda>0, \frac{1}{3} \left( 3-\lambda^{2}\right) <\mu<1$, or

\item $\lambda>0, \mu>\frac{3}{2}$.
\end{enumerate}

It is nonhyperbolic otherwise.

The cosmological observables are $\Omega_{m}(A_{5})=-\frac{2 \mu-3}{3
\lambda^{2}}$,  $w_{tot}(A_{5})=-\frac{2}{3}$. It is always accelerating.
\end{enumerate}

\section{Conclusions}

\label{sec5}

Let us now summarize the main findings of this article. In this work we considered a spatially flat FLRW cosmological model of two
scalar fields in the Jordan frame. Specifically, one of
the fields is the Brans-Dicke field which is nonminimally coupled to gravity, and the
second scalar field is minimally coupled to gravity. The equivalent model in Einstein frame is that of two minimally coupled
scalar fields with an interaction only in the potential part of the fields which effectively represents a quintom cosmological model.

We focused on the evolution of the physical quantities of the
universe by studying the stationary points of the field equations. Such
analysis is important to understand the cosmological viability of the
cosmological model, and also we try to conclude about the differences among the
Einstein and the Jordan frames.

In the vacuum scenario the field equations admit four stationary points. Two
of the points actually, namely $A_{\pm}$, describe two families of stationary
points in the phase space of the field equations. The critical points can
describe ideal gas solutions, or de Sitter universes for various values of the
free parameters.

When we introduce a matter source in the field equations the dynamical
analysis differs and in order to perform a complete study we study separate
cases where the Brans-Dicke parameter $\omega$ is $\omega=1,~\omega=0$ and
$\omega=arbitrary.~$ For $\omega=1$, the dynamical system admits four critical
points, when $\omega=0$, we again find four critical points while for
arbitrary $\omega$ the dynamical system admits six critical points. In all the
cases the physical quantities and the stability of the solutions at the
critical points are discussed.

From our analysis we observe that there are many differences between our
model and the Brans-Dicke theory with one-scalar field. There are new exact
solutions which can describe additional cosmological eras in the evolution of
the universe. Consequently, the cosmological evolution differs from the
two-scalar field models defined in the Einstein frame.

The cosmological model of our consideration can describe the generic
cosmological history, hence, in a future work, we plan to investigate the
cosmological viability of the model by using observational constraints.

\section*{Acknowledgments}

A.P. thanks Dr. Alex Giacomini and the Universidad Austral de Chile for the
hospitality provided while this work was performed. G. L. was funded by
Comisi\'{o}n Nacional de Investigaci\'{o}n Cient\'{\i}fica y Tecnol\'{o}gica
(CONICYT) through FONDECYT Iniciaci\'{o}n grant no. 11180126 and by
Vicerrector\'{\i}a de Investigaci\'{o}n y Desarrollo Tecnol\'{o}gico at
Universidad Cat\'{o}lica del Norte. SP gratefully acknowledges the support through the Mathematical Research Impact-Centric Support Scheme  (MATRICS), File No. MTR/2018/000940, given by the Science and Engineering Research Board
(SERB), Govt. of India.

\end{document}